\documentclass[12pt]{article}

\usepackage{epsfig}
\usepackage{cite}
\usepackage{amsmath, amssymb, amsfonts}
\usepackage{color}
\usepackage{latexsym}
\usepackage{graphicx}
\usepackage[colorlinks,bookmarks]{hyperref}
\hypersetup{pdfpagemode=UseNone, pdfstartview=FitH, linkcolor=blue,
            citecolor=red, urlcolor=blue}

\def\be{\begin{equation}}
\def\ee{\end{equation}}
\def\ba{\begin{eqnarray}}
\def\ea{\end{eqnarray}}

\def\bdm{\begin{displaymath}}
\def\edm{\end{displaymath}}
\def\la{~\mbox{\raisebox{-.6ex}{$\stackrel{<}{\sim}$}}~}
\def\ga{~\mbox{\raisebox{-.6ex}{$\stackrel{>}{\sim}$}}~}
\def\bq{\begin{quote}}
\def\eq{\end{quote}}

 at 10truept
 at 14truept

\newcommand{\p}{\partial}

\newcommand{\Mpl}{M_{\mathrm{Pl}}}
\newcommand{\mps}{M_{\mathrm{Pl}}^2}

\newcommand{\bea}{\begin{eqnarray}}
\newcommand{\eea}{\end{eqnarray}}

\newcommand{\bi}{\begin{itemize}}
\newcommand{\ei}{\end{itemize}}

\newcommand{\beq}{\begin{equation}}
\newcommand{\eeq}{\end{equation}}
\newcommand{\beqa}{\begin{eqnarray}}
\newcommand{\eeqa}{\end{eqnarray}}
\newcommand{\mpl}{\Mpl}

\def\la{~\mbox{\raisebox{-.6ex}{$\stackrel{<}{\sim}$}}~}
\def\ga{~\mbox{\raisebox{-.6ex}{$\stackrel{>}{\sim}$}}~}

\def\ltap{\ \raise.3ex\hbox{$<$\kern-.75em\lower1ex\hbox{$\sim$}}\ }
\def\gtap{\ \raise.3ex\hbox{$>$\kern-.75em\lower1ex\hbox{$\sim$}}\ }
\def\gl{\ \raise.5ex\hbox{$>$}\kern-.8em\lower.5ex\hbox{$<$}\ }
\def\roughly#1{\raise.3ex\hbox{$#1$\kern-.75em\lower1ex\hbox{$\sim$}}}

\begin{document}

\thispagestyle{empty}
\begin{flushright}
February 2022 \\
\end{flushright}
\vspace*{1.5cm}
\begin{center}

{\Large \bf  Pancosmic Relativity and Nature's Hierarchies}  
\vskip.3cm

\vspace*{1.15cm} {\large 
Nemanja Kaloper$^{a, }$\footnote{\tt
kaloper@physics.ucdavis.edu} 
}\\
\vspace{.5cm}
{\em $^a$QMAP, Department of Physics and Astronomy, University of
California}\\
\vspace{.05cm}
{\em Davis, CA 95616, USA}\\

\vspace{1.65cm} ABSTRACT
\end{center}
We define `third derivative' General Relativity, by promoting the integration measure 
in Einstein-Hilbert action to be an arbitrary $4$-form field strength. We project out its local 
fluctuations by coupling it to another $4$-form field strength. This ensures that the 
gravitational sector contains only the usual massless helicity-2 propagating modes. 
Adding the charges to these $4$-forms allows for discrete variations
of the coupling parameters of conventional General Relativity: $G_N, \Lambda, H_0$,  
and even $\langle {\tt Higgs }\rangle$ are all variables which can change by jumps. 
Hence de Sitter is unstable to membrane nucleation. Using this instability 
we explain how the cosmological constant problem can be solved. 
The scenario utilizes the idea behind the irrational axion, but instead of an axion it requires one
more $4$-form field strength and corresponding charged membranes. When the membrane charges 
satisfy the constraint  $\frac{2\kappa_{\tt eff}^2 \kappa^2 |{\cal Q}_i|}{3{\cal T}^2_i} < 1$, the 
theory which ensues exponentially favors a huge hierarchy 
$\Lambda/\mpl^4 \ll 1$ instead of $\Lambda/\mpl^4 \simeq 1$. The discharges 
produce the distribution of the values of 
$\Lambda$ described by the saddle point approximation of the Euclidean path integral. 

\vfill \setcounter{page}{0} \setcounter{footnote}{0}

\vspace{1cm}
\newpage

\vspace{1cm}

\section{Preface}
 
The standard formulation of General Relativity employs diffeomorphism 
invariant second-order partial differential equations, first formulated in 
\cite{Hilbert:1915tx,Einstein:1915ca}\footnote{An interesting historic perspective is 
offered in \cite{Todorov:2005rh}.}. 
Allowing only two derivatives, demanding
diffeomorphism invariance, and restricting dynamical degrees of freedom
to only metric fluctuations is very constraining. Together, these requirements single out General Relativity as a
unique covariant, massless spin-2, second derivative theory \cite{Lovelock:1971yv,Lovelock:1972vz}. 
It has dimensional constants as universal gravitational couplings: 
Newton's constant $G_N = \frac{1}{8\pi \mps}$ and the cosmological constant 
$\Lambda$. In addition, the matter sector couplings, dimensional (e.g. masses) and
dimensionless (e.g. charges and Yukawa couplings),  are determined by flat space
physics, irrespective of gravity. In the minimal approach these parameters are spacetime constants, 
which could care less about whether gravity exists or not. 

The observed great numerical variance between the values
of the gravitational dimensional parameters, and between them and the matter sector masses, however, 
remains mysterious. Attempts to decrypt these mysteries and  
the curiosity to see if General Relativity might be consistently generalized  
have produced a vast diversity of extended theories of gravity which
typically include new degrees of freedom. 

Such models can often be understood as higher derivative theories, since 
higher derivative terms introduce new propagating modes (see e.g. \cite{elsgolts}). 
A tricky aspect of these `generic' modifications of General Relativity is that they  
lead to new long range forces and/or lower UV cutoffs, which can be tightly constrained.
Furthermore, the origin of fundamental scales remains just as mysterious. 

In this article, we will define what may be technically the simplest possible modification of General
Relativity, that nevertheless does extend the phase space of the theory dramatically. 
There are {\it no new local degrees
of freedom}. Hence no new forces arise, and no new perturbative cutoff scales appear. 
Yet the theory predicts variations of Newton's constant, the cosmological constant,
and even the matter sector couplings, throughout spacetime - albeit {\it discontinuously} and 
{\it discretely}. These variations affect cosmology of (extremely large) ``local" regions, and more generally local
particle physics, and may be a link in understanding the origin of the 
observed puzzling hierarchies of particle physics. 

In a sense our formulation of Pancosmic Relativity -- i.e. Pancosmic General Relativity -- 
is reminiscent of Coleman's wormhole approach \cite{Coleman:1988tj}. However we 
work in the semiclassical limit where the mediators of the transitions altering the local
values of theory's couplings do not require direct deployment of full blown nonperturbative
quantum gravity. 

Our key new idea is that the action for General Relativity, originally given by Hilbert \cite{Hilbert:1915tx}, 
can be generalized by replacing the covariant integration measure $4$-form $\sqrt{g} \, d^4 x$ 
by a more general $4$-form ${\cal F} = d{\cal A}$, where ${\cal A}$ is an arbitrary $3$-form potential. We 
preempt any new local degrees of freedom in the measure $4$-form ${\cal F}$ by introducing another 
$4$-form ${\cal G}=d{\cal B}$, which we couple to ${\cal F}$ via the action 
$\int {\cal F} \, \frac{\epsilon^{\mu\nu\lambda\sigma}}{\sqrt{g}} {\cal G}_{\mu\nu\lambda\sigma}$. 
This enforces the conservation law for Planck scale $\propto {\cal F}$, promoting it 
into an integration constant. The total action 
also yields another integration constant, via the `conserved dual flux' coupled to ${\cal G}$ 
\cite{Kaloper:2015jra},  which is degenerate with the cosmological constant. 

Thus our, conspicuously third-derivative, Pancosmic General Relativity 
generalizes the so-called ``unimodular" two derivative formulation of General Relativity\footnote{Unimodular formulation of General Relativity simply means
that  the cosmological constant term in the equations of motion contains an additive integration constant which serves as a counterterm for renormalizing
the physical cosmological constant which sources the geometry os spacetime. The properly formulated theory is otherwise equivalent to conventional treatment
of General Relativity \cite{Henneaux:1989zc,Fiol:2008vk}.}  
\cite{Einstein:1919gv,Anderson:1971pn,Aurilia:1980xj,Duff:1980qv,Buchmuller:1988wx,Buchmuller:1988yn,Henneaux:1989zc,Ng:1990xz,Fiol:2008vk}. 
Further generalizations, where the matter sector parameters also get contributions from integration constants, can be 
obtained by allowing the matter sector integral measure\footnote{Alternatives to minimal measure in the action were noted in 
\cite{Guendelman:1996qy,Gronwald:1997ei,Wilczek:1998ea,Kaloper:2015jra,DAmico:2017ngr,Kaloper:2018kma,Benisty:2019znu,Lee:2019efp,Cribiori:2020wch}.}  
to also be controlled, at least in part, either by
${\cal F}$, or by additional $4$-form field strengths like ${\cal F}$. As it will turn out,
such more general theories are more easily formulated using the magnetic duals of the new $4$-forms. 

We will focus on the minimal and `conformal' theories, in dual variables. The reason we focus on these two special cases is the
robustness of their form to the perturbatively generated corrections from matter QFT, to arbitrary order in the loop expansion. 
For other $4$-form/matter couplings, the
quantitative results would depend in principle on the loop expansion truncation, causing issues with 
calculational control. In the general case, the form of the $4$-form/matter couplings could change from loop to loop. 
The minimal and `conformal' theories, however, avoid this complication. 
Although the minimal theory is the simplest-looking one, the conformal theory is actually more straightforward 
to work with since we can devise a simple proof that it can avoid transitions which summon ghosts. 

That the modifications of the measure promote the parameters of the theory into integration
constants  follows from the gauge symmetry of the $4$-forms, invariant under 
${\cal A} \rightarrow {\cal A} + d \omega_A$. 
Thus the summands in the Lagrangian multiplying those specific $4$-forms are the associated 
conserved fluxes \cite{Kaloper:2015jra}. Our observation points the way how to 
add extra dynamics to the theory without including
new local fields. We introduce objects charged under the $4$-forms ${\cal F}$ and ${\cal G}$, which
are membranes with units of charge ${\cal Q}_i$ and tension ${\cal T}_i$. Membranes can spontaneously 
nucleate quantum-mechanically, changing the values of the conjugate variables to 
${\cal F}$ and ${\cal G}$ inside the bubbles of space surrounded
by membranes. As a result in the interior of the bubbles the effective strength 
of gravity and the value of cosmological 
constant, and also the values of couplings and scales of the local matter theory, jump relative to the outside. 

It follows that an outcome of a sequence of bubble nucleations 
are systems of nested expanding bubbles scanning over a range of values of 
parameters. These configurations essentially realize 
a toy model of the multiverse of eternal inflation \cite{Linde:2015edk} already at the
semiclassical level of Pancosmic General Relativity. This may provide a very simple framework for describing
eternal inflation in the semiclassical limit, and in fact could be a toy model which incorporates 
leading order effects of quantum gravity at very large scales and low energies, specifically the effects of
spacetime foam and wormholes \cite{Wheeler:1955zz,Hawking:1978pog,Coleman:1988tj}.

Examples of where such effects may play an important role include cosmological mechanisms to address 
various hierarchies observed in nature (using discretely varying parameters as 
in \cite{Arvanitaki:2016xds,Giudice:2019iwl,Kaloper:2019xfj}). We will discuss in detail the cosmological
constant problem \cite{Zeldovich:1967gd,Wilczek:1983as,Weinberg:1987dv} in this article, and show how it can be solved. 
In a shorter companion paper \cite{Kaloper:2022oqv}, we have 
provided a {\it resum\'e} of the cosmological constant problem and its solution in this approach. 
To solve the problem we will include one more $4$-form, which, on shell, also contributes only to the 
cosmological constant. When the charges of the two $4$-forms have an 
irrational ratio, since their contributions to the effective cosmological constant are degenerate, 
we can invoke a variant of the discretuum of the irrational axion \cite{Banks:1991mb} and use the instability of
the positive cosmological constant to membrane discharges to show that any positive cosmological constant
eventually decays to smaller values. When the charges satisfy $\frac{2\kappa_{\tt eff}^2 \kappa^2 |{\cal Q}_i|}{3{\cal T}^2_i} < 1$ 
(where $\kappa^2$ are linked to the local value of Planck scale),
the membrane discharges are restricted to a subset of nucleation processes, for which the instability 
invariably stops when $\Lambda \rightarrow 0^+$ since their bounce actions have a pole
at $\Lambda \rightarrow 0^+$. In the leading order, the outcome of such dynamical evolution effectively realizes the 
Hawking-Baum distribution of terminal values of $\Lambda$ \cite{Hawking:1981gd,Baum:1983iwr,Hawking:1984hk,Abbott:1984qf}, 
controlled by the semiclassical, saddle point Euclidean action on the background.

We find that when combined together, these ingredients exponentially favor vacua with 
\be
\frac{\Lambda_{total}}{\mpl^4} \rightarrow 0 \ll 1\, .
\ee
A very mild `weak anthropic' determination of Newton's constant, which needs to be near the observed value of 
$G_N = \frac{1}{8\pi \mps} \simeq 10^{-38} \, ({\rm GeV})^{-2}$ 
to ensure that Earth is neither charred nor frozen, is the only cameo of the anthropic reasoning. 
As a result the Pancosmic General Relativity dynamics  
reduces the cosmological constant problem simply to finding the answer to 
the ``{\it Why now?"} question. In other words, we find that effectively the cosmological constant 
is as close to zero as can be, and the question which remains is what is the driver of the current 
epoch of cosmological acceleration. We will comment on how this might be achieved. In the summary, we will also 
briefly comment on the prospects for inflation.

\subsection{Comparison with Past Work}

The use of $4$-forms and their fluxes to formulate contributions to the cosmological constant 
\cite{Aurilia:1980xj,Duff:1980qv,Henneaux:1989zc,Hawking:1981gd,Hawking:1984hk}, 
and screen and cancel the sum total \cite{Brown:1987dd,Brown:1988kg,Duncan:1989ug,Bousso:2000xa,Feng:2000if} has a 
substantial past history as evidenced by the references listed here. We feel that it will be beneficial to a reader if we stress
the main differences between those approaches and the present work. 

While we use the $4$-forms and their fluxes
and charges to reduce the cosmological constant, and also change in a similar manner the Planck scale and
possibly other dimensional parameters in Nature (the latter being mostly ignored in the previous approaches),
we have discovered a very different formulation of the theory where the contributions of the fluxes to the
cosmological constant come as bilinear terms. Those terms in general can be modified by adding higher powers, but
as long as one of the factors in the bilinar is the effective Planck scale -- as we find here -- the additional powers of 
the flux, such as $\propto F^2$ terms common in the literature, are subleading. Thus in our case the 
contribution to the net cosmological constant involves only first powers of the individual fluxes. 

This has dramatic 
consequences for the dynamics. In particular the membrane junction conditions are completely altered from
those derived by Brown and Teitelboim \cite{Brown:1987dd,Brown:1988kg} (which are used by other 
approaches in the literature). Those conditions control which types of instantons can mediate the membrane 
nucleation processes, that in turn control the cosmological constant decay rates. In particular, when the tension
is large, such that $\frac{2\kappa_{\tt eff}^2 \kappa^2 |{\cal Q}_i|}{3{\cal T}^2_i} < 1$, the only possible instanton
transitions are two: one mediating $dS \rightarrow dS$, and one mediating $dS \rightarrow AdS$. Further,
since in these two cases the relevant instantons have bounce actions which feature a pole at $\Lambda \rightarrow 0^+$,
the terminal Minkowski space is absolutely stable, and a quantum dynamical attractor of the evolution. Thus for any
initial value of the cosmological constant in the universe the evolution will bring it to $\Lambda \rightarrow 0^+$, and
stop there. 

This \underbar{\it does not} happen in any of the previously studied cases which have $\propto F^2$ 
terms as dominant fluxes contributing to the cosmological constant without severe fine tunings. When
$\propto F^2$ dominate, other instantons which are dominated by charge contributions instead of tensions will occur,
which have a bounce action without the pole at $\Lambda \rightarrow 0^+$, and which will simply run through
$\Lambda = 0$ and allow the system to evolve to $\Lambda < 0$. For those approaches, one must use
anthropic selection to pick a small positive terminal $\Lambda$. In our case, those instantons are
robustly excluded by the altered junction conditions when the tension is sufficiently big, 
the evolution relaxes $\Lambda$ to $0^+$ by quantum Brownian drift, 
and it stops at $\Lambda \rightarrow 0^+$, favoring tiny cosmological constant 
without any need for anthropics. We carefully and meticulously go over the details in the rest of the manuscript 
showing precisely what it takes to set a system which ensures such new evolution of $\Lambda$.

Again, one might worry that the bilinear dependence of the theory on the flux variables, as opposed to other powers, 
is special, even fine tuned. That is not the case. Even if the higher order corrections are included,
since their weighing is by $\Mpl$, the bilinear terms remain dominant for sub-Planckian fluxes and the same behavior as in the pure bilinear case
remains. Further, the higher-order corrections could come in with different coefficients for the two flux sectors. This may induce
mutually irrational variation of fluxes even if the actual ratio of charges were rational.

Our mechanism also evades naturally the venerated Weinberg's 
no-go theorem \cite{Weinberg:1987dv} for the adjustment of the cosmological
constant, by exploiting loopholes in the assumption of the theorem. Since the adjustment occurs by quantum Brownian drift,
instead of smooth field variation, the semiclassical field theory arguments do not apply. Further since the evolution involves a special point in phase space,
the quantum attractor $\Lambda = 0^+$ where the bubble nucleation stops, Weinberg's premise of smooth and self-similar 
evolution in field space is circumvented. As a result the no-go theorem of \cite{Weinberg:1987dv} does not apply.

\section{Variations on and of the Action(s)}

\subsection{Volumes and 4-Forms}

As noted above, we start with replacing the covariant integration 
measure in the gravitational sector of Einstein-Hilbert action $\sqrt{g} \, d^4 x$ 
with a completely general $4$-form ${\cal F} = d{\cal A}$. Here ${\cal A}$ is an arbitrary $3$-form potential. 
Our motivation is simply that we {\it can} -- there are no symmetries or principles prohibiting it. So we substitute
\be
\int d^4x \sqrt{g} \, \frac{\mps}{2} \, R \rightarrow \int {\cal F} \, R \, ,
\label{action1}
\ee
effectively promoting Planck scale $\mps$ controlling the strength of gravity to 
a single independent component of the spacetime filling flux of the $4$-form ${\cal F}$. This follows
since by antisymmetry, ${\cal F} \propto \sqrt{g} d^4x$. The `ratio' of these two $4$-forms is a completely
arbitrary scalar function, which must be determined by additional dynamics. Since both 
$\sqrt{g} d^4 x$ and ${\cal F}$ transform as scalars under diffeomorphisms, (\ref{action1}) is guaranteed
to be covariant. 

However, since $\frac{{\cal F}}{\sqrt{g} d^4x} = \Phi$ is an {\it \`a priori} arbitrary scalar function, it can fluctuate. 
The field $\Phi$ would behave exactly like the Brans-Dicke scalar field with $w = 0$. Even its engineering 
dimension is mass squared. Since here we restrict
our interest to the framework(s) with only the usual helicity-2 propagating modes in the gravitational sector, 
we project out\footnote{It is interesting to explore what happens if $\Phi$ is
left in, having both local and discrete variations. Some analysis of only local variations can be gleaned in 
\cite{Guendelman:1996qy}.}  all the local fluctuations in $\Phi$ by introducing the 
second $4$-form ${\cal G}=d{\cal B}$, where ${\cal B}$ is another arbitrary $3$-form potential. 
We couple ${\cal G}$ to the measure $4$-form ${\cal F}$ via the action 
\be
S \ni - \frac{1}{4!} \int {\cal F} \, \frac{\epsilon^{\mu\nu\lambda\sigma}}{\sqrt{g}} {\cal G}_{\mu\nu\lambda\sigma} \, .
\label{action2}
\ee
We note that since ${\cal F} = \frac{1}{4!}  {\cal F}_{\mu\nu\lambda\sigma} dx^\mu \ldots dx^\sigma = 
- \frac{d^4x}{4!} \epsilon^{\mu\nu\lambda\sigma}{\cal F}_{\mu\nu\lambda\sigma}$, 
a straightforward manipulation yields 
\ba
{\cal F} \frac{\epsilon^{\mu\nu\lambda\sigma}}{\sqrt{g}} {\cal G}_{\mu\nu\lambda\sigma} &=&
- \frac{ d^4x}{4!}  {\cal F}_{\alpha\beta\gamma\delta} \epsilon^{\alpha\beta\gamma\delta} 
\frac{\epsilon^{\mu\nu\lambda\sigma}}{\sqrt{g}} {\cal G}_{\mu\nu\lambda\sigma} \nonumber \\
&=& - \frac{ d^4x}{4!}  {\cal G}_{\alpha\beta\gamma\delta} \epsilon^{\alpha\beta\gamma\delta} 
\frac{\epsilon^{\mu\nu\lambda\sigma}}{\sqrt{g}} {\cal F}_{\mu\nu\lambda\sigma} = 
{\cal G} \frac{\epsilon^{\mu\nu\lambda\sigma}}{\sqrt{g}} {\cal F}_{\mu\nu\lambda\sigma}  \, , 
\label{mani}
\ea
and hence 
\be
- \frac{1}{4!} \int {\cal F} \, \frac{\epsilon^{\mu\nu\lambda\sigma}}{\sqrt{g}} {\cal G}_{\mu\nu\lambda\sigma} = - \frac{1}{4!} \int {\cal G} \, 
\frac{\epsilon^{\mu\nu\lambda\sigma}}{\sqrt{g}} {\cal F}_{\mu\nu\lambda\sigma} = \int {\cal G} \, \Phi \, .
\label{commutaction}
\ee
As long as we allow ${\cal G}$ only in this term in the full action, to be given shortly, the variation with respect to ${\cal B}$ 
guarantees that on shell, $\partial_\mu \Phi = 0$, which precisely projects out all the local fluctuations of $\Phi$, as 
desired. However, the value of $\Phi$ is left as a completely arbitrary integration constant. 
We note that while $\Phi$ is introduced here heuristically as a `ratio' of two $4$-forms, in 
what follows we will show that it can be interpreted as the magnetic dual of the $4$-form ${\cal F}$.

As the final ingredient we include the matter sector. In principle, we could just add the matter minimally, using 
the action with the standard measure $\int d^4 x \sqrt{g} \, {\cal L}$. However, as long as the total action contains 
the contribution (\ref{action2}), we can replace the measure $d^4x \sqrt{g}$ according to
\be
d^4x \sqrt{g} \rightarrow d^4x \sqrt{g} + {\tt c} \frac{\cal F}{{\cal M}^2} = (1+ {\tt c} \frac{\Phi}{{\cal M}^2}) \sqrt{g} d^4x \, ,
\label{mattmeas}
\ee
where the last equality follows from the definition of $\Phi$, and ${\cal M}^2$ is a new UV scale normalizing the
flux ${\cal F}$. Likewise, we could replace $g^{\mu\nu}$ in the Lagrangian with  $g^{\mu\nu}(\frac{\Phi}{{\cal M}^2})^\alpha$.
On shell, these represent constant rescalings of the matter
sector variables and can be absorbed away by parameter redefinitions and/or wavefunction renormalizations. 
The numbers ${\tt c}$ and $\alpha$ are, in principle, arbitrary. As a special example, we can write down the matter sector as
\be
S_{\tt QFT} = - \int \frac{{\cal F}}{{\cal M}^2} {\cal L}(\Psi, \frac{g^{\mu\nu}}{\sqrt{\Phi/{{\cal M}^2}}}) \, ,
\label{confmatt}
\ee
such that $({\cal F}/{{\cal M}^2})^{1/4}$ plays the role of a conformally coupled spurion on shell, when $\Phi$
is constant by virtue of the field equations. 

In what follows, we will work with two special cases, which preserve their $4$-form/matter couplings
in the quantum field theory (QFT) loop expansion\footnote{We will treat perturbative gravity semiclassically 
only, ignoring graviton loops, as in e.g. \cite{Englert:1975wj,Arkani-Hamed:2000hpr}.}. These two setups are the theory with the minimally 
coupled matter, which does not include any direct $4$-form/matter coupling, and the theory with the conformal coupling
(\ref{confmatt}). For these two special cases the couplings will not be altered by radiative corrections generated in the loop expansion 
as long as the UV regulator of the matter sector depends on ${\cal F}$ in the same way \cite{Englert:1975wj,Arkani-Hamed:2000hpr}. 
In other cases, the couplings will change order-by-order, as it should be obvious from power counting. 

For simplicity's sake, in the mathematical derivations to follow we will mainly use the minimally coupled 
matter action. However our main physical interest will be in
the conformally coupled theory, because it will turn out that we can devise a simple proof that this variant of 
Pancosmic General Relativity has a safe  
behavior in the semiclassical limit, and avoids a potential problem with ghosts. Our singling out
this example is of technical nature, as we will discuss later. Other types of theories may also be ghost-safe,
but we have not found a general argument yet. 

Note that in the case of conformal coupling, the simplest realization is when the ratio of the
matter sector mass scales and the effective Planck scale, set inside each local region of constant $\kappa^2$, does not
change from region to region even if a bubble wall is crossed. I.e. this corresponds to 
${\mps}_{\tt ~eff} = \kappa^2$. Infrared quantities may still change, such as the
sizes of objects, and ultimately bubble sizes measured from the inside and out. We can however
add the standard Einstein-Hilbert term $\propto \mps R$ to the action, so that the effective Planck scale is $\mps{}_{\tt ~eff} = \mps + \kappa^2$.
This will change the mass ratios (${\rm mass}/{\mpl}_{\tt ~eff}$) in the matter sector as a membrane is crossed, and yield different
QFT hierarchies from bubble to bubble. 
Since we treat gravity only semiclassically, the dynamical equations are altered only minimally. 

Working with our simplest total action generalizing Einstein-Hilbert's \cite{Hilbert:1915tx,Einstein:1915ca}, we have  
\be
S = \int {\cal F} \Bigl( R - \frac{1}{4!} \frac{\epsilon^{\mu\nu\lambda\sigma}}{\sqrt{g}} {\cal G}_{\mu\nu\lambda\sigma} \Bigr) 
- \int d^4x \sqrt{g} \, {\cal L}_{\tt QFT} \, .
\label{actionnew}
\ee
Note that this action is formally third-derivative, as ${\cal F} = d{\cal A}$. Nevertheless, this theory 
is locally indistinguishable from General Relativity, as we now show. The simplest way
to proceed is to write down the field equations, extremizing the action (\ref{actionnew}). Varying with respect to
${\cal A}$ and ${\cal B}$ (keeping in mind the identity (\ref{commutaction})) yields
\be
\partial_\mu \Bigl( R - \frac{1}{4!} \frac{\epsilon^{\mu\nu\lambda\sigma}}{\sqrt{g}} {\cal G}_{\mu\nu\lambda\sigma} \Bigr) = 0 \, , 
~~~~~~~~~ \partial_\mu \Bigl(- \frac{1}{4!} \frac{\epsilon^{\mu\nu\lambda\sigma}}{\sqrt{g}} {\cal F}_{\mu\nu\lambda\sigma} \Bigr) = 
\partial_\mu \Phi = 0 \, , 
\label{eqblocks} 
\ee
where we already alerted the reader to the last equation. These two equations are the 
conservation laws for the dual magnetic fluxes of the theory,
which follow from the $3$-form potential gauge symmetries 
${\cal A} \rightarrow {\cal A} + d\omega_A$, ${\cal B} \rightarrow {\cal B} + d\omega_B$, where $\omega_k$ are arbitrary $2$-forms 
(see, e.g. \cite{Kaloper:2015jra}). Since these are the statements that the two zero forms are closed, they can be readily integrated
locally, introducing two integration constants $\lambda$ and $\kappa^2$,
\be
R - \frac{1}{4!} \frac{\epsilon^{\mu\nu\lambda\sigma}}{\sqrt{g}} {\cal G}_{\mu\nu\lambda\sigma} = 2 \lambda \, , 
~~~~~~~~~ - \frac{1}{4!} \frac{\epsilon^{\mu\nu\lambda\sigma}}{\sqrt{g}} {\cal F}_{\mu\nu\lambda\sigma} = \Phi = \frac{\kappa^2}{2} \, . 
\label{eqints} 
\ee
The final set of gravitational sector field equations follows from variations of (\ref{actionnew}) with respect to the metric $g_{\mu\nu}$.
Since the metric now appears only in $R$, in the denominator of the term $\propto \epsilon^{\mu\nu\lambda\sigma} 
{\cal G}_{\mu\nu\lambda\sigma}$, and in the matter sector, the variational equations will differ from their counterpart in
standard General Relativity. The variation of the action is
\be
\delta_g S = \int {\cal F}  \, \Bigl(- R^{\mu\nu} + 
\frac{\epsilon^{\alpha\beta\lambda\sigma}}{2 \cdot 4! \sqrt{g}} {\cal G}_{\alpha\beta\lambda\sigma}
g^{\mu\nu}\Bigr) \, 
\delta g_{\mu\nu} + \frac12 \int d^4x \sqrt{g} \, T^{\mu\nu} \delta g_{\mu\nu} + 
\int \frac{\cal F}{\sqrt{g}} \partial_\mu \Bigl( \sqrt{g}  J^\mu \Bigr) \, ,
\label{variation1} 
\ee
where $\partial_\mu (\sqrt{g} J^\mu)/\sqrt{g} = g^{\mu\nu} \delta_g R_{\mu\nu}$ is the textbook metric variation of the Ricci
tensor, well known to be a local $4$-divergence. Here $T^{\mu\nu}$ is the standard symmetric matter stress energy tensor, 
$T^{\mu\nu} = \frac{2}{\sqrt{g}} \frac{\delta S_{\tt matter}}{\delta g_{\mu\nu}}$, which
is covariantly conserved, $\nabla_\mu T^{\mu\nu} = 0$, by virtue of flat space matter field theory equations which remain 
unchanged\footnote{The story looks more complicated when the theory involves couplings nonlinear in ${\cal F}$. 
However as long as transformations are analytical the dual theory can be formulated readily, and the same conclusions hold.}. 
To proceed with extracting the gravitational field equations from the action, we can use the field equations 
which we already obtained, specifically (\ref{eqblocks}). Using the second of those equations, after integrating by parts and using
$\partial_\mu \Bigl(- \frac{1}{4!} \frac{\epsilon^{\mu\nu\lambda\sigma}}{\sqrt{g}} {\cal F}_{\mu\nu\lambda\sigma} \Bigr) =0$, 
\be
\int \frac{\cal F}{\sqrt{g}} \partial_\mu \Bigl( \sqrt{g}  J^\mu \Bigr) = - \frac{1}{4!} \int d^4x 
\frac{{\epsilon^{\alpha\beta\lambda\sigma}}{\cal F}_{\alpha\beta\lambda\sigma}}{\sqrt{g}} \partial_\mu \Bigl( \sqrt{g}  J^\mu \Bigr) = 
- \frac{1}{4!} \int dS_\mu J^\mu\frac{{\epsilon^{\alpha\beta\lambda\sigma}}{\cal F}_{\alpha\beta\lambda\sigma}}{\sqrt{g}} \, ,
\ee
where the last equality follows from Gauss' theorem. 
Thus since the last term in (\ref{variation1}) is a boundary term, it does not contribute to the field equations and we can drop it.
Further using ${\cal F} = - \frac{1}{4!} \frac{\epsilon^{\mu\nu\lambda\sigma}}{\sqrt{g}} {\cal F}_{\mu\nu\lambda\sigma} d^4x \sqrt{g}$ 
on shell, we obtain that $\delta_g S = 0$ leads to
\be
- \frac{2}{4!} \frac{\epsilon^{\rho\zeta\gamma\delta}}{\sqrt{g}} {\cal F}_{\rho\zeta\gamma\delta}   \Bigl(R^\mu{}_\nu - 
\frac{{\epsilon^{\alpha\beta\lambda\sigma}}{\cal G}_{\alpha\beta\lambda\sigma}}{2 \cdot 4! \sqrt{g}} 
\delta^\mu{}_\nu \Bigr) = T^\mu{}_\nu \, ,
\label{tenseqs}
\ee
where for convenience we are using the mixed tensor representation for $R^\mu{}_\nu$ and $T^\mu{}_\nu$. 

So to recapitulate, our field equations are the set of (\ref{tenseqs})
and the $3$-form variations (\ref{eqblocks}) or equivalently their first integrals (\ref{eqints}), 
which we collect together here for clarity:
\ba
&&~~~~ - \frac{2}{4!} \frac{\epsilon^{\rho\zeta\gamma\delta}}{\sqrt{g}} {\cal F}_{\rho\zeta\gamma\delta}  \Bigl(R^\mu{}_\nu - 
\frac{{\epsilon^{\alpha\beta\lambda\sigma}}{\cal G}_{\alpha\beta\lambda\sigma}}{2 \cdot 4! \sqrt{g}} 
\delta^\mu{}_\nu \Bigr) = T^\mu{}_\nu \, , \nonumber \\
&& R - \frac{1}{4!} \frac{\epsilon^{\mu\nu\lambda\sigma}}{\sqrt{g}} {\cal G}_{\mu\nu\lambda\sigma} = 2 \lambda \, , 
~~~~~~~~~ - \frac{1}{4!} \frac{\epsilon^{\mu\nu\lambda\sigma}}{\sqrt{g}} {\cal F}_{\mu\nu\lambda\sigma} = \frac{\kappa^2}{2} \, . 
\label{alleqs}
\ea
At first glance these equations do not look like General Relativity\footnote{For example, one would think that the structure
of General Relativity field equations is fixed by local gauge invariance, whose first check is provided by Bianchi identities. The equations 
(\ref{alleqs}) nevertheless {\it do} satisfy Bianchi identities, as follows: 
denoting $- \frac{1}{4!} \frac{\epsilon^{\mu\nu\lambda\sigma}}{\sqrt{g}} {\cal F}_{\mu\nu\lambda\sigma} 
= \frac{\kappa^2}{2}$, subtracting and adding $(R/2) \delta^\mu{}_\nu$ in the parenthesis, taking $4$-divergence 
and using $\partial_\mu \lambda = \partial_\mu \kappa^2 =0$ 
indeed yields $\nabla_\mu T^{\mu}{}_\nu = 0$ on shell. Which is why the substitution of the 
equations in (\ref{alleqs}) reproduces (\ref{tenseqsgr}).}. However this is not so: indeed a 
simple substitution of the last two equations into the first ones readily yields
\be
\kappa^2 \Bigl(R^\mu{}_\nu - \frac12 R \, 
\delta^\mu{}_\nu \Bigr) = - \kappa^2 \lambda \, \delta^\mu{}_\nu + T^\mu{}_\nu \, ,
\label{tenseqsgr}
\ee
which are structurally just the field equations of General Relativity, but with one very important new physical ingredient.
In (\ref{tenseqsgr}), both the strength of gravity and the vacuum curvature -- i.e. the effective Planck scale and the
cosmological constant -- are set by two, so far completely arbitrary, integration constants $\kappa^2$ and $\lambda$. 
As they stand the equations (\ref{alleqs}), (\ref{tenseqsgr}) don't describe just one General Relativity, but an
infinity of them, parameterized by the values of $\kappa^2$, $\lambda$. 

When we include modified measures
in the matter sector, the values of the local matter scales and couplings would also vary from one theory to another. 
This means that our third-derivative 
General Relativity is in fact a further extension of the ``unimodular gravity" formulation
of General Relativity, which included an {\it \`a priori} integration constant contribution to only the cosmological constant term 
\cite{Einstein:1919gv,Anderson:1971pn,Aurilia:1980xj,Duff:1980qv,Buchmuller:1988wx,Buchmuller:1988yn,Henneaux:1989zc,Ng:1990xz,Fiol:2008vk}. 

One might be tempted to dismiss this point as a mere curiosity, since after all the integration constants of the ``metatheory" 
given by the action (\ref{actionnew}), or its more general cousins which feature modified matter sector measure as well, 
are constant after all. One picks their values by measurement, fixes the theory, {\it et voil\`a}, the parameters are selected.
In a sense this is even justified by renormalization in QFT, where the UV-sensitive quantities must be regulated,
and their physical values determined by measurement (see, e.g. \cite{Kaloper:2014dqa})\footnote{One may hope that the 
UV completion of the theory might go
beyond the renormalization procedure of QFT and actually predict this value, or at least predict that
the favored values feature a large hierarchy (see, e.g. \cite{giudice}).}. 
Thus different General Relativities governed by the metaaction (\ref{actionnew}) might appear like 
a set of superselection sectors in QFT, which remain forever distinct and separated from each other. 

However,  consider for a moment matter sectors which contain a multiplet of QFT 
vacua, with phase transitions between them.
Such processes link asymptotically different superselection sectors of the metatheory (\ref{actionnew}). Not all physical parameters in the 
(renormalized) Lagrangian will forever remain the same when phase transitions are turned on. 
Common examples are
the transitions which change vacuum energy (and lead to the ideas of 
string landscape \cite{Brown:1987dd,Brown:1988kg,Duncan:1989ug,Bousso:2000xa,Feng:2000if}). In 
quantum gravity in principle all parameters may be subject to such variations 
\cite{Wheeler:1955zz,Hawking:1978pog,Coleman:1988tj,Hawking:1984hk,Giddings:1988wv,Fischler:1988ia,Fischler:1989ka,Banks:1984cw,Polchinski:1989ae}. 
Thus given that the metatheory  (\ref{actionnew}) brings in an infinity of General Relativities, which appear to be classically 
mutually disconnected like universes with different cosmological constant in unimodular formulation of 
General Relativity (or multi-Relativity, \cite{Linde:1988ws}), it is interesting 
to explore possible channels which allow such universes to evolve into each other. 

The generalization of  (\ref{actionnew}) which opens up the channels for the General Relativities with 
different $\kappa^2$ and $\lambda$ to evolve into each other, while retaining their local spectrum of propagating modes, 
turns out to be very straightforward in our case. 
Since $\kappa^2$ and $\lambda$ are conserved dual magnetic fluxes of the gauge fields 
${\cal F} = d{\cal A}$ and ${\cal G}=d{\cal B}$, we can ``unfreeze" them by 
introducing objects which are charged under ${\cal A}$ and ${\cal B}$. 
When the charge carriers nucleate quantum-mechanically, they change discretely the fluxes in their vicinity. 
The fluxes can discharge by charge emission: the charges open 
the possibility that the fluxes can be relaxed by the production of charge carriers. Because ${\cal A}$ and
${\cal B}$ are $3$-forms, the charge carriers must be membranes. So we add membranes charged under
${\cal A}, {\cal B}$ to the action  (\ref{actionnew}):
\ba
S &=& \int {\cal F} \Bigl( R - \frac{1}{4!} \frac{\epsilon^{\mu\nu\lambda\sigma}}{\sqrt{g}} {\cal G}_{\mu\nu\lambda\sigma} \Bigr) 
- \int d^4x \sqrt{g} \, {\cal L}_{\tt QFT} + S_{\tt boundary} \nonumber \\
&-& {\cal T}_A \int d^3 \xi \sqrt{\gamma}_A - {\cal Q}_A \int {\cal A}  - {\cal T}_B \int d^3 \xi \sqrt{\gamma}_B - {\cal Q}_B \int {\cal B}  \, .
\label{actionnewmem}
\ea
Here ${\cal T}_i, {\cal Q}_i$ are the membrane tension and charge, respectively, and $\xi^\alpha$ are the restriction of the membrane 
embedding maps $x^\mu = x^\mu(\xi^\alpha)$ to the membrane worldvolumes. The term $S_{\tt boundary}$ denotes the boundary terms 
which properly covariantize the bulk actions in the presence of boundaries. It is a straightforward generalization of  
Israel-Gibbons-Hawking boundary terms of standard General Relativity \cite{israel,gibbhawk}, 
including also contributions from the $4$-form sector. 
We will give their explicit general form shortly. 

Note that the presence of membranes alters the theory even at the classical level. We would have background geometries which are made 
up of many regions in the huge metaverse, with classical parameters changing discretely from one region to another. In the absence of the
local matter sources those regions would be de Sitter or Anti-de Sitter patches with, 
in general, different strength of gravity in each, and separated by expanding spherical walls. 
The distribution of these regions would be set by the classical `initial conditions' on some 
Cauchy surface, and classically `frozen' forever. 

In quantum mechanics however new membranes can nucleate, changing the number and the distribution of 
bubbles, and also changing how bubble interiors evolve. The various classical `initial surfaces', 
frozen in the limit $\hbar \rightarrow 0$, 
would evolve into each other. The membrane nucleation processes would be described by 
Euclidean instantons, which are subsequently analytically continued to a Lorentzian signature spacetime. 
We will work with this in mind here, using quantum-mechanical effects to leading order to 
understand the dynamics of the space of `vacua' of Pancosmic General Relativity introduced above. 

The gauge couplings $\propto \int {\cal A}$ are integrated over the membrane worldvolumes, 
\be
\int {\cal A} = \frac16 \int d^3 \xi {\cal A}_{\mu\nu\lambda} \frac{\p x^\mu}{\p \xi^\alpha} \frac{\p x^\nu}{\p \xi^\beta} 
\frac{\p x^\lambda}{\p \xi^\gamma} \epsilon^{\alpha\beta\gamma} \, ,
\ee
and likewise for ${\cal B}$. Note that these couplings can describe both positively 
and negatively charged membranes, accommodated by the change of the
winding direction of $x^\mu = x^\mu(\xi^\alpha)$. We will take the tensions ${\cal T}_i$ to be 
strictly positive, however, to enforce local positivity of energy. Our membranes
could be fundamental objects, generalizing electrically charged fundamental particles. 
Alternatively, they could be ``emergent", arising as the composite boundaries, i.e. walls, 
in strongly coupled gauge theories at low energies. We can be agnostic about their microscopic 
nature\footnote{Membranes might arise at low energies as thin wall approximation of 
domain walls in systems with a discrete system of a very large number of vacua 
\cite{Gabadadze:1999pp}.} and imagine that they can be described in the 
thin wall approximation as in (\ref{actionnewmem}) regardless. 

It is now clear that the nucleation of membranes can mediate variation of the `integration constants' $\kappa^2$ and $\lambda$.
To illustrate this, consider membranes with ${\cal Q}_B \ne 0$. Rewriting the second term in the bulk action (\ref{actionnewmem})
as $ -  \frac{1}{4!} \int {\cal F}  \frac{\epsilon^{\mu\nu\lambda\sigma}}{\sqrt{g}} {\cal G}_{\mu\nu\lambda\sigma} = 
-  \frac{1}{4!} \int {\cal G}  \frac{\epsilon^{\mu\nu\lambda\sigma}}{\sqrt{g}} {\cal F}_{\mu\nu\lambda\sigma}$ and varying 
(\ref{actionnewmem}) with respect to ${\cal B}$ now yields
\be
-(\frac{\epsilon^{\mu\nu\lambda\sigma}}{4! \sqrt{g}} {\cal F}_{\mu\nu\lambda\sigma})|_{out} 
+ (\frac{\epsilon^{\mu\nu\lambda\sigma}}{4! \sqrt{g}} 
{\cal F}_{\mu\nu\lambda\sigma})|_{in}  = \frac12 \kappa^2_{out} - \frac12 \kappa^2_{in} =  {\cal Q}_B \, ,
\label{kappajump}
\ee
across a membrane, moving out in the direction of the local normal. In other words, the emission of a membrane with the charge
${\cal Q}_B$ yields a discrete jump of the Planck scale between the exterior ({\it out}) and the interior ({\it in}) by $2 {\cal Q}_B$. Similarly, 
$\lambda$ changes discretely by an emission of a charge ${\cal Q}_A$. In the next section we will consider these processes
in detail, outline the possible transition channels, and estimate their rates.

\subsection{Canonical Transformation to Magnetic Duals}

Before we proceed with the study of general transitions between different `vacua' of three derivative General Relativity (i.e. the
metatheory of General Relativities) given by (\ref{actionnewmem}), it is instructive to rewrite the metaaction in terms of the 
magnetic dual variables to ${\cal F}$ and ${\cal G}$. This transformation is a generalization of canonical 
transformations in classical mechanics trading generalized coordinates and generalized momenta \cite{mercier}. 

Using this formulation we will see even more clearly 
how the parameters of standard General Relativity are 
promoted to dynamical, albeit non-propagating, degrees of freedom. We will also be able to immediately discern the explicit 
form of the boundary terms $S_{\tt boundary}$. Finally this form of the action will come in handy in the calculation of  on shell
Euclidean actions which control the membrane nucleation rates, to be considered below. 

The dualization procedure starts with recasting the $4$-form sector of (\ref{actionnewmem})
into the first order formalism, where each variable in both pairs 
${\cal F}$,  ${\cal A}$ and ${\cal G}$, ${\cal B}$ is treated as an independent dynamical variable to be integrated over in the path
integral. The relations ${\cal F} = d{\cal A}$ and ${\cal G} = d{\cal B}$ are enforced with the help of Lagrange multipliers, 
${\cal P}_A, {\cal P}_B$. 
These Lagrange multipliers are also integrated over in the path integral, 
\be
Z = \int \ldots [{\cal D} {\cal A}] [{\cal D} {\cal B}] [{\cal D}{\cal F}] [{\cal D} {\cal G}] [{\cal D}{\cal P}_A] 
[{\cal D} {\cal P}_B] \, e^{i S({\cal A}, {\cal B}, {\cal F}, {\cal G}, ...) + i\int  {\cal P}_A ( {\cal F} -d {\cal A})
+ i\int  {\cal P}_B ( {\cal G} -d {\cal B})} \ldots \, ,
\label{partf} 
\ee
Then simply changing the order of integration of variables yields different dual pictures. 
This technique was utilized in supergravity \cite{Nicolai:1980td,dvali}, and has been a mainstay in the formulation of flux monodromy models 
of inflation \cite{ks1,ks2,ks3}. Explicitly, the idea is that after transitioning to the first order variables, we 
integrate out  the $4$-form field strengths, and recognize 
that in the resulting action the scalar Lagrange multipliers are in fact precisely the magnetic duals of ${\cal F}$ and ${\cal G}$. This
procedure is the same regardless of the direct $4$-form/matter couplings, although the specifics can complicate the explicit transformation formulas 
(as in, for example, hybrid monodromy inflation models \cite{hybrid}). We will therefore work with the minimal matter action,
and simply generalize the result after the fact in the obvious way. 

To keep track of all the relevant terms in this procedure and reduce the clutter, we will only look at the part of the action (\ref{actionnewmem})
which depends explicitly on ${\cal F}$ and ${\cal G}$, and rewrite it in terms of the components of ${\cal F}$ and ${\cal G}$. Since
$-\frac{\epsilon^{\mu\nu\lambda\sigma}}{\sqrt{g}} {\cal G}_{\mu\nu\lambda\sigma} {\cal F} = - d^4x \sqrt{g} {\cal F}_{\mu\nu\lambda\sigma} 
{\cal G}^{\mu\nu\lambda\sigma}$, we find 
\ba
S &\ni& \int d^4x \sqrt{g} \Bigl( - \frac{1}{4!} {\cal F}_{\mu\nu\lambda\sigma} 
{\cal G}^{\mu\nu\lambda\sigma} - \frac{R}{4!} \frac{\epsilon^{\mu\nu\lambda\sigma}}{\sqrt{g}} {\cal F}_{\mu\nu\lambda\sigma}
- {\cal L}_{\tt QFT} \nonumber \\
&&~~~~~~~~~~~~ + \frac{{\cal P}_A}{4!} \frac{\epsilon^{\mu\nu\lambda\sigma}}{\sqrt{g}} ({\cal F}_{\mu\nu\lambda\sigma} 
- 4\partial_\mu {\cal A}_{\nu\lambda\sigma}) + \frac{{\cal P}_B}{4!} 
\frac{\epsilon^{\mu\nu\lambda\sigma}}{\sqrt{g}} ({\cal G}_{\mu\nu\lambda\sigma} 
- 4\partial_\mu {\cal B}_{\nu\lambda\sigma}) \Bigr) \, ,
\label{firstorder}
\ea
where the second line are the Lagrange multipliers. Defining new independent degrees of freedom
\be
\tilde {\cal F}_{\mu\nu\lambda\sigma} = {\cal F}_{\mu\nu\lambda\sigma} - {{\cal P}_B} {\sqrt{g}} {\epsilon_{\mu\nu\lambda\sigma}} \, ,
~~~~~~~~~ \tilde {\cal G}_{\mu\nu\lambda\sigma} = 
{\cal G}_{\mu\nu\lambda\sigma} - ({{\cal P}_A - R}) {\sqrt{g}} {\epsilon_{\mu\nu\lambda\sigma}} \, ,
\label{trans}
\ee
and recalling that the translational changes of variables as in (\ref{trans}) do not change 
the path integral since the functional
Jacobian is unity, we can rewrite this part of the action as 
\ba
S &\ni&  \int d^4x \Bigl\{\sqrt{g} \Bigl( - \tilde {\cal F}_{\mu\nu\lambda\sigma} 
\tilde {\cal G}^{\mu\nu\lambda\sigma} + {{\cal P}_B} (R - {\cal P}_A) 
- {\cal L}_{\tt QFT} \Bigr) \nonumber \\
&& ~~~~~~~~~~~~~~~~~~~~~~~~~~~~~~~~~
- \frac{{\cal P}_A}{6} {\epsilon^{\mu\nu\lambda\sigma}} \partial_\mu {\cal A}_{\nu\lambda\sigma}
-\frac{{\cal P}_B}{6} {\epsilon^{\mu\nu\lambda\sigma}}\partial_\mu {\cal B}_{\nu\lambda\sigma} \Bigr\}\, .
\label{firstorder2}
\ea
Since $\tilde {\cal F}$ and $\tilde{\cal G}$ do not appear anywhere else, the integration over one of them yields
a functional Dirac $\delta$-function for the other, 
\be
Z = \int \ldots [{\cal D}\tilde {\cal F}] [{\cal D}\tilde {\cal G}] e^{i \int d^4x\sqrt{g} \Bigl( - \tilde {\cal F}_{\mu\nu\lambda\sigma} 
\tilde {\cal G}^{\mu\nu\lambda\sigma} \Bigr) } \ldots = \int \ldots [{\cal D}\tilde {\cal G}] \delta(\tilde {\cal G}) \ldots \, ,
\ee
and then the integration over this one sets the corresponding factor in the path integral to unity.
Further, note that the variables
${\cal P}_A$ and ${\cal P}_B$ are precisely $\propto \kappa^2, \lambda$, respectively. So we can make these substitutions right away:
\be
{\cal P}_A = 2 \lambda \, , ~~~~~~~~~~~~~ {\cal P}_B = \frac{\kappa^2}{2} \, .
\ee

Thus our new dual variables action, with the membrane terms from (\ref{actionnewmem}) included, is
\ba
S &=& \int d^4x \Bigl\{\sqrt{g} \Bigl(\frac{\kappa^2}{2} R - \kappa^2 \lambda 
- {\cal L}_{\tt QFT} \Bigr)- \frac{\lambda}{3} {\epsilon^{\mu\nu\lambda\sigma}} \partial_\mu {\cal A}_{\nu\lambda\sigma}
- \frac{\kappa^2}{12} {\epsilon^{\mu\nu\lambda\sigma}}\partial_\mu {\cal B}_{\nu\lambda\sigma} \Bigr\}  \nonumber \\
&& ~~ + \,\, S_{\tt boundary} - {\cal T}_A \int d^3 \xi \sqrt{\gamma}_A - {\cal Q}_A \int {\cal A}  
- {\cal T}_B \int d^3 \xi \sqrt{\gamma}_B - {\cal Q}_B \int {\cal B}  \, .
\label{actionnewmemd} 
\ea
This action closely resembles the theory of local vacuum energy sequester
\cite{Kaloper:2015jra}, but it is not the same. The main differences are that the independent 
variables here are $\kappa^2$ and $\lambda$ instead of $\kappa^2$ and $\Lambda = \kappa^2 \lambda$, 
and the presence of membranes with charges ${\cal Q}_i$. 
However as we will see in what follows, that will be of no consequence for our considerations here. 
Approaching the cosmological constant problem in Pancosmic General Relativity 
follows a different path. 

This form of the action lays out the framework of Pancosmic General Relativity very transparently. First off, the variables $\kappa^2$ and
$\lambda$ are now principal dynamical variables, which change only discontinuously, by membrane emissions, and in discrete 
amounts controlled by the units of charge ${\cal Q}_B$ and ${\cal Q}_A$, respectively. The local constancy of the $4$-forms 
in the absence of a charged source follows from 
the variations of (\ref{actionnewmemd}) with respect to ${\cal A}$ and ${\cal B}$. The
gravitational sector away from the membranes is identical to that in the standard formulation of General Relativity thanks to the fact that the
new bulk action terms $\propto  \partial_\mu {\cal A}_{\nu\lambda\sigma},  \partial_\mu {\cal B}_{\nu\lambda\sigma}$ are completely
independent of the metric, being purely topological. 

To summarize all this mathematically, we write down the Euler-Lagrange equations
obtained by varying (\ref{actionnewmemd}) with respect to the metric, $\kappa^2$, $\lambda$, ${\cal A}_{\nu\lambda\sigma}$ 
and ${\cal B}_{\nu\lambda\sigma}$, in that order: 
\ba
\kappa^2 G^\mu{}_\nu  &=& - \kappa^2 \lambda \, \delta^\mu{}_\nu + T^\mu{}_\nu + \ldots \, , ~~~~~~ 
\hat {\cal F}_{\mu\nu\lambda\sigma} = \frac{\kappa^2}{2} \sqrt{g} \, {\epsilon_{\mu\nu\lambda\sigma}} \, ,  ~~~~~~ 
\hat {\cal G}_{\mu\nu\lambda\sigma} = \frac{2 \lambda-R}{4} \sqrt{g} \, {\epsilon_{\mu\nu\lambda\sigma}} \, , \nonumber \\
&& 2 n^\mu \partial_\mu \lambda  =  {\cal Q}_A \delta(r-r_0) \, , ~~~~~~~~~ 
\frac12 n^\mu \partial_\mu \kappa^2 = {\cal Q}_B \delta(r-r_0) \, . ~~~~
\label{tenseqsgrmagdual}
\ea
The ellipsis in the first equation designate the generalization of Israel-Gibbons-Hawking boundary terms. 
Here we have reintroduced the `spectator' $4$-forms $\hat {\cal F} = d{\cal A}$ and $\hat{\cal G} = d{\cal B}$ to 
utilize a more compact notation, and used Einstein's tensor $G^\mu{}_\nu$ in the first line. 
The vector $n^\mu$ is the outward normal to a membrane, and $r$ the coordinate along the axis in the 
direction of that normal. 

We cannot stress enough here that although $\kappa^2$ and $\lambda$ look like fixed Lagrangian parameters
in the action (\ref{actionnewmemd}), they are {\it not}. The variables $\kappa^2$ and $\lambda$ are discrete dynamical degrees
of freedom, and are completely arbitrary until one picks their numerical values by solving the first order differential equations
in the second line of (\ref{tenseqsgrmagdual}). The variations of these variables
will be quantized, taking values which are integer multiples of the charge, by which they 
change by membrane emission. This is similar to flux monodromy models \cite{ks1,ks2,ks3}. 

In the magnetic dual form of the action, the third derivative in the original formulation of the 
theory (\ref{actionnewmem}) seems to have disappeared from (\ref{actionnewmemd}). 
However the arbitrariness of
$\kappa^2$ is its legacy: the reason the derivative seems to have gone away is that the duality transformation which we carried out
starting with (\ref{firstorder}) is a {\it canonical transformation} in the dynamical sense \cite{mercier}, exchanging the 
canonical `electric' field momentum variable $\pi_A \sim \partial_0 {\cal A}_{123}$ with the dual `magnetic' conjugate field
variable $\phi_B \sim {\cal P}_B$, and correspondingly for  $\pi_B, \phi_A$.  
Since the gauge symmetries of ${\cal A}$ and ${\cal B}$ are linearly realized, the action does not
directly depend on those variables -- they are {\it cyclic}, yielding the conserved magnetic 
fluxes of Eq. (\ref{eqblocks}), and so concealing the derivative -- as in a Legendre transformation. 
In more general frameworks, that may exist, where gauge symmetries would be realized
nonlinearly, one would expect both sides of the dual theory to feature extra derivatives \cite{dvali,ks1,ks2}.

One may wonder which of these sets of variables is more ``natural" or ``physical". The simple answer is,
neither -- they are all equivalent. Perhaps the most comforting example illustrating this is the linear harmonic oscillator, with
the Hamiltonian $H = p^2/2 + q^2/2$. Clearly, the transformation $(q,p) \rightarrow (P,-Q)$ preserves both the form of $H$ and
the Poisson brackets, meaning either pair $(q,p)$ or $(Q,P)$ (or any symplectic rotation of them in the $Q,P$ plane) is just as good.
Thus we are free to pick any of these as our dynamical basis. 

On the other hand, note that employing the `electric' formulation (\ref{actionnewmem}), motivated by the recognition that the measure
of integration chosen by Hilbert in \cite{Hilbert:1915tx} is but a special case of a more general set of possibilities, immediately led
to the way of introducing the discrete dynamics that can change Planck scale and the cosmological constant by membrane emission.
As a consequence both standard General Relativity \cite{Hilbert:1915tx,Einstein:1915ca} and its unimodular formulation 
\cite{Einstein:1919gv,Anderson:1971pn,Aurilia:1980xj,Duff:1980qv,Buchmuller:1988wx,Buchmuller:1988yn,Henneaux:1989zc,Ng:1990xz,Fiol:2008vk} 
are merely special limits of our theory (\ref{actionnewmem}),
(\ref{actionnewmemd}). They arise in the limit when the membranes decouple, which 
happens\footnote{Note that making the charges infinitesimally small would correspond to 
making the variables $\kappa^2$ and $\lambda$ change almost continuously. Making tensions very large however
seizes membrane nucleations and freezes $\kappa^2$ and $\lambda$. This is just an example of the standard realization of decoupling.} 
when $ {\cal T}_A/\kappa^3, {\cal T}_B/\kappa^3 \rightarrow \infty$.

Finally, by inspection of  (\ref{actionnewmemd}), we can determine the boundary terms in addition to 
the tension and charge terms. First off, the non-gravitating, topological ``spectator" terms $\frac{\lambda}{3} 
{\epsilon^{\mu\nu\lambda\sigma}} \partial_\mu {\cal A}_{\nu\lambda\sigma}$
and $\frac{\kappa^2}{12} {\epsilon^{\mu\nu\lambda\sigma}}\partial_\mu {\cal B}_{\nu\lambda\sigma}$ in the action (\ref{actionnewmemd}) 
are there to enforce that the magnetic dual degrees of freedom $\lambda$ and $\kappa^2$ satisfy their field equations,
given in the second line of Eq. (\ref{tenseqsgrmagdual}). Once these equations are solved -- i.e. 
$\lambda,\kappa^2$ are
chosen to satisfy them -- the spectators automatically reduce to 
boundary terms, very much like the $4$-form boundary 
terms considered
in \cite{Duncan:1989ug,ks1,ks2,ks3}. To see it, we rewrite the spectator terms in Eq. (\ref{actionnewmemd}) as
\ba
&&- \int d^4x \Bigr(\frac{\lambda}{3} {\epsilon^{\mu\nu\lambda\sigma}} \partial_\mu {\cal A}_{\nu\lambda\sigma}
+ \frac{\kappa^2}{12} {\epsilon^{\mu\nu\lambda\sigma}}\partial_\mu {\cal B}_{\nu\lambda\sigma} \Bigr) = 
- \int d^4x \partial_\mu \Bigr( \frac{\lambda}{3} {\epsilon^{\mu\nu\lambda\sigma}} {\cal A}_{\nu\lambda\sigma} 
+ \frac{\kappa^2}{12} {\epsilon^{\mu\nu\lambda\sigma}} {\cal B}_{\nu\lambda\sigma} \Bigr) \nonumber \\
&& ~~~~~~~~~~~~~~~~~~~~~~~~~~~~~~~~~~~~~~~~~~ +
\int d^4x \Bigr( \frac{\partial_\mu \lambda}{3} {\epsilon^{\mu\nu\lambda\sigma}} {\cal A}_{\nu\lambda\sigma} 
+ \frac{\partial_\mu \kappa^2}{12} {\epsilon^{\mu\nu\lambda\sigma}} {\cal B}_{\nu\lambda\sigma} \Bigr) \, .
\label{formbcs}
\ea
It is now obvious that the terms in the second line precisely cancel the charge terms in (\ref{actionnewmemd}). The total derivatives 
integrate -- by Gauss' law -- to a boundary term which needs to be subtracted from the total action to ensure the correct variational
behavior of the $4$-forms on the boundary, generalizing similar terms encountered in massless and massive
``canonical" $4$-form theories in \cite{Duncan:1989ug,ks1,ks2}. Thus the $4$-form induced boundary term, evaluated on 
the membrane worldvolumes, is
\be
S_{\tt boundary}^{4-{\rm forms}} = \int d^3 \xi \Bigr( [\frac{\lambda}{3} {\epsilon^{\alpha\beta\gamma}} {\cal A}_{\alpha\beta\gamma}]
+ [\frac{\kappa^2}{12} {\epsilon^{\alpha\beta\gamma}} {\cal B}_{\alpha\beta\gamma}] \Bigr) \, .
\label{boundactf}
\ee
Here $[...]$ designates the discontinuity across a membrane (a.k.a. the difference of the exterior and interior limits 
of the bracketed quantity). Note that $\lambda, \kappa^2$ reside inside $[...]$ since both can jump if a
charge ${\cal Q}_i$ is emitted, as shown in Eq. (\ref{kappajump}). 
Also note that since membranes are compact and smooth, the integrals like $\sim \int {\cal A}$ remain gauge invariant.
The ``job" of these boundary terms is to cancel the total derivatives in (\ref{formbcs}), which would have remained 
after the membrane charge terms $\sim {\cal Q}_{i}$ are cancelled 
by the $4$-form and $\lambda, \kappa^2$ equations in (\ref{tenseqsgrmagdual}). 
In practice, when computing the Euclidean action for the on shell solutions, 
we can drop both the charge terms and the 
``spectators". Of course, this is nothing else but an analogue of Gauss' laws for a 
system of charges in usual electromagnetism. We will keep these
terms in the action for completeness sake, but bear in mind that they drop out on shell 
when it comes to actually computing
the Euclidean bounce actions, to follow in the next section. 

Further we see that the boundary action $S_{\tt boundary}$ must be precisely 
Israel-Gibbons-Hawking action, but with a different $\kappa^2$ normalizing  
Israel-Gibbons-Hawking integrand on each side of
a membrane:
\be
S^R_{\tt boundary} = - \int d^3 \xi \sqrt{\gamma} [\kappa^2 K ] \, ,
\label{gibbhawkbc}
\ee
where $\xi^\alpha$ are intrinsic coordinates on the membrane, $\gamma$ the induced metric, $K$ the extrinsic curvature
computed relative to the outward normal, defined as the trace of $K_{\alpha\beta} = - \nabla_\alpha n_\beta$. The covariant derivative 
here is with respect to the induced metric on the membrane. 
With wisdom after the fact, this form of (\ref{gibbhawkbc}) is inevitable, since the purpose of  
Israel-Gibbons-Hawking terms is to cancel the canonical momentum-dependent terms on the boundaries which arise from integrations
by parts of the variations of Einstein-Hilbert action. In other words, (\ref{gibbhawkbc}) precisely cancels the discontinuity
in $R$ generated by the tension source on the membrane, and prevents the over-counting of the tension contributions.
This of course is just Gauss' law for gravity. 
Since we have generalized the action to $\int {\cal F} R$ here, and allowed ${\cal F}$ to jump
across a boundary, we must slightly generalize the boundary action to allow for the jump of $\kappa^2$ -- as stated above -- and 
properly compensate for it. Ergo (\ref{gibbhawkbc}). 

One important point which should be borne in mind is that for non-compact geometries we should also include boundary terms
accounting for the flux of various fields at infinity. In Lorentzian signature, where we only care about the field equations, such
terms are irrelevant. However in Euclidean signature when we interpret the total Euclidean action as a measure of probability,
or the rate of a process, retaining such terms is critical, since we may be dealing with regulated divergent integrals. Indeed, one 
starts by imposing an infrared cutoff on a Euclidean geometry to regulate the integral, covariantizing it with boundary
terms at the cutoff, and then taking the limit where the cutoff is removed. This means that at infinity we retain the
``inside" contribution to (\ref{gibbhawkbc}), meaning the single $\propto \kappa^2 K$ contribution to the boundary integral with
an overall ``+" sign, residing on the ``interior" of the regulator wall. This is the source ``at the end of the world",
conserving the total ``charge". We will encounter this in the computation of some of the bounce actions in the next section. 

The total boundary action is, with all the features elaborated above accounted for, 
\be
S_{\tt boundary} = S_{\tt boundary}^{4-{\rm forms}} + S^R_{\tt boundary} \, .
\label{totalba}
\ee
With this, we have completely fixed all the dynamical conditions controlling the evolution of the theory
on and off the membrane sources in the case of the minimal matter/gravity couplings given by the $4$-form action
(\ref{actionnewmem}) or equivalently its magnetic dual (\ref{actionnewmemd}). 

Before we turn to analyzing  
the geometric transitions catalyzed by the membrane emissions, however, 
let us quickly sketch out the ingredients of the theory for 
the conformal $4$-form/matter case as well. This generalization of (\ref{actionnewmemd}) is straightforward. 
The idea is to start with the magnetic dual action, where all the terms in (\ref{actionnewmemd}) except the
matter Lagrangian are the same. The matter Lagrangian is replaced by 
\be
\sqrt{g} {\cal L}_{\tt QFT}(g^{\mu\nu}) \rightarrow \sqrt{\hat g} {\cal L}_{\tt QFT}(\hat g^{\mu\nu}) \, ,
\label{confcoup}
\ee
where $\hat g_{\mu\nu} = g_{\mu\nu} \sqrt{\frac{\kappa^2}{{\cal M}^2}}$ using the notation of the previous section,
and, as noted, ${\cal M}$ is a UV scale controlling the perturbative expansion of the full effective action in the powers of
${\cal F}$. It is now manifest that the matter loop corrections preserve this form of the action, as long as 
the regulator depends on $\kappa/{\cal M}$ in the same way as the matter Lagrangian 
\cite{Englert:1975wj,Arkani-Hamed:2000hpr}. In other words, all matter 
sector operators include powers of $({\frac{\kappa}{{\cal M}}})^{1/2}$ controlled by their engineering dimension. On the other hand,
in general we can also add to the action the pure Einstein-Hilbert term, replacing
\be
\frac{\kappa^2}{2} R \rightarrow \frac{\mps + \kappa^2}{2} R \, .
\label{efterm}
\ee
We can think of this as the semiclassical effective gravity Lagrangian term which includes matter 
sector loop corrections in this specific theory. Even if  $\propto \kappa^2$ terms were absent to start, the conformally coupled matter sector
would induce them via renormalization\footnote{Notice that this action does not have a global scale symmetry. It shouldn't,
if it is to have a chance of linking to quantum gravity \cite{banksseiberg}.} of $\mps$. Thus the full action is
\ba
S &=& \int \Bigl\{\sqrt{g} \Bigl(\frac{\mps+\kappa^2}{2} R-\kappa^2 \lambda 
-\frac{\kappa^2}{{\cal M}^2} {\cal L}_{\tt QFT}(\frac{{\cal M}}{\kappa}{g^{\mu\nu}})\Bigr)
-\frac{\lambda}{3} {\epsilon^{\mu\nu\lambda\sigma}} \partial_\mu {\cal A}_{\nu\lambda\sigma}
- \frac{\kappa^2}{12} {\epsilon^{\mu\nu\lambda\sigma}}\partial_\mu {\cal B}_{\nu\lambda\sigma} \Bigr\}  \nonumber \\
&&~~~~ + \,\, S_{\tt boundary} - {\cal T}_A \int d^3 \xi \sqrt{\gamma}_A - {\cal Q}_A \int {\cal A}  
- {\cal T}_B \int d^3 \xi \sqrt{\gamma}_B - {\cal Q}_B \int {\cal B}  \, . 
\label{actionnewmemdconf}
\ea
Note that we could have written this action in terms of the original electric $4$-forms ${\cal F}$ and ${\cal G}$ and their components. 
We could still do this, by performing the inverse Legendre map to the one we defined in the beginning of this section. It
clearly exists. However it would be quite cumbersome due to a variety of nonlinear terms which appear in the matter
sector Lagrangian; yet the answers would be the same as when 
we work with the magnetic variables. Thus we will ignore
this step and simply reset to starting right away with (\ref{actionnewmemdconf}). 

Again, away from the membranes the 
gravitational sector is identical to standard General Relativity. The variational equations
obtained from (\ref{actionnewmemdconf}) with respect to the metric, $\kappa^2$, $\lambda$, ${\cal A}_{\nu\lambda\sigma}$ 
and ${\cal B}_{\nu\lambda\sigma}$, in that order, are, after some manipulation of the functional derivatives in the matter 
sector (where $(\kappa/{\cal M})^{1/2}$ coincides with the ``stiff dilaton" of \cite{kalseq}), 
\ba
&&~~~~~~~~~~~~~ (\mps + \kappa^2) G^\mu{}_\nu =  - \kappa^2 \lambda \, \delta^\mu{}_\nu + T^\mu{}_\nu + \ldots\, , \nonumber  \\
&& 
\hat {\cal F}_{\mu\nu\lambda\sigma} = \frac{\kappa^2}{2} \sqrt{g} \, {\epsilon_{\mu\nu\lambda\sigma}} \, ,  ~~~
\hat {\cal G}_{\mu\nu\lambda\sigma} = \frac{2 \kappa^2\lambda- \kappa^2 R - T/4}{4\kappa^2} 
\sqrt{g} \, {\epsilon_{\mu\nu\lambda\sigma}} \, , \nonumber \\
&&~~~~ 2 n^\mu \partial_\mu \lambda  = {\cal Q}_A \delta(r-r_0) \, , ~~~~~ 
\frac12 n^\mu \partial_\mu \kappa^2 = {\cal Q}_B \delta(r-r_0) \, . ~~~~
\label{tenseqsgrmagdualconf}
\ea
As before, the ellipsis in the first equation denote the generalization of Israel-Gibbons-Hawking boundary terms. 
Comparing to (\ref{tenseqsgrmagdual}), the only difference is the $\propto \mps$ term in the first equation, and the $\sim T$ term
in the third (where $T = T^\mu{}_\mu$). As a consequence, one can easily 
check that the $4$-form boundary terms remain exactly the 
same as in the previous case with minimal matter couplings. In particular, the equation (\ref{boundactf}) 
does not change. Our generalization of Israel-Gibbons-Hawking action changes a little, by 
replacing $\kappa^2$ in (\ref{gibbhawk}) with 
\be
\kappa^2_{\tt eff} = \mps + \kappa^2 \, .
\label{kappanew}
\ee
With this in mind, 
\be
S^R_{\tt boundary} = - \int d^3 \xi \sqrt{\gamma} [\kappa^2_{\tt eff} K ] \, ,
\label{gibbhawk}
\ee
and we can finally turn to the nonperturbative membrane dynamics.

\section{{\it Sic Transit} ... } 

The presence of membranes with nonvanishing charges and tensions facilitates transitions in the spectrum of values of 
$\kappa^2, \lambda$. 
In any geometry which is locally described by a solution of (\ref{alleqs}), 
with some values of $\kappa^2$, $\lambda$ and the
matter sources, a membrane can nucleate quantum-mechanically with some probability. As long as the net energy density in the region
where nucleation occurs is smaller than $(\kappa^2)^2$, the region can be described as a locally Minkowski space, and the formalism of 
Euclidean bubble nucleation, with the bubble surrounded by a thin membrane, which was originally developed 
by Coleman and collaborators 
\cite{Coleman:1977py,Callan:1977pt,Coleman:1980aw},  can be deployed to compute the nucleation rates. Then  
Euclidean bubbles can be analytically continued back to Lorentzian metric, and their interior geometry can be determined by matching
conditions on a membrane, provided by Israel junction conditions. 

In this section, we focus on determining the membrane nucleation rate and the matching of the exterior (parent) and
interior (offspring ) geometries, in the simplest possible cases. We imagine that both the parent and the offspring  geometries are locally
maximally symmetric, with the symmetry broken only by membrane nucleation. So we assume that the only nontrivial sources of 
the gravitational field are the various contributions to the cosmological constant and the membrane charges and tensions. This will suffice
to sketch out the evolution of a spacetime in the leading order approximation. 

To this end, we will use the actions 
(\ref{actionnewmem}), (\ref{actionnewmemd}), Wick-rotated to Euclidean space, determine 
Euclidean geometries describing
various possible parent-offspring pairs, and compute Euclidean actions 
of these configurations. Our goal is to get an estimate of the rate
of a nucleation processes, $\Gamma \sim e^{-S_{bounce}}$ 
\cite{Coleman:1977py,Callan:1977pt,Coleman:1980aw}, which should be reliable at least in 
the thin-wall, slow nucleation rate regime. 

\subsection{Euclidean Action and Field Equations}

Let us first Wick-rotate the action. At this point it is easier to work with the magnetic dual action (\ref{actionnewmemd}), which we need
to analytically continue to Euclidean space. To analytically continue 
the time, we use $t = - i x^0_E$, which yields $- i \int d^4x \sqrt{g} {\cal L}_{\tt QFT} = - \int d^4x_E \sqrt{g} {\cal L}^E_{{\tt QFT}}$. With the convention 
${\cal A}_{0 jk} = {\cal A}^{E}_{0jk}$, 
${\cal A}_{jkl} =  {\cal A}^{E}_{jkl}$ we have ${\cal F}_{\mu\nu\lambda\sigma} = {\cal F}^{E}_{\mu\nu\lambda\sigma}$, and so on for ${\cal B}$.
Further $\epsilon_{0ijk} = \epsilon^{E}_{0ijk}$ and $\epsilon^{0ijk} = -\epsilon_E^{0ijk}$. The tension and charge terms transform to
$- i {\cal T}_i \int d^3 \xi \sqrt{\gamma} = - {\cal T}_i \int d^3 \xi_E \sqrt{\gamma}$ and $i {\cal Q}_i \int {\cal A}_i = - {\cal Q}_i \int {\cal A}_i$. 
The scalars do not change (but if they include time derivatives, those terms change accordingly). Now, we will be working with 
backgrounds which are locally maximally symmetric, meaning that $\langle {\cal L}^E_{\tt QFT} \rangle = \Lambda_{\tt QFT}$, 
where $\Lambda_{\tt QFT}$ is a matter sector cosmological constant, that includes contributions to an arbitrary order 
in the loop expansion. 

Defining the Euclidean action by $i S = - S_E$, this yields, using $\kappa_{\tt eff}^2 = \mps+\kappa^2$, 
\ba
S_E&=&\int d^4x_E \Bigl\{\sqrt{g} \Bigl(-\frac{\kappa^2_{\tt eff}}{2} R_E + \kappa^2 \lambda 
+ \Lambda_{\tt QFT} \Bigr)- \frac{\lambda}{3} {\epsilon^{\mu\nu\lambda\sigma}_E} \partial_\mu {\cal A}^E_{\nu\lambda\sigma}
- \frac{\kappa^2}{12} {\epsilon^{\mu\nu\lambda\sigma}_E}\partial_\mu {\cal B}^E_{\nu\lambda\sigma} \Bigr\} \nonumber \\
\label{actionnewmemeu}
&& +~S_{\tt boundary} + {\cal T}_A \int d^3 \xi_E \sqrt{\gamma}_A - \frac{{\cal Q}_A}{6} \int d^3 \xi_E \, {\cal A}^E_{\mu\nu\lambda} \, 
\frac{\p x^\mu}{\p \xi^\alpha} \frac{\p x^\nu}{\p \xi^\beta} 
\frac{\p x^\lambda}{\p \xi^\gamma} \epsilon_E^{\alpha\beta\gamma} \\
&& ~~~~+~  {\cal T}_B \int d^3 \xi_E \sqrt{\gamma}_B  - \frac{{\cal Q}_B}{6} \int d^3 \xi_E \, 
{\cal B}^E_{\mu\nu\lambda} \, \frac{\p x^\mu}{\p \xi^\alpha} \frac{\p x^\nu}{\p \xi^\beta} 
\frac{\p x^\lambda}{\p \xi^\gamma} \epsilon_E^{\alpha\beta\gamma} \, . \nonumber
\ea

It is important now to stress the difference between the theories with the minimally coupled matter and the 
conformal $4$-form/matter coupling. In the case of the minimally coupled theory, $\Lambda_{\tt QFT}$ 
is independent of the discrete variable $\kappa^2$. On the other hand, 
for the theory with the conformal $4$-form/matter coupling, 
\be
\Lambda_{\tt QFT} = \frac{\kappa^2}{{\cal M}^2} \bigl({\cal M}_{\tt UV}^4 + \ldots) = \kappa^2 {\cal H}_{\tt QFT}^2 \, , 
\label{cftcc}
\ee
where, as before, ${\cal M}_{\tt UV}^4$ plays the role of the locally flat space QFT cutoff. 
This is because the regulator 
depends on $\kappa^2$ in exactly the same way as the dimensional parameters of ${\cal L}_{\tt QFT}$. 
The ellipsis stand in for subleading corrections. From here on we will simply absorb them into the cutoff. 
As a result, if we define the total cosmological constant,
\be
\Lambda = \Lambda_{\tt QFT} + \kappa^2 \lambda \, , 
\label{cc}
\ee
for both of our theories $\Lambda$ is a linear function of $\kappa^2$. The distinction  is that in the minimal 
case $\Lambda_{\tt QFT}$ is $\kappa^2$ independent, whereas in the conformal $4$-form/matter coupling 
$\Lambda_{\tt QFT} = \kappa^2 \frac{{\cal M}_{\tt UV}^4}{{\cal M}^2} + \ldots$. 
Thus, in what follows we will have the total cosmological constant as 
\be
\Lambda = \begin{cases}
\kappa^2 \lambda + \Lambda_{\tt QFT}\, , 
~~~~~~~~~~~ {\rm minimal~coupling}    \, ;\\
\kappa^2 \Bigl(\lambda + {\cal H}^2_{\tt QFT}\Bigr) \, ,  ~~~~~~~{\rm conformal~coupling}   \, . \\
\end{cases}
\label{ccgen}
\ee

We will look for transitions between geometries with $\kappa^2_{out/in}, \Lambda_{out/in}$, 
where the subscripts {\it out/in} denote parent and offspring geometries (exterior and interior of 
a membrane, respectively). Both of the {\it out/in} geometries may be described with the metrics
\be
ds^2_E =  dr^2 + a^2(r) \, d\Omega_3 \, ,
\label{metricsmax}
\ee
where $d\Omega_3$ is the line element on a unit $S^3$. The Euclidean 
scale factor $a$ is the solution of the Euclidean ``Friedmann equation", 
\be
3 \kappa^2_{\tt eff} \Bigl( \bigl(\frac{a'}{a}\bigr)^2 - \frac{1}{a^2} \Bigr) = -(\Lambda_{\tt QFT} + \kappa^2 \lambda) = - \Lambda \, ,
\label{fried}
\ee
which follows because the bulk 
metric-dependent part of (\ref{actionnewmemeu}) is structurally the same as in standard General Relativity. 
The prime designates an  $r$-derivative\footnote{ 
We won't need the explicit form of the solutions, although they are easy to obtain:
$$a(r) =  a_0 \sin(\frac{r+\delta}{a_0}) \, ,\hfill ~~ {\rm for} ~~ \Lambda > 0 \, ; ~~~~~ 
a(r) = r+\delta  \, , \hfill ~~ {\rm for} ~~ \Lambda = 0 \, ; ~~~~~ 
a(r) = a_0 \sinh(\frac{r+\delta}{a_0}) \, , \hfill ~~ {\rm for} ~~ \Lambda< 0 \, .
$$%
}.
We are focusing on at least $O(4)$-invariant configurations and their 
complex extensions since they have minimal Euclidean action. Hence they describe most likely processes
in this approximation \cite{Coleman:1977py,Callan:1977pt,Coleman:1980aw}. 

The idea now is to assemble together two patches of geometry, each with a local metrics given by
(\ref{metricsmax}) but with different $\kappa^2, \Lambda$, and then use 
the junction conditions to connect the patches into a quilt. 
Since we are working with geometries which have three-spheres $S^3$ as subspaces, we keep 
only the $S^3$ invariant $3$-forms ${\cal A}_{123}$, ${\cal B}_{123}$. 
The magnetic dual field boundary conditions induced on a membrane  -- analogous  
to the boundary conditions for the electric field on the interface between two dielectrics --  
follow from  (\ref{actionnewmemeu}) by varying with respect to ${\cal A}$ and ${\cal B}$. 

The variations
give (where for the notational economy we write them as if both a membrane
charged under ${\cal A}$ and under ${\cal B}$ are co-located; in general, of course they won't be)
\ba
\lambda_{out} - \lambda_{in}  &=& \frac12 {\cal Q}_A \, , \nonumber \\
\kappa^2_{out} - \kappa_{in}^2 &=& 2 {\cal Q}_B \, .
\label{lagrangejc}
\ea
As stated above, {\it out/in} denote a relevant quantity just to the right or to the left of the 
membrane, in the coordinate system where membrane is at rest, and 
where the outward membrane normal vector is oriented in the direction of the 
radial coordinate, and $r$ measures the distance in this direction. 

The metric boundary conditions come from the tension-induced curvature jump on the membrane,
and can be obtained by using Israel junction conditions. Alternatively we can write 
down Einstein's equations in the rest frame of the membrane and determine the 
discontinuity of the second derivative. 
Either way, and again writing the condition as if both ${\cal A}$ and 
${\cal B}$ membranes are co-located, we find 
\ba
 && ~~~~~~~~~~~~~~~~~~~~ a_{out} = a_{in} \, , ~~~~~~~ \nonumber \\
 && \kappa^2_{{\tt eff}~{out}} \frac{a_{out}'}{a} - \kappa^2_{{\tt eff}~{in}} \frac{a_{in}'}{a} 
 = -\frac12 \Bigl( {\cal T}_A + {\cal T}_B \Bigr) \, .
 \label{metricjc}
\ea

Note that we can think of the first of these two equations as just a ``Gaussian pillbox" 
integral of $a' = \pm \sqrt{1-  \frac{\Lambda a^2}{3 \kappa_{\tt eff}^2}}$ obtained by solving (\ref{fried}) for $a'$. Here
$\pm$ in $a'$ allows for either branch of the square root. 

It is important to stress that even though $\kappa^2$ and $\lambda$ 
in this equation are discontinuous across a membrane, since the discontinuity is finite and the membrane is thin, the
metric variable $a$ remains continuous. Similarly, $a'$ jumps because the tension sources are Dirac $\delta$-functions in the thin wall limit. 

Finally, the spectator $4$-forms $\hat {\cal F}$ and $\hat {\cal G}$, given in the first line of Eq.
(\ref{tenseqsgrmagdual}), 
may also experience a discontinuity. On shell they are set by the geometric quantities 
which jump. These discontinuities do not control the geometry matching, but {\it do} contribute to Euclidean
actions -- by generating boundary terms 
(\ref{boundactf}) in Euclidean action which precisely cancel the charge terms and 
the (Euclideanized) spectator terms in (\ref{actionnewmemeu}). So as a result, on shell (dropping the index $``E"$ from here on)
\ba
&&S_{\tt boundary}^{4-{\rm forms}} - \int d^4x \Bigl(\frac{\lambda}{3} 
{\epsilon^{\mu\nu\lambda\sigma}} \partial_\mu {\cal A}_{\nu\lambda\sigma}
- \frac{\kappa^2}{12} {\epsilon^{\mu\nu\lambda\sigma}}\partial_\mu {\cal B}_{\nu\lambda\sigma} \Bigr) \nonumber \\
&&~~~ - \frac{{\cal Q}_A}{6} \int d^3 \xi \, {\cal A}_{\mu\nu\lambda} \, 
\frac{\p x^\mu}{\p \xi^\alpha} \frac{\p x^\nu}{\p \xi^\beta} 
\frac{\p x^\lambda}{\p \xi^\gamma} \epsilon^{\alpha\beta\gamma} - \frac{{\cal Q}_B}{6} \int d^3 \xi \, 
{\cal B}_{\mu\nu\lambda} \, \frac{\p x^\mu}{\p \xi^\alpha} \frac{\p x^\nu}{\p \xi^\beta} 
\frac{\p x^\lambda}{\p \xi^\gamma} \epsilon^{\alpha\beta\gamma} = 0 \, . \, ~~~
\label{eucl4formba}
\ea
Thus in fact correctly evaluated spectator terms cancel out in the action. 
Nevertheless we will write the $3$-potential discontinuities here for completeness, before we 
ignore them once for all thanks
to (\ref{eucl4formba}). It turns out that since the discontinuity of $\kappa^2$ is finite and the metric is continuous, the discontinuity
of $\hat {\cal F}$ is also finite, and hence ${\cal A}_{\mu\nu\lambda}$ is continuous. On the other hand,
since $R$ has a Dirac $\delta$-function divergence induced by the jump of $a'/a$, the $3$-form potential ${\cal B}_{\mu\nu\lambda}$ is
discontinuous, because the Gaussian integral enclosing the membrane is 
\be
\oint d {\cal B}= \oint \hat {\cal G} = - \frac14 \oint d^4x \sqrt{g} R \, .
\label{gaussian}
\ee
Other terms appearing in the equation for ${\cal G}$ are all continuous and therefore drop out from 
the integral here. 
Using $R = - 6 a''/a + \ldots$, and integrating we find that for all cases of interest to us,
\ba
&& ~~~~~~~~~~~ {\cal A}_{\mu\nu\lambda~out} = {\cal A}_{\mu\nu\lambda~in} \, , ~~~~~~~  \nonumber \\
 &&{\cal B}_{\mu\nu\lambda~out} - {\cal B}_{\mu\nu\lambda~in} = -9 \bigl(\frac{a'_{out}}{a} - \frac{a'_{in}}{a}\bigr) \, .
 \label{ABjc}
\ea

\subsection{{The Spectrum of Instantons}}

We can now consider ``elementary transitions" mediated by the emission of a single membrane, with either ${\cal Q}_A$ or ${\cal Q}_B$ 
charge. More general cases are realized by multiple emissions, which generically occur 
consecutively. In any case, those transitions are combinations of the elementary ones, and 
their rates are controlled by linear combinations of Euclidean actions of the ``elementary transitions". 

In determining the `spectrum' of possible instantons, we will closely follow the excellent
expose of \cite{Brown:1987dd,Brown:1988kg}. Much of our analysis, especially in subsection 3.2.1., overlaps with the details
of those works. {\it However there are some crucial changes in results and conclusions 
due to the structural differences between the field equations here and in}  \cite{Brown:1987dd,Brown:1988kg}. 
This will come up shortly, and we will pay particular attention to them, and highlight the differences as we go.  

Since we are working with
several theories simultaneously, we will try to deploy universal notation and analysis whenever possible. In particular the exploration of the
instantons ${\cal T}_A, {\cal Q}_A \ne 0$ is essentially independent on the $\kappa^2$ dependence (which can vary 
$\kappa^2$ dependence between theories) 
and so we will be able to present the results in a 
general fashion. For ${\cal T}_B, {\cal Q}_B \ne 0$ we will look at the 
specific cases separately, since the $\kappa^2$ dependence makes the analysis simpler in one of those cases. 

\subsubsection{ \texorpdfstring{${\cal T}_A, {\cal Q}_A \ne 0$}{Lg} }

The first case, with ${\cal T}_A, {\cal Q}_A \ne 0$ and 
${\cal T}_B = {\cal Q}_B = 0$ obviously is similar to the thin wall bubble nucleation in standard General
Relativity, and to theories with membrane discharge of the flux screened cosmological constant. 
However there are important technical differences when we compare to those models since in our theory 
the bulk cosmological constant depends on the $4$-form dual magnetic fluxes 
(bi)linearly, as opposed to quadratically \cite{Hawking:1984hk,Brown:1987dd,Brown:1988kg,Duncan:1989ug,Bousso:2000xa,Feng:2000if}, 
as is clear from Eq. (\ref{fried}). This will lead to interesting new features, breaking up the spectrum of instantons describing allowed transitions into two 
separate, disjoint sectors. 

In any case, the relevant boundary conditions we found in the previous section on a membrane are
\ba
&&~~~~~~ a_{out} = a_{in} = a \, , ~~~~~~ \kappa^2_{{\tt eff}~out} = \kappa_{{\tt eff}~in}^2 
= \kappa_{\tt eff}^2\, ,~~~~~~  {\cal A}_{\mu\nu\lambda~{out}} = {\cal A}_{\mu\nu\lambda~{in}} \, , \nonumber \\
&&\frac{a_{out}'}{a} - \frac{a_{in}'}{a} = -\frac{{\cal T}_A}{2\kappa_{\tt eff}^2} \, ,
~~~~~~  \lambda_{out} - \lambda_{in} =  \frac12 {\cal Q}_A \, , 
~~~~~~ {\cal B}_{\mu\nu\lambda~{out}} - {\cal B}_{\mu\nu\lambda~{in}} = \frac{9{\cal T}_A}{2\kappa_{\tt eff}^2} \, .
\label{bcsA}
\ea
Let us very briefly review the meaning of these boundary conditions. The point here is that to find the solution
we must allow $\lambda$ to jump across the membrane, since it is a dual magnetic the flux to ${\cal G}$,
which changes due to the $A$-membrane charge. The other jump, in $a'$, is accommodated by arranging for the membrane
to reside at just the right value of $a$, which scans the range of the parent geometry until it settles to the right value.

Clearly, for compact geometries, either parent or offspring, the range of $a$ is bounded, and thus for many values of 
parameters $a$ will not exist. In the case of noncompact geometries, on the other hand, the Euclidean bounce may involve
infinite volume contributions, which are positive. This will infinitely suppress the configuration, even if it is not excluded `kinematically'. 
Thus only a subset of transitions will be physically relevant. 

Solving Eq. (\ref{fried}) for $a' = \zeta_j \sqrt{1-  \frac{\Lambda a^2}{3 \kappa_{\tt eff}^2}}$, with 
$\zeta_j = \pm 1$ designating the two possible branches of the square root, 
we rewrite the first two equations in the second line of (\ref{bcsA}) as 
\ba
\zeta_{out} \sqrt{ 1-  \frac{\Lambda_{out} a^2}{3 \kappa_{\tt eff}^2}} 
- \zeta_{in}  \sqrt{1-  \frac{\Lambda_{in} a^2}{3 \kappa_{\tt eff}^2}} &=& -\frac{{\cal T}_A a}{2\kappa_{\tt eff}^2}\, , \nonumber \\
\zeta_{out} \sqrt{1-  \frac{\Lambda_{out} a^2}{3 \kappa_{\tt eff}^2}} 
+ \zeta_{in} \sqrt{1-  \frac{\Lambda_{in} a^2}{3 \kappa_{\tt eff}^2}} &=& \frac{\kappa^2 {\cal Q}_A a}{3{\cal T}_A} \, . 
\label{balanceA}
\ea
The first equation is obvious. To get the second,  start with 
$a'^2_{out} - a'^2_{in} =- a^2 \frac{\kappa^2}{\kappa_{\tt eff}^2} (\lambda_{out}-\lambda_{in})/3$ 
which follows from (\ref{fried}) and the second equation on the second
line of (\ref{bcsA}), factorize the difference of squares, and use the first equation to replace $a_{out}'-a_{in}'$. Importantly, the 
second equation does {\it not} involve the background $4$-form flux on the R.H.S. due to the linear 
dependence of $\Lambda$ on $\lambda$ (as is clear from the fact that R.H.S. depends on ${\cal Q}_A$ linearly, as opposed to 
quadratically). This leads to differences in solutions when compared to \cite{Brown:1987dd,Brown:1988kg}. 

\begin{figure}[htb]
    \centering
    \includegraphics[width=12cm]{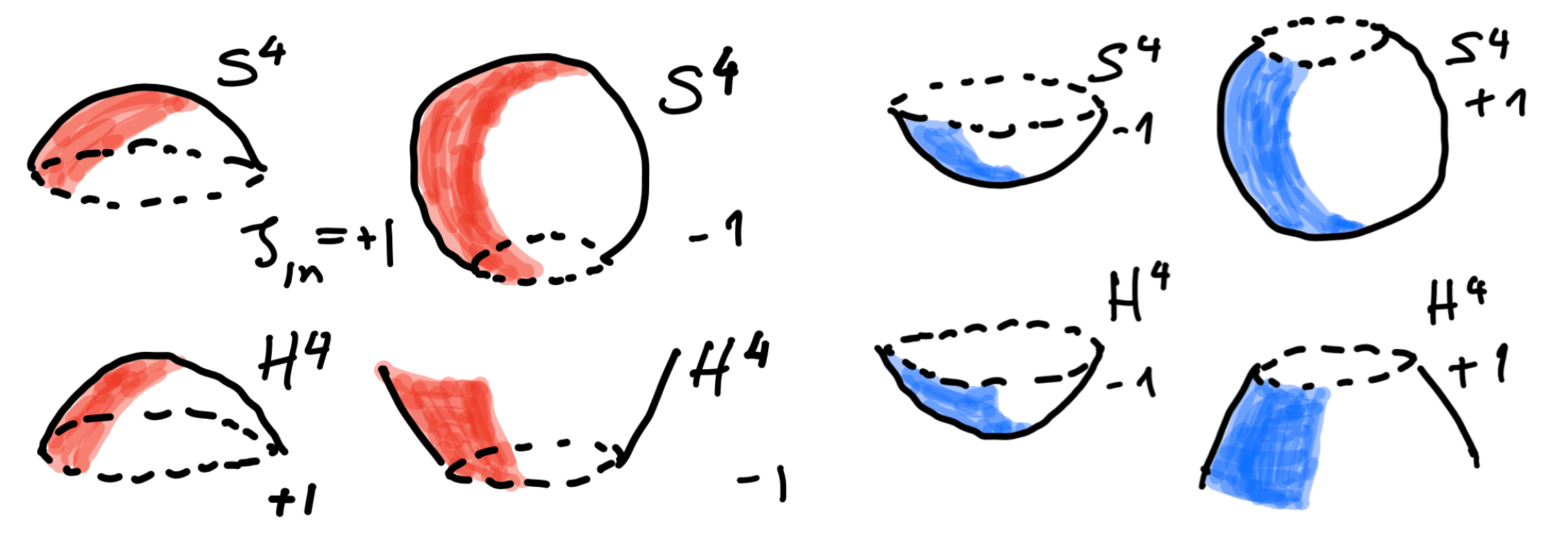}
  \caption{Spherical ($S^4$, top row) and horospherical (a.k.a. hyperbolic; $H^4$, bottom row) 
  sections which are glued together to form instantons. Red ones are the interiors and
    the blue ones the exterior geometries of the instanton. The $\pm$ are the values of $\zeta_{in/out}$.}
    \label{sections}
\end{figure}

The possible configurations which can be obtained by gluing together sections of exterior and 
interior metrics (\ref{metricsmax}) are counted by the variations of the sign of $\Lambda$ and the 
branches of solutions ($\zeta_j = \pm 1$) of Euclidean Friedmann equation (\ref{fried}). They must 
satisfy the Eqs. (\ref{balanceA}), however. The ``sections" of Euclidean space, which should be 
sewn together to construct the complete instanton configuration are qualitatively 
the same as those taxonomized by
\cite{Brown:1987dd,Brown:1988kg}. We sketch them in Fig. (\ref{sections}). 

Here the ``red" shaded sections correspond to the possible interior patches of the geometry, and the ``blue" ones
to the exterior patches. The spherical sections $S^4$ arise when $\Lambda > 0$ and the horospherical sections
$H^4$ when $\Lambda < 0$. After Wick rotation back to Lorentzian signature, $S^4$ become patches of de Sitter 
and $H^4$ turn into Anti-de Sitter. The sign $\zeta_i$, controlling which branch of the square root we pick, controls 
geometrically whether the circumference of the latitude circle on the section near the cut (the location
of the membrane, represented by the dashed circle in Fig. (\ref{sections})) increases ($\zeta_i = +1$) or
decreases ($\zeta_i = -1$) by parallel transport increasing the arc length $a$ in the direction of the positive normal
to the membrane (directed outwards) -- ie away from the coordinate origin at the center 
of the "inside" section, which we will take to be the North Pole (see below). 

The equations (\ref{balanceA}) restrict the possible combinations of these sections
already kinematically. In fact we can simplify the 
Eqs. (\ref{balanceA}) by adding and subtracting them:
\ba
\zeta_{out} \sqrt{ 1-  \frac{\Lambda_{out} a^2}{3 \kappa_{\tt eff}^2}} 
&=& -\frac{{\cal T}_A}{4\kappa_{\tt eff}^2}\Bigl(1 -  \frac{2\kappa_{\tt eff}^2 \kappa^2{\cal Q}_A}{3{\cal T}^2_A} \Bigr)\, a \, , \nonumber \\
\zeta_{in} \sqrt{1-  \frac{\Lambda_{in} a^2}{3 \kappa_{\tt eff}^2}} 
&=& \frac{{\cal T}_A }{4\kappa_{\tt eff}^2}\Bigl(1 + \frac{2\kappa_{\tt eff}^2 \kappa^2 {\cal Q}_A}{3{\cal T}^2_A} \Bigr) \, a \, . 
\label{diffroots}
\ea
Since ${\cal T}_A>0$, the signs $\zeta_{out/in}$ are completely controlled by the ratio 
\be
q = \frac{2\kappa_{\tt eff}^2 \kappa^2 |{\cal Q}_A|}{3{\cal T}^2_A} \, .
\label{qbound}
\ee
Exploring the possibilities for the ``assembly" of the instanton solutions we find
\begin{itemize}
\item
if $q <1$, the only allowed combination of $\zeta$'s is $\zeta_{out} = -1, \zeta_{in}=+1$. All other
options are excluded;
\item
if $q>1$, then we can have two combinations:
$\zeta_{out} = -1, \zeta_{in}=-1$ for ${\cal Q}_A < 0$ and $\zeta_{out} =+1, \zeta_{in}=+1$ for ${\cal Q}_A > 0$; the other
two combinations are excluded.
\end{itemize}
The listed cases might not be automatically completely disjoint: $q>1$ might evolve to $q<1$, and vice versa, 
{\it iff} $\kappa_{\tt eff}^2$ changes from 
bubble to bubble by the emission of ${\cal Q}_B \ne 0$ membranes. Crucially, however, the processes
which could flip $q<1$ to $q>1$ can be completely blocked off. We will discuss this issue in much more
detail further along. For now, we merely note that in a given bubble, the 
membrane emissions will only yield one of the two cases here. This is a direct consequence of the
fact that $\Delta \Lambda$ depends on ${\cal Q}_A$ linearly and not quadratically, as in  \cite{Brown:1987dd,Brown:1988kg}.

Therefore, kinematically allowed combinations $(\zeta_{out}, \zeta_{in})$ are $(-,+)$ for $q<1$ and $(+,+)$, $(-,-)$ for
$q>1$. The combination $(+,-)$ is kinematically completely prohibited for any signs and values of $\Lambda_{out/in}$ by ${\cal T}_A > 0$.
In addition one can check by examination of Eqs. (\ref{diffroots}) that the instantons mediating transitions 
$\Lambda_{out} \le 0, \zeta_{out}=+1 \rightarrow \Lambda_{in} > 0, \zeta_{out}=+1$
and  $\Lambda_{out} > 0, \zeta_{out}=-1 \rightarrow \Lambda_{in} \le 0 , \zeta_{out}=-1$ are also kinematically
prohibited. This is identical to what was found in \cite{Brown:1987dd,Brown:1988kg}. 
\begin{figure}[ht]
    \centering
    \includegraphics[width=11cm]{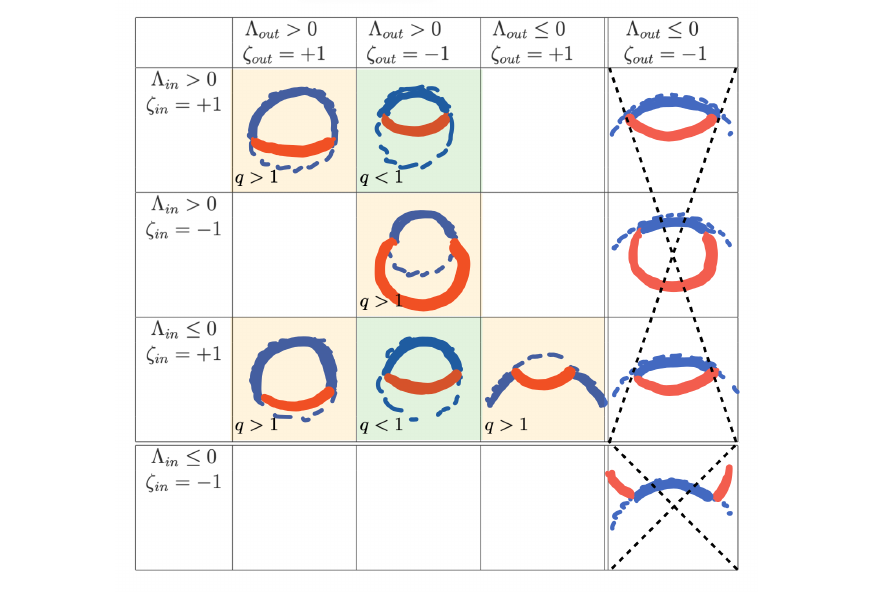}
    \caption{The instanton `Baedeker'.  The instantons fall into four types, divided by 
    double lines in the table, and counted clockwise from the top corner \cite{Brown:1988kg}.  The transitions corresponding to
    empty squares are ruled out kinematically by  Eqs. (\ref{balanceA}), (\ref{diffroots}). The top nine are further split by
    $q = \frac{2\kappa_{\tt eff}^2 \kappa^2 |{\cal Q}_A|}{3{\cal T}^2_A}<1$ (pale green) or $q>1$ (pale gold). We keep both 
     since $\kappa_{\tt eff}^2$ {\it might} vary independently (we will suppress those variations later on). 
    The ``ogre"-like configurations in the right column which are crossed out are allowed kinematically, 
    but are suppressed dynamically since their bounce action is huge and positive, $S_{\tt bounce} \gg 1$, diverging when Anti-de Sitter 
    sections are non-compact (see the text). } 
    \label{allcases}
\end{figure}

The list of the possible instantons 
is given in the instanton `Baedeker' of Fig. (\ref{allcases}). We taxonomize 
the allowed possibilities of $(\Lambda_{out},\zeta_{out},\Lambda_{in},\zeta_{in})$ which are solutions 
of Eqs. (\ref{diffroots}). The classification of the possible solutions in \cite{Brown:1987dd,Brown:1988kg}
is extremely convenient. The tabular representation of Fig. (8) of that work sums the options very concisely, 
and we adopt it here as well. {\it A key qualitative 
difference in our case is that the so-called type 1 instantons, comprising the top
nine examples, separated by the two double lines from the rest in Fig. (\ref{allcases}), are additionally divided
into two subsets depending on the local value of $q$. If $q<1$, only  $(\zeta_{out}, \zeta_{in})=(-,+)$ are allowed.
If $q>1$, only  $(\zeta_{out}, \zeta_{in})=(+,+)$ or $(-,-)$ can occur.} 

In Fig. (\ref{allcases}) the dashed contours depict the initial, exterior geometry, given by $S^4$ 
(depicted by spherical cross sections) or $H^4$ (the hyperbolic cross sections). The solid contours show the
cross sections of the instantons, the blue being the retained section of the parent in the exterior,
and the red the offspring in the interior. The empty squares are kinematically prohibited, 
such as e.g. all cases ${\zeta_{out}=+1}, \zeta_{in}=-1$,
by Eqs. (\ref{balanceA}), (\ref{diffroots}). An important
feature to pay attention to, which is a particularly useful aspect of the taxonomy of \cite{Brown:1987dd,Brown:1988kg}, 
is the manifest difference of the exterior and interior geometries seen when comparing the solid red contours
with the dashed blue ones. In most cases when the initial exterior geometry is not compact, the bounce 
action is divergent. Positivity of the action then implies those instantons are impossible 
dynamically, as we are about to
see explicitly shortly. The instantons are divided into four types by the double lines, 
1 through 4, counting clockwise from the top corner.

\begin{figure}[thb]
    \centering
    \includegraphics[width=7cm]{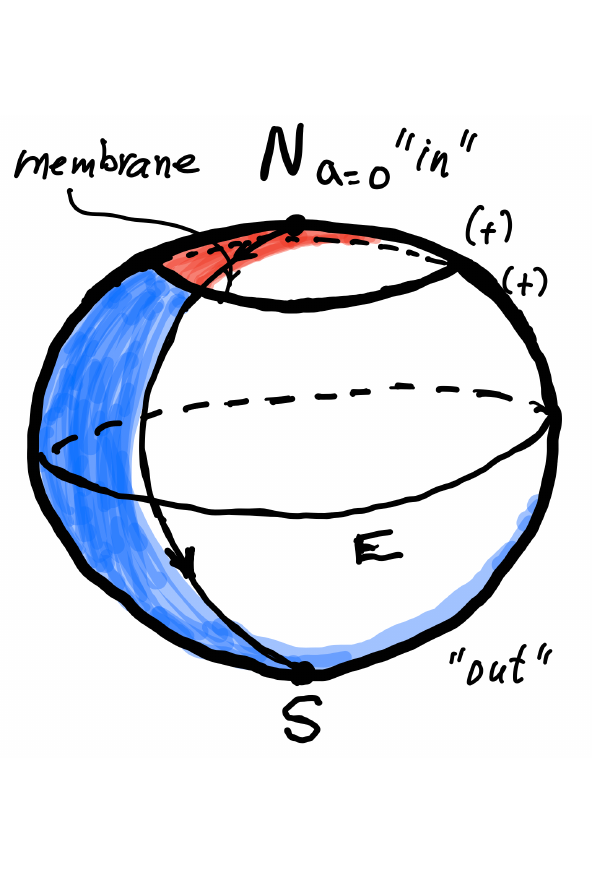}
    \caption{A cartoon of an instanton comprised of two sections of $S^4$. 
    The region around the North Pole, shaded red, has a larger curvature radius because ${\cal T}_A > 0$, by 
    Israel junction conditions \cite{israel}.}
    \label{fig1}
\end{figure}

To illustrate how to patch the instantons together and ensure they are
solutions of (\ref{fried}), (\ref{bcsA})--(\ref{diffroots}), let's 
consider a special case when both the exterior and the interior solutions have $\Lambda > 0$,
so that each is locally a section of a $4$-sphere $S^4$. Let's also consider the configuration 
$\zeta_{out/in} =+1$. To coordinatize the geometry, we can start with the interior solution, a section of $S^4$ 
with the radius $\kappa_{\tt eff}\sqrt{3/\Lambda_{in}}$. Choosing as the origin of coordinates the North Pole, we 
proceed away from it along a fixed longitude, parameterizing the distance from the pole by the 
arc length $a$, which is zero at the North Pole. At the value of $a$ which satisfies (\ref{balanceA}) 
for given parameters, we terminate the interior by placing the membrane along the latitude `circle' $S^3$. 
Crossing the membrane at this latitude, we are in the exterior region, which is locally also an $S^4$, of the radius 
$\kappa_{\tt eff} \sqrt{3/\Lambda_{out}}$, and we continue to move along a longitude until we reach the 
South Pole. The signs $\zeta_{out} =+1, \zeta_{in}=+1$ control the location of the latitude $S^3$, along 
which the membrane resides, relative  to each pole. If the section of 
the $S^3$ on the interior does not include the equator between the North Pole and the membrane latitude, 
we choose $\zeta_{in} = +1$, since the perimeter of the latitude increases with the arc length from the pole. 
On the exterior section, 
the assignment for $\zeta$ is reversed: if the Southern cap does not include the equator, the radius of the 
latitudes is decreasing along a longitude as $a$ grows, reversing $\zeta_{out}$ to $-1$, 
and vice versa if the equator is included. 
And so on for other cases. We depict our chosen example $\zeta_{out} =+1, \zeta_{in}=+1$ in Fig. (\ref{fig1}).  

It is straightforward to compute Euclidean action of the solution, and also the bounce action. The bounce action
is defined as the difference 
of the membrane-induced instanton and the Euclidean action of the parent geometry, 
\be
S({\tt bounce}) = S({\tt instanton}) - S({\tt parent}) \, .
\label{bounceact}
\ee
The `decay rate' is then \cite{Coleman:1977py,Brown:1987dd,Brown:1988kg}
\be
\Gamma \sim e^{-S({\tt bounce})} \, .
\label{decrate}\ee

One can easily see that the bounce actions of instantons of type 2 and 3 are divergent. In the type 2 cases, the reason is that 
the outside, parent geometry, is non-compact, and has negative curvature. Thus the contribution to the parent Euclidean action from 
the exterior geometry to the membrane is, after integrating over the angular variables 
on $S^3$ (which yields a factor of $V_{S^3} = 2\pi^2$), regulating the 
exterior geometry with the infrared cutoff $L$ and including the exterior curvature term on the inside of the boundary 
at $a=L$, recalling that for all type 2 instantons $\zeta_{out} = -1$, $\Lambda_{out} < 0$ and
$K_{out} = 3(a'/a)_{out} = 3 \frac{\zeta_{out}}{a} {\sqrt{1+\frac{|\Lambda_{out}| a^2}{3\kappa_{\tt eff}^2}}}$, 
\ba
S_{out}({\tt parent}) &=& - 2 \pi^2 |\Lambda_{out}| \int_{membrane}^L 
\frac{da\, a^3}{\zeta_{out} \sqrt{1+\frac{|\Lambda_{out}| a^2}{3\kappa_{\tt eff}^2}}}
+ 2\pi^2  \kappa_{\tt eff}^2 \bigl(a^3 K\bigr)|_L  \, , \nonumber \\
&=& 2\pi^2 \kappa^2_{\tt eff} \Bigl( \sqrt{\frac{|\Lambda_{out}|}{3\kappa^2_{\tt eff} }} L^3 \bigl(1+ {\cal O}(1/L) + \ldots \bigr) -  
3 {L}^2 {\sqrt{1+\frac{|\Lambda_{out}| L^2}{3\kappa_{\tt eff}^2}}} \Bigr)  \nonumber \\
&=& - 4\pi^2 \kappa_{\tt eff} \sqrt{\frac{|\Lambda_{out}|}{3}} L^3 
\bigl(1+ {\cal O}(1/L) + \ldots \bigr)  \rightarrow - \infty|_{L \rightarrow \infty} \, .\label{gibbhawkfully}
 \ea
The special case of $\Lambda_{out} = 0$ is also divergent due to the divergent area of the regulator boundary.
Thus the bounce action $S({\tt bounce})$ picks up the contribution from $- S_{out}({\tt parent}) \rightarrow + \infty$, and so 
$\Gamma_{\rm type\,2} \rightarrow 0$. 

Similarly, in the case of type 3 instantons (rightmost bottom corner of Fig. (\ref{allcases})), the bounce action receives divergent contributions 
from both the divergent exterior and interior sections of the geometry. Again, we need to regulate the divergences covariantly, introduce
the appropriate boundary terms with the cutoffs and then take the limit when the boundaries are sent to infinity. Eqs. (\ref{diffroots}) show 
that in this case $|\Lambda_{out}| > |\Lambda_{in}|$ and as a result, $S_{\tt bounce} \rightarrow \infty$ and so also 
$\Gamma_{\rm type\,3} \rightarrow 0$. The only dynamically allowed transitions are those mediated by the instantons of type 1, same as 
in \cite{Brown:1987dd,Brown:1988kg}. And as there, only the ones whose ``squares" aren't blank. 

Note, that this conclusion about type 2 and 3 instantons rests on the assumption that the Anti-de Sitter sections are not compact. If they
were compactified, the bounce actions need not be divergent. However, they would still be very large and positive, proportional to the 
volume of the compact region. This would suppress them relative to the other instantons. Our discussion assumes this \cite{Hawking:1984hk}. 

For type 1 instantons, the contributions in (\ref{bounceact}) coming from the exterior of the membrane exactly cancel against the 
corresponding parent action contribution, as is obvious from Fig. (\ref{allcases}), and we need to 
only integrate over the interior, up to and including membrane terms, but bearing in mind that 
the spectator terms, the membrane charges and $4$-form boundary terms mutually 
cancel as per our discussion above. Then, substituting\footnote{A subtlety in this step concerning relative signs of 
contributions to $R$ was pointed out in \cite{duff}, where it was noted 
that replacing solutions back into the Euclidean action on shell must be done carefully. In our case, since 
the variables $\kappa_{\tt eff}^2$ and 
$\lambda$ are discrete dynamical variables, instead of Lagrangian parameters, and the boundary terms and total derivatives combine into their 
field equations, the sign we obtain is the correct one. This was also noticed in, e.g. \cite{Duncan:1989ug,hawktur}.} 
$\kappa_{\tt eff}^2 R= 4(\kappa_{\tt eff}^2 \lambda+\Lambda_{\tt QFT}) = 4 \Lambda$ in the bulk integrals in (\ref{actionnewmemeu}),
\ba
S({\tt bounce}) &\equiv& -\int_{\tt instanton}  d^4x  \sqrt{g} \Lambda  
+ \int d^3\xi  \sqrt{\gamma} {\cal T}_A  + \int d^3 \xi \sqrt{\gamma} [\kappa_{\tt eff}^2 K ] 
 + \int_{\tt parent}  d^4x \sqrt{g} \Lambda  ~~ \nonumber \\
&=&\, -\int d^3\xi  \int_{North~Pole}^{membrane} dr\sqrt{g}|_{in} \, \Lambda_{in} 
+ \int d^3\xi  \sqrt{\gamma} {\cal T}_A  + \int d^3 \xi \sqrt{\gamma} [\kappa_{\tt eff}^2 K ] \nonumber \\
&~&-\int d^3\xi  \int^{South~Pole}_{membrane} dr\sqrt{g}|_{out} \, \Lambda_{out}  
 + \int d^3\xi \int_{North~Pole}^{South~Pole} dr\sqrt{g}|_{out} \Lambda_{out}  \, .
\label{bounceactcompact}
\ea
We rewrote the first term in the first line of (\ref{bounceactcompact}) 
splitting it into two pieces, as the the first term in second and
third lines, to make manifest the partial cancellation between the last two terms, in the last line. This leaves us with a very 
simple final expression for the bounce action. Integrating over the remainder of $S^3$ coordinates
covering the interior section,
\ba
S({\tt bounce}) &=&  - 2\pi^2 \Lambda_{in}\int_{North~Pole}^{membrane} da \Bigl(\frac{a^3}{a'}\Bigr)_{in}  
+ 2\pi^2 \Lambda_{out}\int_{North~Pole}^{membrane} da \Bigl(\frac{a^3}{a'}\Bigr)_{out} \nonumber \\ 
&&  + 2 \pi^2 a^3 {\cal T}_A + 6\pi^2 \kappa_{\tt eff}^2 a^2 a'_{out} - 6\pi^2 \kappa_{\tt eff}^2 a^2 a'_{in} \, ,
\label{bounce1d}
\ea
where the domain of integration is over the interval of $a$ which 
covers the interior of any of the type 1 instantons from the table in Fig. (\ref{allcases}), 
from the North Pole to the ``seam" where the parent and offspring geometries are sewn together, 
as depicted in Fig. (\ref{fig1}). The boundary terms are evaluated at the latitude $a$ where the membrane is to be located. 
We used here
$\int d^3 \xi \sqrt{\gamma} [\kappa_{\tt eff}^2 K]= 6\pi^2 a^3\kappa_{\tt eff}^2 \bigl((a'/a)_{out} - (a'/a)_{in}\bigr)$ 
since $\kappa^2_{{\tt eff}~out} = \kappa^2_{{\tt eff}~in} = \kappa_{\tt eff}^2$. 

Note that the cancellation of the ``outside" terms, which
are the contributions of the parent geometry to the instanton and the `parent reference' actions, 
means we retain the integral over the complement of the outside geometry of the instantons. 
This is the residual part of the parent Euclidean action after the outside volume contributions cancelled
between the instanton and the parent actions \cite{Coleman:1980aw,Brown:1987dd,Brown:1988kg}.

Now, clearly, when the seam 
coincides with the location of the membrane solving the Eqs. (\ref{fried}), (\ref{bcsA}), we have 
$(a'/a)_{out} - (a'/a)_{in} = - {\cal T}_A/2\kappa_{\tt eff}^2$, combining the last term in (\ref{bounce1d})
with the tension term in the on-shell bounce action. However, before computing this action
on shell, it is instructive to let $a$ move off the membrane latitude, and consider the
bounce action as a variational principle for it \cite{Brown:1987dd,Brown:1988kg}. This is a braney variant of
d'Alembert's principle of virtual works. In that case, 
\ba
S({\tt bounce}, a) &=& 2 \pi^2 a^3 {\cal T}_A - 2\pi^2 \Bigl( \Lambda_{in}\int_{North~Pole}^{a} da \Bigl(\frac{a^3}{a'} ~ \Bigr)_{in}  
+ 3 \kappa_{\tt eff}^2 a^2 
\zeta_{in} \sqrt{1 - \frac{\Lambda_{in} a^2}{3\kappa_{\tt eff}^2}} ~ \Bigr) 
 \nonumber \\ 
&&~~ + 2\pi^2 \Bigl( \Lambda_{out}\int_{North~Pole}^{a} da \Bigl(\frac{a^3}{a'}\Bigr)_{out}  
+ 3 \kappa_{\tt eff}^2 a^2  \zeta_{out} \sqrt{1 - \frac{\Lambda_{out} a^2}{3\kappa_{\tt eff}^2}} ~ \Bigr) \, , ~~~
\label{bounce1dact}
\ea
plugging $a'_j =  \zeta_{j} \sqrt{1 - \frac{\Lambda_{j} a^2}{3\kappa_{\tt eff}^2}}$ in the boundary terms, as we explained above. 
As a check, this expression coincides with the 
bounce action of \cite{Brown:1987dd,Brown:1988kg}. 

The minimum of $S({\tt bounce}, a)$, after solving $\partial_a S({\tt bounce}, a) = 0$, is  
precisely at the value of $a$ which satisfies the first of the Eqs. (\ref{balanceA}).
The junction conditions pick exactly the latitude of the membrane such that the effective ``energy" of the configuration
given by $S({\tt bounce}, a)$ is minimized. As noted by Coleman et 
al \cite{Coleman:1977py,Callan:1977pt,Coleman:1980aw}, the problem of gluing together 
two geometric patches with different intrinsic curvature 
along a membrane is physically equivalent to the problem of emergence of a bubble wall separating two 
different phases of a medium. The bubble can only emerge if the energy cost due to the surface tension is compensated by the
energy gain of changing the excess latent heat in the interior of the bubble. This is precisely why the
integration in  $S({\tt bounce}, a)$ is over the interior, complement volume: the integral 
in (\ref{bounce1d}) $\propto  \Lambda_{out}$ is not over the region
occupied by the outside of the bubble, which is still the original parent phase, but over its interior complement (including 
the corresponding flip of the sign $\zeta_{out}$). The 
integrals in (\ref{bounce1d}), (\ref{bounce1dact}) comprise the energy difference in the bulk which balances
the energy of the `areal' tension term. 

Conversely, a membrane cannot nucleate when the bulk energy gain is insufficient. 
The energy bound can be understood geometrically as a condition that the membrane latitude $a$ must be a real
number, if a solution is to exist \cite{Brown:1987dd,Brown:1988kg}. We can easily solve Eqs.
(\ref{diffroots}) for $a$, 
\be
\frac{1}{a^2} = \frac{\Lambda_{out}}{3\kappa_{\tt eff}^2} + \Bigl(\frac{{\cal T}_A}{4 \kappa_{\tt eff}^2}\Bigr)^2 
\Bigr(1 - {\frac{2\kappa_{\tt eff}^2 \kappa^2{\cal Q}_A}{3 {\cal T}_A^2}}\Bigr)^2 
= \frac{\Lambda_{in}}{3\kappa_{\tt eff}^2} + \Bigl(\frac{{\cal T}_A}{4 \kappa_{\tt eff}^2}\Bigr)^2 
\Bigr(1 + {\frac{2\kappa_{\tt eff}^2 \kappa^2 {\cal Q}_A}{3 {\cal T}_A^2}}\Bigr)^2 \, .
\label{radii}
\ee
We see that the transitions will stop for kinematic reasons if $\Lambda_{j}$ 
are too negative. The real solutions for $a$ will disappear. This is the reason
behind the empty squares in the `Baedeker' of Fig. (\ref{allcases}). 

We also see that for fixed ${\cal T}_A$, 
${\cal Q}_A$ and $\kappa^2_{\tt eff}$, the size of a nucleating bubble is a monotonically decreasing 
function of $\Lambda$. For $q<1$, its minimal value, set by $\Lambda/\kappa^4_{\tt eff} \la 1$, 
is therefore never much smaller than
$a_{min} \simeq 4 \frac{\kappa^2_{\tt eff}}{{\cal T}_A} = 4 \frac{\kappa^3_{\tt eff}}{{\cal T}_A} \kappa^{-1}_{\tt eff}$
as long as $\Lambda \ge 0$. This 
is much larger than the effective Planck length $1/\kappa_{\tt eff}$ for $\frac{{\cal T}_A}{\kappa^3_{\tt eff}} \ll 1$. 
The dynamics of the bubbles  $q < 1$ is therefore safely separated from the quantum gravity regime 
when $\frac{{\cal T}_A}{\kappa^3_{\tt eff}} \ll 1$ and at least one of the two cosmological terms $\Lambda$ is nonnegative. 

We can use the equations in (\ref{radii}) to express $a$ in terms of
$\Lambda_{out}$, $\Lambda_{in}$ and the membrane tension ${\cal T}_A$, eliminating ${\cal Q}_A$. 
The boundary condition for $\lambda_j$ -- or equivalently, the subtraction of the two equations in (\ref{radii}) --
yields $\Lambda_{out} - \Lambda_{in} = \kappa^2 {\cal Q}_A/2$. The sum of the two equations
lets us express $1/a^2$ as their arithmetic mean. Then eliminating $\kappa^2 {\cal Q}_A = 2 \Delta \Lambda$
and manipulating the equation yields
\be
a^2 = \frac{9 {\cal T}_A^2}{{\bigl( \Lambda_{out} + \Lambda_{in} + \frac{3{\cal T}_A^2}{4 \kappa^2_{\tt eff}} \bigr)^2
- 4\Lambda_{out} \Lambda_{in} }} \, .
\label{radiusa}
\ee
This is merely the thin wall formula for the radius, derived in \cite{Coleman:1980aw}, translated to our case. 
However one should take the square root of this equation carefully, in order to follow the `Baedeker' structure
of Fig. (\ref{allcases}), to satisfy the `superselection' rules imposed on 
the `square root' equations (\ref{diffroots}) which take into account the signs $\zeta_i$. 
This subtlety is somewhat obscured with the procedure of calculating (\ref{radiusa}) in 
\cite{Coleman:1980aw}, where it is obtained by minimizing the bounce action (\ref{bounce1dact}).
In the limit of taking $\Lambda_{out}$ below zero, the bounce action computed after the fact remains 
the same, but the prescription for $\zeta_{out}$ jumps discontinuously, since otherwise the
bounce action would have diverged. We 
will analyze these bounds in detail below, since their implications are quite consequential. 

We can finally write down the bounce action for the type 1 instantons in its explicit form. Evaluating the 
boundary terms in Eq. (\ref{bounce1dact}) using the junction conditions in Eqs. (\ref{diffroots}), we find
\be
S({\tt bounce}) 
=  2\pi^2 \Bigl\{ \Lambda_{out}\int_{North~Pole}^{a} 
da \Bigl(\frac{a^3}{a'}\Bigr)_{out}  -\Lambda_{in}\int_{North~Pole}^{a} da \Bigl(\frac{a^3}{a'}\Bigr)_{in}  \Bigr\} - \pi^2 a^3 {\cal T}_A \, .
\label{bounce1dshell}
\ee
The integrals are straightforward to compute, recalling that they are combinations of
various definite integrals of the primitive function 
$\int^{\frac{\Lambda a^2}{3\kappa_{\tt eff}^2}} \frac{xdx}{\sqrt{1-x}}$, and that integrals may cross over the equator 
$\frac{\Lambda a^2}{3\kappa_{\tt eff}^2} = 1$, where branches change, 
in which case they have to be split into two terms. 

Direct evaluation gives,
irrespective of the sign of $\Lambda_{in}$, but bearing in mind that the integral is
over the inside of the instanton volume,
\be
2\pi^2 \Lambda_{in} \int^a_{North~Pole} da \Bigl(\frac{a^3}{a'} \Bigr) = 18 \pi^2 
\frac{\kappa_{\tt eff}^4}{\Lambda_{in}} \Bigl( \frac23 -  \zeta_{in} \bigl({1 - \frac{\Lambda_{in} a^2}
{3\kappa_{\tt eff}^2}}\bigr)^{1/2} + \frac{ \zeta_{in}}{3} \bigl(1 - \frac{\Lambda_{in} a^2}{3\kappa_{\tt eff}^2}\bigr)^{3/2} \Bigr) \, ,
\label{integrals}
\ee
where different branches are reproduced with the sign assignment of $\zeta$, while the total dimensionless
volume factor always remains $\frac12 \int_0^1 \frac{xdx}{\sqrt{1-x}} = \frac23$. This follows from the 
total volume formula for a unit $S^4$, which is $24\pi^2$, such that each $S^4$ hemisphere has the volume
$18 \pi^2 \times \frac23 = 12 \pi^2$. 

For the outside contribution, we have to do the integral $\int^a_{North~Pole} da$ over the complement of the 
outside volume. We must flip signs when crossing the (imaginary or real) equator, and account 
for the local signs of $\zeta_{out}$ which coincide with the sign on the (cancelled) outside volume 
to the membrane. In the end this produces formally the same expression as in (\ref{integrals}), with
$\Lambda_{in} \rightarrow \Lambda_{out}$, $\zeta_{in} \rightarrow \zeta_{out}$. 
Indeed, as a quick check, note that for $\zeta_{out} = -1$, the ``outside" region for small $a$ is a polar cap around the South Pole.
As the radius goes to zero, the integral over the outside region vanishes. Hence 
the complement must max out. And indeed,
plugging $\zeta =-1$ and $a = 0$ in (\ref{integrals}) produces 
$24 \pi^2 \kappa^4_{\tt eff}/\Lambda_{out}$, and so the Euclidean action
coincides with the parent entropy, given by de Sitter horizon area divided by $4G_N$, as expected \cite{gibbhawk,gibbhawkcosm}.

Various terms in 
(\ref{bounce1dshell}), (\ref{integrals}) can be directly evaluated by substituting again Eqs. (\ref{diffroots}). 
We will discuss these terms shortly, when we turn to physical and phenomenological implications of the
various transitions as a function of the background and the membrane parameters. 

\subsubsection{\texorpdfstring{${\cal T}_B, {\cal Q}_B \ne 0$}{Lg} }

The instantons mediated by membranes with ${\cal T}_B, {\cal Q}_B \ne 0$ 
are a new feature, and to the best of our knowledge have never  
been considered previously in the literature. Nevertheless, the analysis is quite straightforward, and it proceeds as in the 
previous case. The full set of the boundary conditions describing the jumps on a membrane are
\ba
\label{bcsB}
&&\,\, a_{out} = a_{in} = a \, , ~~~~~~ \lambda_{out} = \lambda_{in} = \lambda \, , 
~~~~~~  {\cal A}_{\mu\nu\lambda~{out}} = {\cal A}_{\mu\nu\lambda~{in}} \, , \nonumber \\
&&~~~ \kappa^2_{{\tt eff}~out} \frac{a_{out}'}{a} - \kappa^2_{{\tt eff}~in} \frac{a_{in}'}{a} = -\frac12 {\cal T}_B   \, ,
~~~~~~  \kappa^2_{out} - \kappa_{in}^2 = 2 {\cal Q}_B \, , \\
&& ~~~~~~~~~~~~~~~~
 {\cal B}_{\mu\nu\lambda~out} - {\cal B}_{\mu\nu\lambda~in} 
 = -9 \bigl(\frac{a'_{out}}{a} - \frac{a'_{in}}{a}\bigr) \, . \nonumber
\ea
The bulk geometry is still given by Eq. (\ref{fried}). 
However now the analysis of the kinematics of cutting and pasting solutions
is complicated by the $\kappa^2$ dependence of the 
bulk solutions and the fact that this variable jumps across the wall.
Using (\ref{fried}) and (\ref{bcsB}) we can obtain the equivalent of 
Eqs. (\ref{balanceA}) and (\ref{diffroots}) by a straightforward manipulation.
To make the notation more compact, let us define first 
\be
{\cal R}_j =  \sqrt{ 1-  \frac{\Lambda_{j} a^2}{3 \kappa_{{\tt eff}~j}^2}} \, .
\label{Rnotation}
\ee
Then after some manipulation, the analogue of Eqs. (\ref{balanceA}) are
\ba
\label{balanceB}
\zeta_{out} \kappa_{{\tt eff}~out}^2 {\cal R}_{out} - \zeta_{in} \kappa_{{\tt eff}~in}^2{\cal R}_{in} &=& -\frac{{\cal T}_B a}{2}\, , \\
\zeta_{out} \kappa_{{\tt eff}~out}^2 {\cal R}_{out} + \zeta_{in} \kappa_{{\tt eff}~in}^2 {\cal R}_{in} &=&
\frac{2a}{3{\cal T}_B} \bigl( \kappa_{{\tt eff}~out}^2 \Lambda_{out} 
- \kappa_{{\tt eff}~in}^2 \Lambda_{in} \bigr)- 
\frac{4{\cal Q}_B}{{\cal T}_B} \frac{\kappa_{{\tt eff}~out}^2 + \kappa_{{\tt eff}~in}^2}{a}\, .  \nonumber 
\ea
The first of these equations is just the fourth of Eqs. (\ref{bcsB}). The second is a bit more complicated, and it is obtained by
starting with $\kappa^4_{{\tt eff}~out} a_{out}'^2 - \kappa^4_{{\tt eff}~in} a_{in}'^2$, evaluating it using (\ref{fried}) and then factoring it
and using the first of (\ref{balanceB}). We can now add and subtract the two equations of (\ref{balanceB}) to get individual expressions
for ${\cal R}_j$, as before. Although the notation looks cumbersome, the formulas 
disentangle somehwat 
after substituting $\Lambda = \Lambda_{\tt QFT}  + \kappa^2 \lambda$, since both $\Lambda_{\tt QFT}$ and
$\lambda$ are formally independent of $\kappa^2$. 

We note a potential danger with the transitions catalyzed by ${\cal Q}_B$. The equation for the jump
in $\kappa^2$ shows that in principle a transition inducing a {\it negative} $\kappa^2$ might be possible. Indeed,
$\kappa^2_{in} = \kappa^2_{out} - 2 {\cal Q}_B$ and thus for a sufficiently small $\kappa^2_{out}$ the offspring
Newton's constant could switch sign. Inside such a bubble this would wreak havoc on local physics since it
would make perturbative gravity repulsive, leading to spin-2 ghosts. Even if this does not happen suddenly, 
if the evolution favors a succession of $\kappa^2$ discharges, this could be an option. 

This is a dreadful prospect. In addition to possible large sinks, that occur when a bubble of Anti-de Sitter 
is nucleated, inside which any kind of normal matter population triggers a black hole formation, we might have to reckon
with massless spin-2 ghosts as well. Thus the question arises 
how could the ghosts be kept at bay and prevented from 
crossing over. 

A clue comes from noting that decreasing $\kappa^2$ while holding $\Lambda$ fixed is 
analogous to increasing $\Lambda$ at $\kappa^2$ fixed. Thus one expects processes that might
flip the sign of $\kappa^2$ to be suppressed at smaller $\Lambda$, and 
so such processes might end up being highly suppressed, and perhaps even impossible. 
The technical problem is clearly with controlling the smallness of $\Lambda$. If it fluctuates, 
by either a variation of $\kappa^2$ or a variation of $\lambda$, or due to the QFT corrections, it may be difficult to 
control the conditions which dictate the membrane dynamics. 
 
The control can be improved with scale {\it covariance}. 
In the theory with the conformal $4$-form/matter coupling and a 
UV regulator which does not break it (e.g, dim reg), vacuum energy corrections come in the form 
$\Lambda_{\tt QFT} = \kappa^2 {\cal H}^2_{\tt QFT}$, as in Eq. (\ref{cftcc}) and the second of Eqs. (\ref{ccgen}). 
So the cosmological constant to
any loop order is $\kappa^2 (\lambda + {\cal H}^2_{\tt QFT})$. We can absorb ${\cal H}^2_{\tt QFT}$ into
$\lambda$, and set $\Lambda_{\tt QFT} = 0$ and completely forget it from here onwards.
So if we define membrane charges and tensions relative to some value of $\lambda = \Lambda/\kappa^2$
such that the transitions to the regime with ghosts are excluded, the subsequent dynamics will preserve these conditions. 

Let us show that this expectation is borne out. There is a simple and straightforward proof that in the physically relevant cases this
limit of our theory is safe from ghosts. 
We underline that the proof might exist for more general cases as well, 
but at this point we have found the conformally
coupled $4$-form/matter theory to be simpler to manage, and will 
keep with it from now on. It would be of interest to explore the general case separately. 

First of all, in this case after a straightforward algebra we can rewrite Eqs. (\ref{balanceB}) 
as, using $\kappa_{{\tt eff}~out}^2 - {\cal Q}_B = \kappa_{{\tt eff}~in}^2 + {\cal Q}_B$, 
\ba
\label{balanceBtwoeqs}
\zeta_{out} \kappa_{{\tt eff}~out}^2 {\cal R}_{out} &=& -\frac{{\cal T}_B a}{4} 
- \frac{4{\cal Q}_B}{{\cal T}_Ba} (\kappa^2_{{\tt eff}~out} - {\cal Q}_B) \bigl( 1 - \frac{\lambda a^2}{3} \bigr)\, , \nonumber \\
\zeta_{in} \kappa_{{\tt eff}~in}^2 {\cal R}_{in} &=&
\frac{{\cal T}_B a}{4} - \frac{4{\cal Q}_B}{{\cal T}_Ba} (\kappa^2_{{\tt eff}~out} - {\cal Q}_B) \bigl( 1 - \frac{\lambda a^2}{3} \bigr)\, . 
\ea
Since $\Lambda = \kappa^2 \lambda$, and $\kappa_{{\tt eff}~j}^2 = \mps + \kappa^2_j$ this means
\be
{\cal R}_j =  \sqrt{ 1-  \frac{\kappa^2_j}{\mps + \kappa^2_j}\frac{\lambda a^2}{3}} 
= \sqrt{ 1-  \frac{1}{1+\mps/\kappa^2_j}\frac{\lambda a^2}{3}}\, .  
\label{rshort}
\ee
Because we are mainly interested in transitions from parent de Sitter spaces, we take $\lambda > 0$. 
Next, we want to first explore transitions which reduce $\kappa^2$. We are interested in 
(precluding) transitions for which $\kappa^2_{in} < 0$ for  
initial de Sitter geometries. Now, $\kappa^2_{in} < \kappa^2_{out}$
could only be facilitated with positive membrane charges ${\cal Q}_B>0$, 
as seen from the fifth of Eqs. (\ref{bcsB}).  This inequality also implies that 
$1+\mps/\kappa^2_{out} < 1 + \mps/\kappa^2_{in}$, and therefore,
\be
{\cal R}_{out} < {\cal R}_{in}\, .  
\label{rshorto}
\ee
So when compared to the previous case with ${\cal T}_A, {\cal Q}_A \ne 0$, the transitions
which reduce $\kappa^2$ are qualitatively similar to the transitions which {\it increase} the local
value of the cosmological constant. 

Having established this, we can now turn our attention to (\ref{balanceB}), which 
after plugging $\kappa_{{\tt eff}~out}^2 - \kappa_{{\tt eff}~in}^2 = 2{\cal Q}_B$ we can rewrite as
\ba
\Bigl(\zeta_{out}  {\cal R}_{out} - \zeta_{in} {\cal R}_{in} \Bigr) \kappa_{{\tt eff}~in}^2 &=& -\frac{{\cal T}_B a}{2} 
- 2{\cal Q}_B \zeta_{out} {\cal R}_{out} \, , \nonumber \\
\Bigl(\zeta_{out}  {\cal R}_{out} + \zeta_{in} {\cal R}_{in} \Bigr) \kappa_{{\tt eff}~in}^2 
&=&- \frac{8{\cal Q}_B}{{\cal T}_Ba} (\kappa^2_{{\tt eff}~out} - {\cal Q}_B) \bigl( 1 - \frac{\lambda a^2}{3} \bigr)
+ 2{\cal Q}_B \zeta_{in} {\cal R}_{in}  \, . 
\label{rbalance}
\ea
Now we impose ${\cal Q}_B >0$ -- which must be true to 
reduce $\kappa^2$ in the offspring de Sitter -- and check what happens for various combinations $(\zeta_{out}, \zeta_{in})$. 

It is straightforward to see
that as long as $\kappa^2_{{\tt eff}~out} \gg {\cal Q}_B$, transitions resulting in $\kappa^2<0$ are blocked off. 
The argument is as follows:
\begin{itemize}
\item $(\zeta_{out}, \zeta_{in}) = ++$: in this case (\ref{rbalance}) is 
$\kappa^2_{{\tt eff}~in} ({\cal R}_{out} - {\cal R}_{in})= - \bigl(\frac{{\cal T}_B a}{2} + 2{\cal Q}_B {\cal R}_{out} \bigr)$.
Both sides are negative, and hence $\kappa^2_{{\tt eff}~in} > 0$. The second equation (\ref{rbalance}) then shows that
small values of $a$ are excluded, since they are incompatible with $\kappa^2_{{\tt eff}~in} > 0$. 
\item $(\zeta_{out},\zeta_{in}) = --$: now, (\ref{rbalance}) is 
$\kappa^2_{{\tt eff}~in} ({\cal R}_{out} - {\cal R}_{in})= \bigl(\frac{{\cal T}_B a}{2} - 2{\cal Q}_B {\cal R}_{out} \bigr)$ due to the sign flips. 
If $\frac{{\cal T}_B a}{2} > 2{\cal Q}_B {\cal R}_{out}$, $\kappa^2_{{\tt eff}~in} < 0$. However, this cannot occur when  
$\kappa^2_{{\tt eff}~out} \gg 2{\cal Q}_B$, implying such solutions are prohibited kinematically. 
The second equation then favors small bubbles. \item $(\zeta_{out}, \zeta_{in}) = +-$: now (\ref{rbalance}) is 
$\kappa^2_{{\tt eff}~in} ({\cal R}_{out} + {\cal R}_{in})= - \bigl(\frac{{\cal T}_B a}{2} + 2{\cal Q}_B {\cal R}_{out} \bigr)$.
Since the right hand side is positive, the only possible solution is $\kappa^2_{{\tt eff}~in} < 0$, but it cannot exist for 
$\kappa^2_{{\tt eff}~out} \gg 2 {\cal Q}_B$. 
\item $(\zeta_{out}, \zeta_{in}) = -+$: in this case (\ref{rbalance}) reduces to 
$\kappa^2_{{\tt eff}~in} ({\cal R}_{out} + {\cal R}_{in})= \bigl(\frac{{\cal T}_B a}{2} - 2{\cal Q}_B  {\cal R}_{out} \bigr)$.
As both sides are positive, $\kappa^2_{{\tt eff}~in} > 0$ for 
$\kappa^2_{{\tt eff}~out} \gg 2{\cal Q}_B$. In this limit the second equation favors larger bubbles. 
\end{itemize}
Bottomline is that $\kappa^2$ will not suddenly dip below zero, and more importantly 
neither will $\kappa^2_{\tt off}$. The emission of ${\cal Q}_B > 0$ may reduce the effective Planck scale, but it will do it ever so slowly.
Since these processes are analogous to the increase in the value of the offspring cosmological constant, we
can expect that they will be suppressed
by the large bounce action, drawing on the results of the previous section. We will see this is borne out shortly. 
Thus the dominant direction of evolution
will be to increase $\kappa^2_{\tt eff}$, which means, to weaken the gravitational force inside the offspring bubbles. 

The increase of $\kappa^2_{\tt eff}$ -- i.e. the reduction of gravitational strength -- 
should also be very slow. We can arrange for it by choosing 
${\cal T}_B$ and ${\cal Q}_B$. This is a necessary condition to have a chance to fit our universe in some of these 
bubbleworlds. A hint for how to achieve this goal comes from our previous analysis of ${\cal T}_A, {\cal Q}_A$ membrane dynamics. We have seen
there that requiring $\frac{2 \kappa^2_{\tt eff} \kappa^2 {\cal Q}_A}{{\cal T}_A^2} < 1$ greatly restricts the instanton processes which 
can occur, singling out the pale green-shaded ones in the `Baedeker' of Fig. (\ref{allcases}). Inspecting 
Eqs. (\ref{balanceBtwoeqs}), we can easily identify the key source of potential problems: the term
$\sim \frac{4{\cal Q}_B}{{\cal T}_B a} \kappa^2_{{\tt eff} ~out}$. When this term is small equations are
qualitatively similar to the $q<1$ case of ${\cal T}_A, {\cal Q}_A$ membrane dynamics. However, 
for small bubbles this term might 
even overwhelm the tension terms in (\ref{balanceBtwoeqs}), thanks to $a$ in the denominator. Since the tension, due to its positivity,
is the barrier which protects the low energy dynamics from problems in the ${\cal T}_A, {\cal Q}_A$ case, 
as well as in the case of domain walls in GR, we should ensure 
that it retains the same role everywhere in the domain of interest in Pancosmic General Relativity. This means, we require that
\be
\frac{4|{\cal Q}_B|}{{\cal T}_B a} \kappa^2_{{\tt eff} } \ll \frac{{\cal T}_B a}{4} \, ,
\label{Bbound1}
\ee
for all bubbles which can form. Since $a$ is the size of the bubble when it nucleates, the bound is under greatest threat from 
the smallest bubbles that might nucleate. Therefore for the semiclassical theory to remain under control, 
this inequality must be true for the smallest bubbles which can be consistently described in the local region.
Since the smallest bubbles are $a \sim 1/\kappa_{\tt eff}$, this finally yields our strong form of the bound:
\be
16 \, \frac{\kappa^4_{{\tt eff} }|{\cal Q}_B|}{{\cal T}^2_B}  \ll 1 \, . 
\label{Bboundfin}
\ee
If (\ref{Bboundfin}) is satisfied, then (\ref{Bbound1}) will hold for any bubble of size $a > 1/\kappa_{\tt eff}$. 
Additionally, the regions of space where (\ref{Bboundfin}) holds 
will not become infested with ghosts - since this will also ensure that the processes decreasing 
$\kappa^2$ are highly suppressed: the 
regions which might be at risk of becoming ghost infested will remain separated from those which are ghost-free. 

To check that this is a self-consistent regime, we can solve explicitly the equations (\ref{balanceBtwoeqs}) 
for $1/a^2$ to obtain expressions which are an analogue of (\ref{radii}). After straightforward algebra, 
using Taylor expansion in $16 \, \frac{\kappa^4_{{\tt eff} }|{\cal Q}_B|}{{\cal T}^2_B}$, we find\footnote{Each of the 
equations (\ref{balanceBtwoeqs}) is a quadratic equation for $a^2$, with two branches of solutions. Here we only keep the solution 
which is perturbative in ${\cal Q}_B$, and ignore the other solution which has an essential singularity when ${\cal Q}_B \rightarrow 0$
because it gives $a^2 < 0$ in the regime we consider. This rules it out on physical grounds.}
\ba
\frac{1}{a^2} &=& \kappa^2_{{\tt eff}~out}
\bigl(\frac{{\cal T}_B}{4\kappa^3_{{\tt eff}~out}} \bigr)^2 \Bigl(
\frac
{1- {\frac{2\lambda}{3 \kappa^2_{{\tt eff}~out}}
\frac{16 \kappa^4_{{\tt eff}~out}  {\cal Q}_B}{{\cal T}^2_B}
(1-\frac{{\cal Q}_B}{\kappa^2_{{\tt eff}~out}})
 }}
 {1- (\frac{{\cal T}_B}{4\kappa^3_{{\tt eff}~out}})^2 
{\frac{16 \kappa^4_{{\tt eff}~out}  {\cal Q}_B}{{\cal T}^2_B}
(1-\frac{{\cal Q}_B}{\kappa^2_{{\tt eff}~out}})
}} 
+ {\cal O}\bigl( (\frac{\kappa^4_{{\tt eff}~out}  {\cal Q}_B}{{\cal T}^2_B})^2 \bigr) \Bigr) \, , \nonumber \\
&=&
\kappa^2_{{\tt eff}~in}
\bigl(\frac{{\cal T}_B}{4\kappa^3_{{\tt eff}~in}} \bigr)^2 \Bigl(
\frac
{1+ {\frac{2\lambda}{3 \kappa^2_{{\tt eff}~in}}
\frac{16 \kappa^2_{{\tt eff}~in}  {\cal Q}_B}{{\cal T}^2_B}
(1+\frac{{\cal Q}_B}{\kappa^2_{{\tt eff}~in}}) 
}}
{1+ (\frac{{\cal T}_B}{4\kappa^3_{{\tt eff}~in}})^2 
{\frac{16 \kappa^2_{{\tt eff}~in}  {\cal Q}_B}{{\cal T}^2_B}
(1+\frac{{\cal Q}_B}{\kappa^2_{{\tt eff}~in}}) 
}} 
+ {\cal O}\bigl( (\frac{\kappa^4_{{\tt eff}~in}  {\cal Q}_B}{{\cal T}^2_B})^2\bigr) \Bigr) \, . ~~~~
\label{radiib}
\ea
So indeed, we see that when (\ref{Bboundfin}) holds, in the regime of consistent semiclassical theory with $\lambda/\kappa^2_{\tt eff} < 1$, 
$\frac{{\cal T}_B}{4\kappa^3_{{\tt eff}}} < 1$, $\frac{{\cal Q}_B}{\kappa^2_{{\tt eff}}}<1$, 
which keep the dynamics of the theory 
below the local Planckian cutoff, the transitions which may change the local value of the Planck scale,
if possible, occur via the bubbles whose size converges to 
\be 
a \simeq 
\frac{4\kappa^3_{{\tt eff}}}{{\cal T}_B} \kappa^{-1}_{{\tt eff}} \gg \kappa^{-1}_{{\tt eff}} \, , 
\label{bubblesizes}
\ee
blocking Planckian scales precisely as we claimed above. Basically, the reason for it is the terms $\propto 1/a$ in 
Eqs. (\ref{balanceBtwoeqs}) which suppress the transitions that are mediated both by big bubbles and small bubbles: 
the effective membrane charge is $\propto {\cal Q}_B/a$, Hence big bubble transitions 
occur via the tiny effective membrane charges, which barely scratch the backgrounds. Small bubble transitions,
on the other hand, always involve cis-Planckian bubbles which are much larger than $ \kappa^{-1}_{{\tt eff}}$, 
because of (\ref{Bboundfin}). 

If the effective charges are small,
so are the variations of the 
inverse curvature radius squared, $\lambda = \Lambda/\kappa^2$. 
Moreover, the bubble nucleation 
processes can only occur if the argument of the square roots in (\ref{rshort}) is a
nonnegative number. For $\lambda>0$, this imposes the constraint 
\be
\frac{\kappa^2 \lambda}{3\kappa^4_{{\tt eff}}} =  \frac{\Lambda}{3\kappa^4_{{\tt eff}}}  < \Bigl(\frac{{\cal T}_B}{4\kappa^3_{{\tt eff}}}\Bigr)^2
\, .
\label{ccbound}
\ee
For larger local values of the positive cosmological constant 
$\Lambda/\kappa^4_{{\tt eff}} > 3\Bigl(\frac{{\cal T}_B}{4\kappa^3_{{\tt eff}}}\Bigr)^2$ 
the effective Planck constant remains frozen. In particular, the faster processes 
which can occur in the discharge of $\lambda$ when the cosmological constant is large are completely blocked off for $\kappa^2_{\tt eff}$.
 
Again, the only
threat to the bound (\ref{Bboundfin}) comes from an {\it increase} of $\kappa_{eff}^2$.  However these processes will be very 
slow; Eq. (\ref{Bboundfin}) is very similar to the bound on (\ref{qbound}), $q < 1$, 
which controls the kinematics of the instantons. 
So where (\ref{Bboundfin}) holds the transitions will also be restricted 
to the green-shaded instantons of the `Baedeker' of Fig. (\ref{allcases}). 
Since the charges and tensions between the two kinds of membranes are not correlated, we can arrange them so that the $B$-wall dynamics 
is much slower -- when allowed -- than the $A$-wall one. We will assume this is the case for the remainder of this work.
In the limit $\lambda \rightarrow 0$ these conclusions remain: the `blockade' of the transitions reducing $\kappa^2_{\tt eff}$ only gets 
stronger and stiffer near the flat space, as it follows from the properties of the green-shaded instantons of Fig. (\ref{allcases}). 

\section{... {\it Gloria Mundi!}}

In contrast to standard General Relativity, where de Sitter space is totally stable thanks to 
Bianchi identities, and Newton's constant is a fixed input parameter, in our
generalization of General Relativity, not only does the cosmological constant change discretely, but so do Planck scale and the 
QFT parameters, like in the original wormhole approach \cite{Coleman:1988tj,Coleman:1989ky}. 
The discharge is quantum-mechanical and nonperturbative, it ceases in the classical limit, and it is different from the
instability to black hole formation of \cite{Ginsparg:1982rs}. This fits with ideas that an eternal, stable de Sitter space may not
exist in a UV complete theory 
\cite{Banks:2000fe,Banks:2001yp,Witten:2001kn,Goheer:2002vf,Dvali:2017eba,Obied:2018sgi,Susskind:2021dfc}.

The picture of the emergent dynamical spacetime is reminiscent of the picture advocated 
in the wormhole approach to Euclidean quantum gravity 
\cite{Hawking:1978pog,Coleman:1988tj,Fischler:1988ia,Fischler:1989ka,Banks:1984cw,Polchinski:1989ae}.
That program attempted to uncover nonperturbative instability of de Sitter space which could
be intrinsic to quantum gravity, which might follow from 
the properties of the semiclassical approximation of Euclidean
path integral \cite{Hawking:1981gd,Baum:1983iwr,Hawking:1984hk}, 
\be
Z = \int e^{-S_E} \simeq e^{-S_{classical}} = 
\begin{cases}
e^{24 \pi^2 \frac{\kappa_{\tt eff}^4}{\Lambda} } = 
e^{\frac{A_{\tt horizon}}{4G_N}} \, , \, ~~~~~~~~ {\Lambda > 0}    \, ;\\
e^{\Lambda \int d^4x \sqrt{g}  }  = 1 \, , \, ~~~~~~~~~~~~~~~{\Lambda = 0}  \, ; \\
e^{-|\Lambda | \int d^4x \sqrt{g} } \rightarrow 0 \, ,  ~~~~~~~~~~~~\Lambda < 0 \, , ~~  {\rm noncompact}   \, .  \\
\end{cases}
\label{bekensteinhawking}
\ee
The function $Z$ has an essential singularity at vanishing $\Lambda$, diverging 
as $\Lambda \rightarrow 0^+$. It is clearly tempting to think of $Z$ as a partition function 
and use this divergence to argue that cosmological constant must be vanishingly small
\cite{Hawking:1981gd,Baum:1983iwr,Hawking:1984hk,Coleman:1988tj}. 

To argue that $Z$ is a partition function which favors any value of $\Lambda$ \cite{Horava:2000tb}, however, 
one needs to decide {\it what} it is a
partition function of. More directly, what are the dynamical degrees of freedom controlling $\Lambda$,
which $Z$ might be counting? The approach to the cosmological constant problem 
based on wormholes \cite{Coleman:1988tj} ran into 
problems with decoupling \cite{Fischler:1988ia,Fischler:1989ka,Banks:1984cw,Polchinski:1989ae}. 
Given the notorious subtleties with the definition and interpretation 
of $Z$ \cite{Gibbons:1978ac,Linde:1991sk,Carlip:1989qq,Carlip:1992wg,Anderson:2003js}, 
and even its restriction to only compact Euclidean spaces (a.k.a. 
the Hartle-Hawking wavefunction \cite{Hartle:1983ai}), other approaches were also pursued. 

Here we follow the approach which resembles to some extent the ideas of  \cite{Coleman:1988tj,Coleman:1989ky},
but with different ingredients. We have defined a semiclassical picture where the theory contains well-defined
`rigid' objects -- the charged membranes -- whose nucleation and dynamics lead to changes in the 
parameters of the theory. At least in the 
semiclassical limit, they automatically obey decoupling, and can be consistently included -- as Euclidean saddle points -- 
in the action, and therefore in $Z$. Our task is to outline the structure of spacetime which membranes can 
seed, and see what happens.

In the next subsection, we consider quantitatively the nucleation rates and stability of solutions in 
certain limits of the theory.  Following it, and using those results, we survey the effect of membrane sources  
and membrane nucleation on the spacetime in the semiclassical limit. Subsequently, in the last subsection we 
outline how the emerging picture of the spacetime can solve the cosmological 
constant problem, by driving it to extremely small values in the units of the effective Planck scale. 

\subsection{Decay Rates}   

At this point we need to explore quantitative aspects of membrane emission transitions and the changes to an initial 
background geometry which the transitions induce. We are particularly interested in geometries which start
as sections of de Sitter space, since they feature more relevant dynamics. From the consideration of the instanton `Baedeker' of Fig. (\ref{allcases}),
the definition of the bounce action (\ref{bounceact}) and the transition rate (\ref{decrate}), as well as the formulas 
for the evaluation of the various contributions to the bounce action, given in Eqs. (\ref{bounce1dshell}), (\ref{integrals}), 
it is clear that in general the fastest possible processes are mediated by the instanton in the top left corner of the
`Baedeker' (\ref{allcases}), for both ${\cal T}_A, {\cal Q}_A \ne 0$ and ${\cal T}_B, {\cal Q}_B \ne 0$ cases. 
We will start with reviewing this case, which is actually the most commonly encountered case in the literature, 
and then move to other channels. 

\subsubsection{\texorpdfstring{$q > 1$}{Lg} } 

To warm up, we now consider the fastest instantons, $(\zeta_{out}, \zeta_{in}) = (++)$ in more detail. 
The reason these are the fastest channels is that the ``outside" geometry contribution to the bounce action for 
this configuration is the smallest, which follows because in the bounce action the ``outside" contribution is over
the {\it complement} of the parent geometry which defines the instanton. This can also be discerned from the
sign assignment $(\zeta_{out}, \zeta_{in}) = (++)$ in this case, which when inserted in (\ref{bounce1dshell}), Eq. (\ref{integrals})
ensures the largest cancellations between various terms in the equation. Their ``time reversed" process,
$(\zeta_{out}, \zeta_{in}) = (--)$, can be understood straightforwardly by reversing the order of $\Lambda_j$ and the
signs of $\zeta_j$. These processes are described by the pale gold shaded configurations in Fig. (\ref{allcases}), which require
$q = \frac{2 \kappa_{\tt eff}^2 \kappa^2 |{\cal Q}_A|}{{\cal T}_A^2} > 1$. Note that $(++)$ processes imply ${\cal Q}_A>0$, while
$(--)$ use ${\cal Q}_A<0$ -- meaning, $(++)$ lower $\Lambda$ and $(--)$ raise it. 

Now, from Eqs (\ref{radii}) we see that the membrane radius at nucleation is
$a^2 < 3\kappa^2/\Lambda_j$ for both the parent and the offspring  geometries. When $a^2$ is comparable
to the outer and inner de Sitter radii, however, Eqs. (\ref{radii}) 
show that the terms $\sim (1-\frac{\Lambda_j a^2}{3 \kappa^2_{\tt eff}})^{1/2}$
are much smaller than unity, and the bounce action 
(\ref{bounce1dshell}) is approximated by the difference of the one half of the
parent and offspring  horizon areas divided by $4G_N$, 
\be 
S_{\tt bounce} \simeq - \frac{12\pi^2 \kappa^4_{\tt eff} \Delta \Lambda}{\Lambda_{out} \Lambda_{in}} \, , ~~~~~~~~ 
\Delta \Lambda = \Lambda_{out} - \Lambda_{in} = \frac12 \kappa^2  {\cal Q}_A \, .
\label{fastbounce}
\ee
Therefore as long as 
$\Lambda_{out} \gg 3 \kappa^2_{\tt eff} \Bigl(\frac{{\cal T}_A}{4 \kappa_{\tt eff}^2}\Bigr)^2 
\Bigr(1 - {\frac{2\kappa_{\tt eff}^2 \kappa^2 {\cal Q}_A}{3 {\cal T}_A^2}}\Bigr)^2$, 
the initial discharge of the cosmological constant is very fast, since $S_{\tt bounce}$ is negative. 
Note, that the reverse processes of increasing the cosmological constant, $\Delta \Lambda < 0$,
can also occur. However their bounce action is the negative of the action (\ref{fastbounce}). Therefore 
these processes are more rare, and so the overall trend is the decrease of $\Lambda$. 
The cosmological constant is repelled down from Planckian densities. This regime will persist until
$\Lambda_{out} \sim \frac{\kappa_{\tt eff}^2 \kappa^4 {\cal Q}^2_A}{12 {\cal T}_A^2}$. 

An interesting feature of the transitions in this regime is that the 
membrane radius is comparable to the background de Sitter radii. Hence 
the dynamics automatically 
caps the ``birth rate" at one offspring  for each parent. No more. The decay 
rate is fast, but not prolific. 

In any case, a large cosmological constant will be discharged, on the average, at 
a fast rate, in steps $\Delta \Lambda = \kappa^2 {\cal Q}_A/2$, 
until its value reduces to 
\be
\Lambda < 3 \kappa^2_{\tt eff} \Bigl(\frac{{\cal T}_A}{4 \kappa_{\tt eff}^2}\Bigr)^2 
\Bigr(1 - {\frac{2\kappa_{\tt eff}^2 \kappa^2 {\cal Q}_A}{3 {\cal T}_A^2}}\Bigr)^2 \sim 
\frac{\kappa_{\tt eff}^2 \kappa^4 {\cal Q}^2_A}{12 {\cal T}_A^2} = 
\frac{\kappa_{\tt eff}^2 \kappa^2{\cal Q}_A}{12 {\cal T}_A^2} \kappa^2 {\cal Q}_A \, .
\ee
At this point the discharge rate slows down. For such values of the cosmological 
constant, the radius of a membrane at nucleation is much smaller than the 
parent and offspring  radii, $a^2 \ll 3\kappa^2/\Lambda_j$. 
We can then compute $S_{\tt bounce}$ in this regime to the leading order in $\Delta \Lambda$, finding 
$S_{\tt bounce} \simeq \frac{\pi^2}{6} a^4 \Delta \Lambda$. Evaluating this using (\ref{radii}) gives 
\be
S_{\tt bounce} \simeq \frac{27\pi^2}{2} 
\frac{{\cal T}_A^4\Delta \Lambda}{\Bigl[{\bigl( \Lambda_{out} 
+ \Lambda_{in} + \frac{3{\cal T}_A^2}{4 \kappa^2_{\tt eff}} \bigr)^2
- 4\Lambda_{out} \Lambda_{in} }\Bigr]^2} \, .
\label{familiars}
\ee 

To compute this action, however, we now must pay more attention to the details of the nucleation 
dynamics. 
Since the $(\zeta_{out},\zeta_{in}) = (++)$ instanton requires $q > 1$, and 
since $\Delta \Lambda = \kappa^2{\cal Q}_A/2$, 
at least one of $\Lambda_j$ must be larger than 
$\frac{3{\cal T}_A^2}{4 \kappa^2_{\tt eff}}$. Hence (\ref{familiars}) 
should be treated perturbatively in $\frac{3{\cal T}_A^2}{4 \kappa^2_{\tt eff}}$
and the smaller of the two $\Lambda_j$. The correct limiting expression is 
\be
S_{\tt bounce} \simeq \frac{27\pi^2}{2} \frac{{\cal T}_A^4}{(\Delta \Lambda)^3}\, ,
\label{familiarso}
\ee 
which is the familiar result from the literature, giving the limit for the nucleation rate 
when the gravitational effects are negligible, and field theory controls the processes 
(see \cite{Coleman:1977py,Callan:1977pt,Coleman:1980aw} and many other papers). 
The reverse processes, mediated by $(--)$ instantons, still occur, but now 
they are more suppressed. 
Substituting  $\Delta \Lambda = \kappa^2 {\cal Q}_A/2$, we finally find, using 
$q  > 1$, 
\be
S_{\tt bounce} \simeq 108 \pi^2 \frac{{\cal T}_A^4}{\kappa^6 {\cal Q}_A^3} < 
\frac{144 \pi^2}{3}\frac{\kappa^4_{\tt eff}}{ \kappa^2{\cal Q}_A}\, ,
\label{familiarfast}
\ee 
This bounce action can still be quite big and these processes may be slow. In this regime
the nucleated bubbles are quite small, and in fact are much smaller than the gravitational radii of the
parent and offspring. Thus multiple processes of nucleating bubbles can happen in different regions 
of the parent geometry -- if the parent geometry is big to start with. 

The real problem with the regime where $q > 1$  
however is the transitions to $\Lambda \le 0$. Those will inevitably occur since 
the limiting bounce action is finite, and the space continues to bubble. All that needs to happen is that 
$\Lambda_{out}$ dips below $\kappa^2 {\cal Q}_A/2$, and the next nucleation process will
lead to the formation of a bubble with $\Lambda_{in} < 0$. The nucleation does not stop even then, since
there is a $(++)$ instanton mediating decay of $\Lambda \le 0$ available, given by the bottom right
of the type 1 instantons in `Baedeker' of Fig. (\ref{allcases}). 
At this point the nucleations can end since in such regions, even a small amount of 
compressible matter will lead to the collapse of the bubble into a black hole. Only then does the nucleation 
of bubbles cease. Regions like this behave like sinks where the evolution is irreversible \cite{Linde:2006nw}.

\subsubsection{\texorpdfstring{$q < 1$}{Lg} } 

\begin{figure}[thb]
    \centering
    \includegraphics[width=8cm]{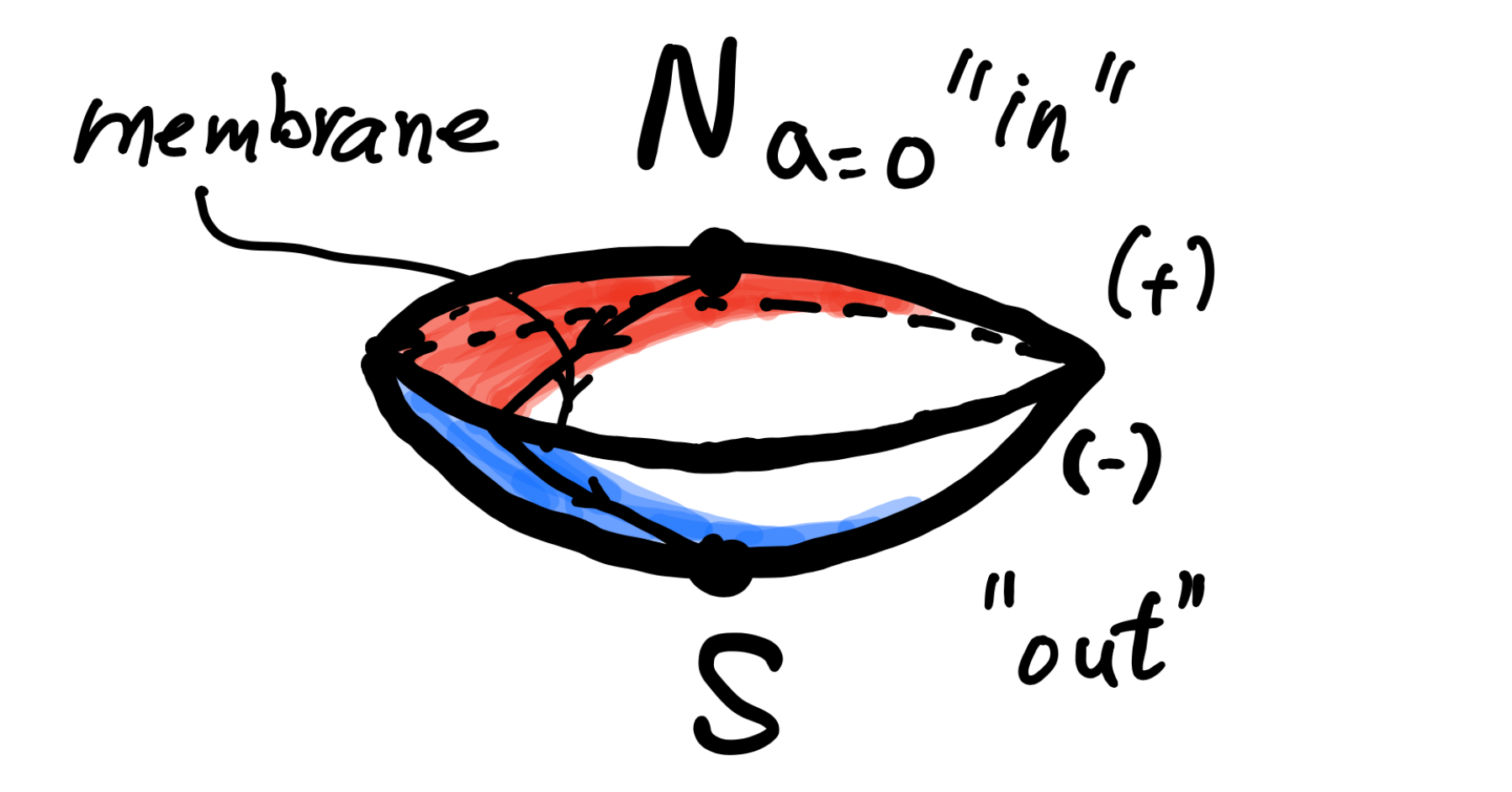}
    \caption{The only $q<1$ instanton mediating $dS \rightarrow dS$.}
    \label{figl}
\end{figure}

The case  $q = \frac{2 \kappa_{\tt eff}^2 \kappa^2 |{\cal Q}_A|}{{\cal T}_A^2} < 1$ is a {\it lot} more interesting. First of all,
as is clear from the instanton `Baedeker' of Fig. (\ref{allcases}), the nucleation processes, now in pale green, are more restricted. 
We examine them in more detail. The $dS \rightarrow dS$ transitions are controlled by the instanton of Fig. (\ref{figl}). At large $\Lambda_j > \kappa^2  {\cal Q}_A/2$, the processes involve 
a single large bubble, with $a^2 \sim 3 \kappa^2_{\tt eff}/ \Lambda$. To the leading order  this stage is 
almost the same as the large $\Lambda_j$ stage for $q>1$. The transition rate is controlled by the bounce action
(\ref{fastbounce}) and the proliferation rate is limited to one offspring  per parent. As before, the reverse transitions are also
allowed, but are more suppressed. 

In this case, however, this stage ends when 
\be
\Lambda < 3 \kappa^2_{\tt eff} \Bigl(\frac{{\cal T}_A}{4 \kappa_{\tt eff}^2}\Bigr)^2 \, ,
\label{goodbound}
\ee
because $q<1$. Subsequent nucleations continue via production of  small bubbles, whose rate is controlled
by the bounce action
\be
S_{\tt bounce} \simeq \frac{24\pi^2 \kappa^4_{\tt eff}}{\Lambda_{out}} 
- \frac{36 \pi^2 \kappa^2_{\tt eff} {\cal T}_A^2}{{\bigl( \Lambda_{out} 
+ \Lambda_{in} + \frac{3{\cal T}_A^2}{4 \kappa^2_{\tt eff}} \bigr)^2
- 4\Lambda_{out} \Lambda_{in} }} \simeq \frac{24\pi^2 \kappa^4_{\tt eff}}{\Lambda_{out}} \Bigl(1
- \frac{8}{3} \frac{\kappa^2_{\tt eff} \Lambda_{out}}{ {\cal T}_A^2} \Bigr)\, ,
\label{familiars2}
\ee 
where we used (\ref{goodbound}) to get the very last equation. Inside the family tree which started at large $\Lambda$ 
the bubble progeny is still limited to one
per `region' since the progenitor started out small. If the original initial bubble were large however, multiple 
bubble nucleations can also occur. In any case, when we continue to the Lorentzian regime, the proliferation rate
can be maintained by repeated 
successive bubble nucleations. 

It is now quite clear that $S_{\tt bounce} > 0$ 
because of (\ref{goodbound}). Further, the bounce action for this class of processes has a {\it pole} at $\Lambda_{out} \rightarrow 0$. 
In turn, the nucleation rate has an essential singularity at $\Lambda_{out} \rightarrow 0$, where the rate {\it vanishes}.
Thus in this regime, the small values of the cosmological constant are metastable, and any 
locally Minkowski  space becomes
absolutely stable to membrane nucleation processes. 

\begin{figure}[htb]
    \centering
    \includegraphics[width=10cm]{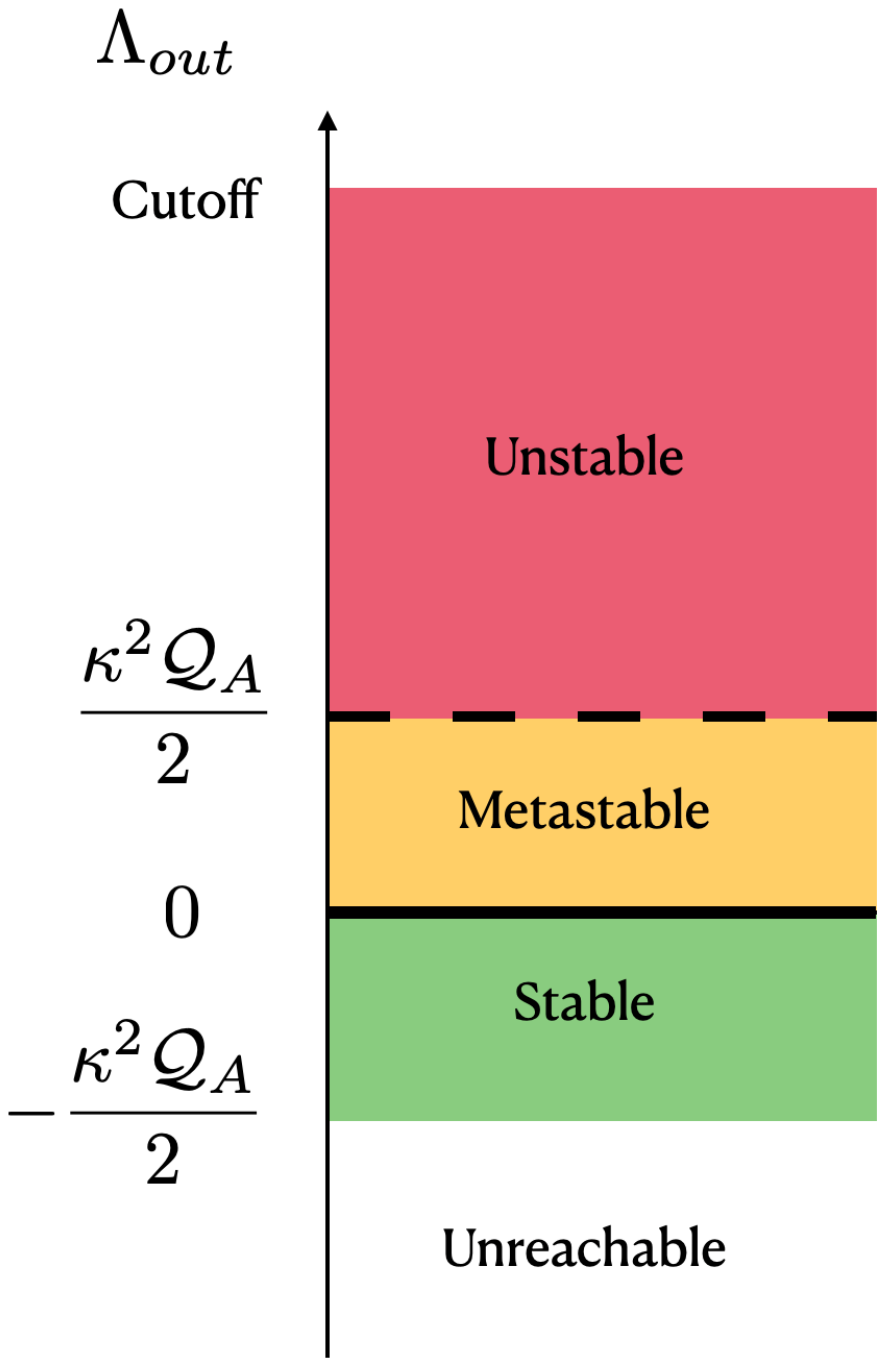}
    \caption{The cosmological constant spectral bands for $q = \frac{2\kappa_{\tt eff}^2 \kappa^2 |{\cal Q}_A|}{3{\cal T}^2_A} < 1$.}
    \label{spectrumpic}
\end{figure}

Although the process of decay of a de Sitter parent to an Anti-de Sitter
offspring  is possible, as per the presence of the second pale green-shaded instanton in the `Baedeker' (\ref{allcases}),
this can only happen if $\Lambda_{out}$ is initially in the window of values $0 < \Lambda_{out} < \kappa^2 {\cal Q}_A/2$. 
Even so, such de Sitter spaces will be long lived. We outline the structure of the spectrum of 
instantons\footnote{Anti-de Sitter could be destabilized by the nucleation of compact locally AdS spaces via the ``ogre" instantons in
the `Baedeker' of Fig. (\ref{allcases}), but we ignore those processes since they would be highly suppressed.} for this branch in
Fig. (\ref{spectrumpic}). The colored regions are depicting the stability zones - if a value of the cosmological constant
of the parent is in the red, it decays by a faster bubble nucleation, if it is in gold it may decay by one more bubble nucleation,
but more slowly, and if it is in the green, it is stable to bubble nucleation. The top of the green zone is Minkowski space, 
$\Lambda = 0$.

\subsubsection{\texorpdfstring{$\Delta \kappa^2_{\tt eff}$}{Lg} Transitions} 

It remains to discuss the `sustainability' of the regime $q <1$ in more detail. 
Since  $q = \frac{2\kappa_{\tt eff}^2 \kappa^2 |{\cal Q}_A|}{3{\cal T}^2_A} < 1$, 
the processes which increase $\kappa^2_{\tt eff}$ could violate this condition. In turn this would yield transitions to the regime $q>1$, where
decay of de Sitter to Anti-de Sitter could become easier, and perhaps even rampant. 
However, as we noted above, the regime $16 \frac{\kappa^4_{{\tt eff} }|{\cal Q}_B|}{{\cal T}^2_B}  \ll 1$ is sustainable. 
The effective Planck scale $\kappa^2_{{\tt eff}}$ remains frozen at least until 
\be
\frac{\kappa^2 \lambda}{\kappa^4_{{\tt eff}}} = \frac{\Lambda}{\kappa^4_{{\tt eff}}} <  3\Bigl(\frac{{\cal T}_B}{4\kappa^3_{{\tt eff}}}\Bigr)^2 \, .
\label{Bbrane}
\ee 
In regions where cosmological constant is larger, the large bubbles which must be nucleated to change 
Planck scale are blocked off. This might seem slightly surprising at first, 
but we recall that the effective charge is $\sim (1-\frac{\lambda a^2}{3}){\cal Q}_B/a$. So in the 
decoupling limit of gravity, $\kappa^2_{\tt eff} \rightarrow \infty$, the processes 
with a fixed and large $\Lambda$
are equivalent to the limit ${\cal Q}_{eff} \rightarrow 0$, with the membrane tension being held fixed. 
So, unsurprisingly, if we fix ${\cal T}_B$ and decouple gravity by sending $\kappa^2_{\tt eff} \rightarrow \infty$, 
we cannot possibly change Planck scale by a membrane nucleation with a tiny charge. 

In the regime where (\ref{Bbrane}) holds transitions can happen. However since $16 \frac{\kappa^4_{{\tt eff} }|{\cal Q}_B|}{{\cal T}^2_B}  \ll 1$ 
the only relevant processes are the pale green-shaded instantons of Fig. (\ref{allcases}). Effective Planck scale may
change, but the leading order bounce action, with ${\cal Q}_B \ll \kappa^2_{\tt eff}$, will be
\be
S_{\tt bounce} \simeq \frac{24\pi^2 \kappa^4_{\tt eff}}{\Lambda_{out}} \Bigl(1
- \frac{8}{3} \frac{\kappa^2_{\tt eff} \Lambda_{out}}{ {\cal T}_B^2} \Bigr) > 0 \, ,
\label{familiars22}
\ee 
which for ${\cal T}_B \gg {\cal T}_A$ leads to a rate which is 
much slower than the cosmological constant relaxation. 
This also implies that all variation of $\kappa^2_{\tt eff}$ 
must cease when  $\frac{{\Lambda_{out}}}{\kappa^4_{\tt eff}} \ll 1$. 
It is therefore possible to arrange for ${\cal T}_B, {\cal Q}_B$ so that 
$\lambda$ dynamics plays the main role in controlling the evolution. Some variation of $\kappa^2$ may occur, but it is extremely slow 
for positive $\Lambda > 0$, either large or small. In fact, in the subsequent article \cite{Kaloper:2022jpv} we have completely decoupled 
the $\kappa^2$ variation, by taking the limit ${\cal T}_B \rightarrow \infty$, in order to focus on the cosmological constant adjustment alone. 
Hence when 
its initial values are large, $\kappa^2_{\tt eff} \gg {\cal Q}_B$, the theory remains in the safe zone, $\kappa^2_{\tt eff} > 0$, 
far from the realm of ghosts, and it protects $q < 1$ throughout. This is the `safe stratus' of the theory's vacua.

\subsection{Fractal Vacua}

An interesting picture emerges. In the leading order approximation we can describe the 
full `phase space' of the Euclidean theory in Eq. (\ref{actionnewmemeu}) by the system of 
saddle points, each of which
extremizes the action (\ref{actionnewmemeu}), with the solutions 
of the Euclidean field equations (\ref{tenseqsgrmagdualconf}) 
classified by the membrane sources. These classical solutions are then interpreted as a 
Wick rotation of the Lorentzian spacetime theory (if one exists!), 
where the membrane sources are the boundaries of the bubbles of new
spacetime nucleating in a parent geometry, changing the values of  
Planck scale, the cosmological constant and even
the QFT parameters upon membrane wall crossing.

\begin{figure}[tbh!]
    \centering
    \includegraphics[width=7.9cm]{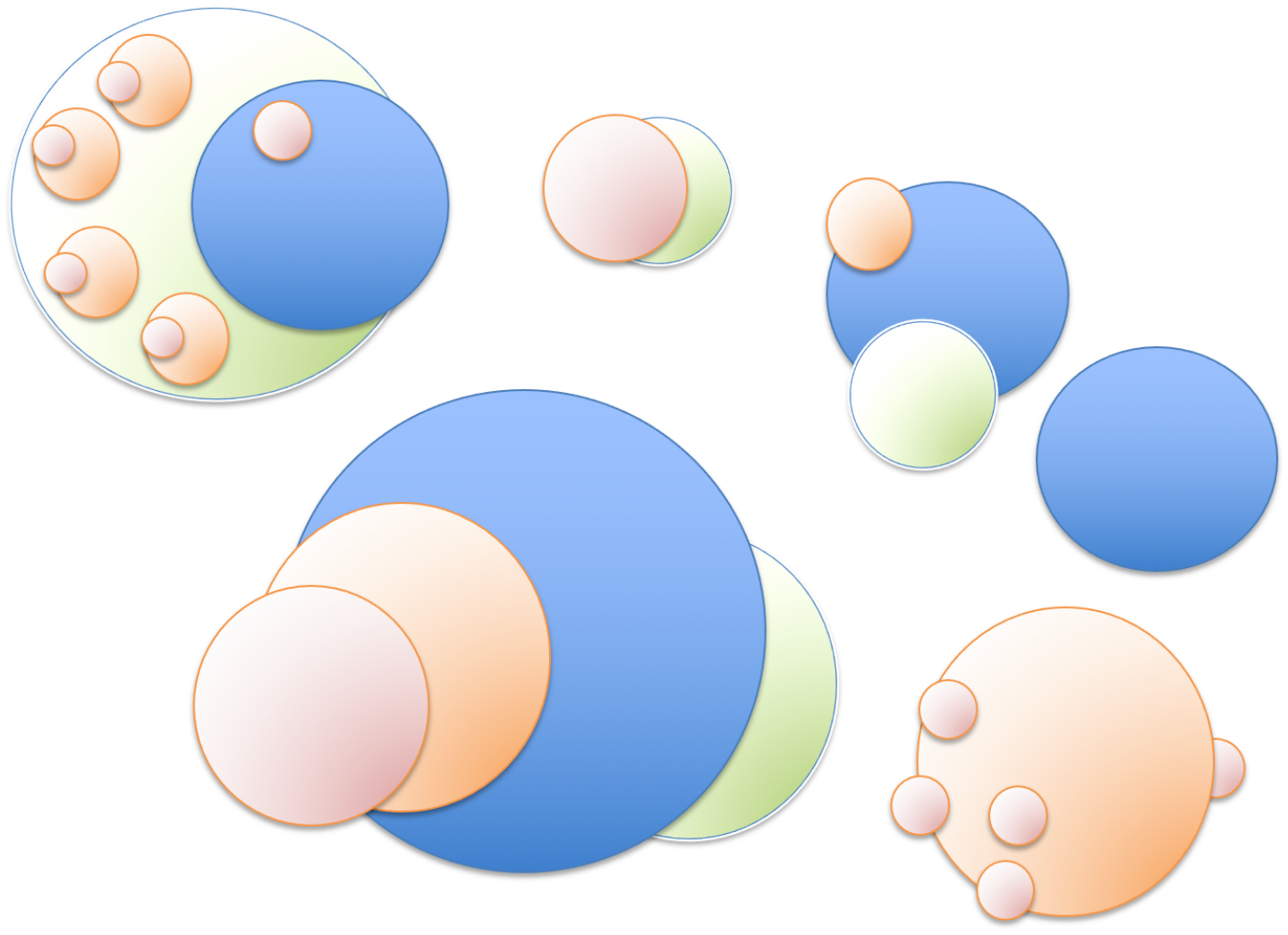} 
        \includegraphics[width=8.4cm]{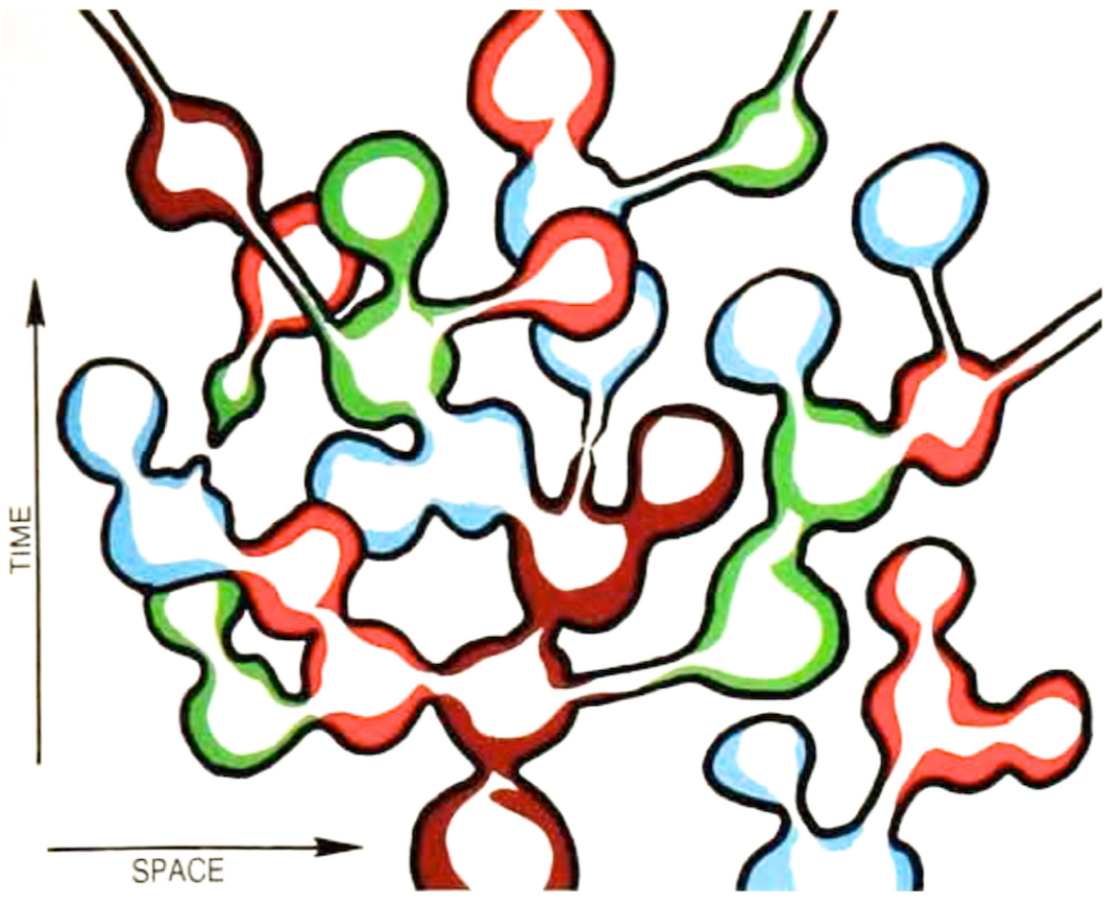}
    \caption{Left panel: a cartoon Euclidean space ``boiling" in Pancosmic General Relativity. Each monochromatic pastille is a universe with a locally
    fixed Planck scale, cosmological constant and the QFT parameters. Those change from one pastille to another. Right panel: 
    a cartoon of a borough of a multiverse of eternal inflation \cite{Linde:2015edk} (in Lorentzian signature).}  
     \label{cookies}
\end{figure}

From this viewpoint, the gravitational field is treated purely classically, and the 
membrane charges and tensions are chosen to ensure that the relevant semiclassical dynamics stays well below the
local Planckian cutoff. Thus the theory remains within its domain of validity, and the only quantum effect
is the process of changing the spacetime geometry by membrane discharge/bubble nucleation. The solutions are depicted 
in Fig. (\ref{cookies}). 

For comparison, we also include a depiction of the multiverse of eternal inflation from 
\cite{Linde:2015edk}. The pictorial depictions, however cartoonish, invite the analogy between the membrane walls 
in the left panel and the wormholes connecting various `baby' universes in the right panel. 

The main bonus of our approach is the simplicity of describing the transitions, since we `separate' the membranes
and the spacetimes they link from the quandaries of full blown quantum gravity. 
In fact we may take an attitude that whatever quantum gravity might be, it still needs 
to obey decoupling to reproduce the classical limit. In this case we could be agnostic about it and consider the
bubbles of spacetime bounded by membranes as at least a reasonable toy model of the deeper theory -- be it a 
theory of spacetime foam \cite{Wheeler:1955zz,Hawking:1978pog}, wormholes 
\cite{Banks:1984cw,Coleman:1988tj,Giddings:1988wv,Fischler:1988ia,Fischler:1989ka,Polchinski:1989ae} 
or whatever else. 

But at least at this level of calculations, we don't have to contend with the problems the deeper 
formulations entail. We have our semiclassical vacua, they are described by the solutions of field equations, which are 
well within their domain of validity, and they are interpreted as a leading order description of quantum transitions
in the spacetime. So we can compare them and count them (at least, schematically).

\subsection{How to Solve the Cosmological Constant Problem}

The discussion in the previous sections showed clearly that in our framework, de Sitter space is unstable. 
Once the cosmological constant is positive, and membranes are present, the bubble nucleation in the 
parent geometry is inevitable. Inside the bubbles -- on the average -- the cosmological constant will be reduced. Thus 
global de Sitter spaces cannot exist. They ``decay" by the discharge of the cosmological constant. 
Subsequently, at least in the case when $q<1$, as $\Lambda$ decreases, the production rate of a single 
bubble slows, and it completely ceases 
for $\Lambda \rightarrow 0$. Note that an initially large de Sitter (with small $\Lambda$) might also decay into 
many other de Sitter spaces with smaller cosmological constant more efficiently. It depends on the process, the channel, 
and the initial condition. But the end result is the trend toward $\Lambda \rightarrow 0$.

This is good news, given the lore that the presence of event horizons, which are unavoidable in eternal de Sitter, 
obstructs the formulations of QFT in de Sitter space. 
It still remains unclear how to define asymptotic free states and the scattering S-matrix in eternal de Sitter geometry 
\cite{Banks:2001yp,Witten:2001kn,Goheer:2002vf,Dvali:2017eba}. 

It is then natural to ask if the instability of de Sitter 
space offers a path for solving the cosmological constant problem.
The usual formulation of the cosmological constant problem in standard General Relativity
is that
\begin{itemize}
\item 1) QFT vacuum energy contributions are big, of the order of the cutoff ${\cal M}_{\tt UV}^4$, and 
\item 2) unless they are
cancelled, order by order in perturbation theory, 
\item 3) the resulting cosmological constant will be huge and {\it eternal}. 
\end{itemize}
The cancellation involves a counterterm whose value must be precisely arranged 
to one part in as much as $\sim 10^{120}$, which in the absence of a symmetry can only be
done by fine tuning \cite{Zeldovich:1967gd,Wilczek:1983as,Weinberg:1987dv}. 
Thus the problem: the conclusion conflicts with the observations in the absence of nearly infinite fine tuning. 
This obviously cannot stand in our generalization of General Relativity, since membranes catalyze the decay
of the cosmological constant source, making it it merely `almost-constant' at best -- but not eternal. 

However in its simplest form our theory does not yet have the capability to solve 
the cosmological constant problem naturally even if we choose a small charge to tension ratio $q<1$. Briefly
the reason is the following: with our conformal $4$-form/matter coupling, the total cosmological 
constant is (see Eqs. (\ref{cftcc}), (\ref{ccgen})) 
\be
\Lambda_{\tt total} = \kappa^2 \Bigl( \frac{{\cal M}_{\tt UV}^4}{{\cal M}^2} + \frac{V}{{\cal M}^2} + \lambda \Bigr) \, ,
\label{totcc}
\ee
where we have now included the QFT UV contributions $\sim {\cal M}_{\tt UV}^4 + \ldots$, any nonvanishing QFT (or inflaton) 
potential $\sim V$, as well as our dynamical contribution $\sim \lambda$.
Furthermore, the actual physical observable is the effective curvature of the background geometry, which we
can define by Friedmann equation,
\be
H^2 = \frac{\kappa^2}{3 \kappa_{\tt eff}^2} \Lambda_{\tt total} =  \frac{\kappa^2}{3 \kappa_{\tt eff}^2}
\Bigl(\frac{{\cal M}_{\tt UV}^4}{{\cal M}^2} + \frac{V}{{\cal M}^2}+ \lambda \Bigr) \, .
\label{totH}
\ee
and here $\kappa^2_{\tt eff} = \mps + \kappa^2$. The variables $\lambda$ and $\kappa^2$ change discretely 
(\ref{lagrangejc}), $\Delta \lambda = {\cal Q}_A/2$, $\Delta \kappa^2 = 2{\cal Q}_B$, which means that we can
write them as
\be
\lambda = \lambda_0 + N \frac{{\cal Q}_A}{2} \, , ~~~~~~~~~~~~~~~  \kappa^2 = \kappa^2_0 + 2 {\cal N} {\cal Q}_B \, ,
\label{lkquant}
\ee
where $N$ and ${\cal N}$ are two integers. 

Note that in \cite{Bousso:2000xa} the $4$-form fluxes screening the
cosmological constant were argued to be quantized in the units of charge, amounting to setting the terms analogous
to our $\lambda_0$ and $\kappa^2_0$ to zero. We do not have any direct reasons to do so here. We could do it 
without loss of generality by absorbing those terms into $\frac{{\cal M}_{\tt UV}^4}{{\cal M}^2}$ and
$\mps$, respectively. However we will keep them here explicitly, since their presence does not affect the argument. 

The values of the cosmological constant term and the curvature 
in some ``ancient parent" geometry can be written as 
\be
\Lambda_{\tt total} = \bigl( \kappa^2_0 + 2 {\cal N} {\cal Q}_B \bigr)\Bigl( \frac{\Lambda_0}{{\cal M}^2} + N \frac{{\cal Q}_A}{2} \Bigr) \, ,
~~~ H^2 = \frac{\kappa_0^2 +  2 {\cal N} {\cal Q}_B}{3 (\mps + \kappa^2_0 + 2 {\cal N} {\cal Q}_B)}
\Bigl(\frac{\Lambda_0}{{\cal M}^2} + N \frac{{\cal Q}_A}{2}  \Bigr) \, ,
\label{totccquant}
\ee
where $\Lambda_0 = {\cal M}_{\tt UV}^4 + V + {\cal M}^2 \lambda_0$. Now, through a sequence of membrane
emissions, the system can change both $\kappa^2_{\tt eff}$ and $\Lambda_{\tt total}$, by gradually changing 
$N$ and ${\cal N}$, up or down, until $\Lambda_{\tt total}/\kappa^4_{\tt eff}$ approaches zero as close as it can,
given the initially fixed $\Lambda_0/{\cal M}^2$ and the initial values of $N,  {\cal N}$. In light of our discussion above,
this will predominantly occur by a change of $N$. 

This brings into the forefront the deficiency of the theory, as it stands at this point. The cosmological constant changes only in 
discrete steps $\Delta \Lambda_{\tt total} = \bigl( \kappa^2_0 + 2 {\cal N} {\cal Q}_B \bigr){{\cal Q}_A}/{2}$. To make 
$\Lambda_{\tt total}/\kappa^4_{\tt eff} < 10^{-120}$, we must either fine tune $\bigl( \kappa^2_0 + 2 {\cal N} {\cal Q}_B \bigr){\Lambda_0}/{{\cal M}^2}$,
or pick an absolutely tiny value for ${\cal Q}_A$, and deal with huge fluxes in the units of membrane charges. 
The fact that $\kappa^2_{\tt eff}$ can also vary does not help, since we can't suppress the curvature of the universe
without simultaneously tremendously suppressing the force between two 
hydrogen atoms, or a Sun and a planet. In a 
sense, this is the avatar of the cosmological 
constant `no go' by Weinberg, in this context \cite{Weinberg:1987dv,Kaloper:2014dqa}. 

\begin{figure}[htb]
    \centering
    \includegraphics[width=9cm]{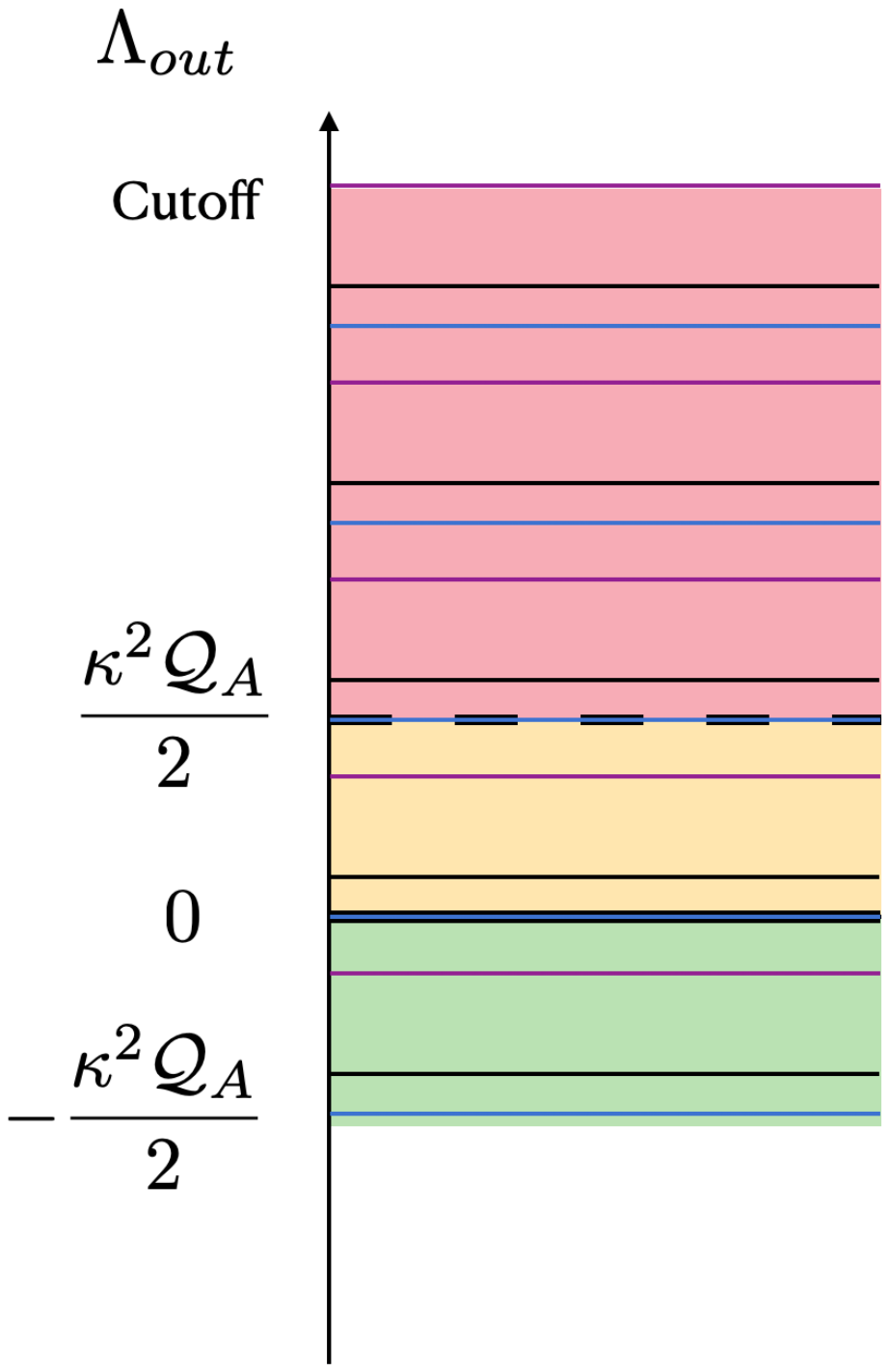}
    \caption{The cosmological constant spectrum with many superselection sectors, depicted by differently colored spectral levels.
    The `blue' spectrum is tuned to get close to $\Lambda \simeq 0$, the `black' and `purple' spectra are not.}
    \label{spectrumdisc}
\end{figure}

The obstruction we are encountering here is that the theory we have studied so far has cosmological constant values which fill out the painted bands
of the spectrum in Fig. (\ref{spectrumpic}) discretely, with fixed finite gaps between the levels. 
The theory splinters into infinitely many ``superselection sectors" 
parameterized by the ``initial value" $\Lambda_0$, which follows from the discrete variation of $\Lambda$. This is depicted in Fig. (\ref{spectrumdisc}),
where the set of bands of the same color belong to the same superselection sector. Unless ${\cal Q}_A$ is extremely small,
the ``terminal" value of the cosmological constant will be in the observationally allowed window only in special superselection
sectors with finely tuned ``initial" vacuum energies. 

This problem is a straightforward one to resolve, however. We simply add to the theory one more $4$-form, and arrange for it 
such that its magnetic dual is degenerate on shell with $\lambda$ in the bulk. It nevertheless couples to a different 
membrane with the tension and charge ${\cal T}_{\hat A}, {\cal Q}_{\hat A}$. Concretely, we take the action
of Eq. (\ref{actionnewmemdconf}), and extend it to 
\ba
{\cal S} = S - \int d^4x \sqrt{g} \Bigl(\kappa^2 \hat \lambda 
+ \frac{\hat \lambda}{3} {\epsilon^{\mu\nu\lambda\sigma}} \partial_\mu \hat {\cal A}_{\nu\lambda\sigma} \Bigr) 
- {\cal T}_{\hat A} \int d^3 \xi \sqrt{\gamma}_{\hat A} - {\cal Q}_{\hat A} \int \hat {\cal A}  \, . 
\label{actionnewext}
\ea
The system of fluxes and membranes $\hat A$ behaves exactly as the system $A$, and all of the analysis to this point which
we carried our for the $A$-sector dynamics applies to $\hat A$. In particular we demand $\hat q<1$, just as $q<1$. 
With the new sector included, however, the cosmological constant and the
curvature `quantization' laws (\ref{totccquant}) are now generalized to 
\ba
&&~~~\Lambda_{\tt total} = \bigl( \kappa_0^2 + 2 {\cal N} {\cal Q}_B \bigr)\Bigl( \frac{\Lambda_0}{{\cal M}^2} + N \frac{{\cal Q}_A}{2} + 
\hat N \frac{{\cal Q}_{\hat A}}{2} \Bigr) \, , \nonumber \\
&&H^2 = \frac{\kappa_0^2 +  2 {\cal N} {\cal Q}_B}{3 (\mps + \kappa^2_0 + 2 {\cal N} {\cal Q}_B)}
\Bigl(\frac{\Lambda_0}{{\cal M}^2} + N \frac{{\cal Q}_A}{2} + \hat N \frac{{\cal Q}_{\hat A}}{2}\Bigr) \, .
\label{totyccquant}
\ea

Now we borrow a trick from the irrational axion proposal \cite{Banks:1991mb} (see also \cite{Kaloper:2018kma}) 
and take the ratio of the charges ${\cal Q}_A$ and ${\cal Q}_{\hat A}$ to be an irrational number $\omega$,
\be
\frac{{\cal Q}_{\hat A}}{{\cal Q}_A} = \omega \, . 
\label{irrcharge}
\ee
We can then rewrite the top line of (\ref{totyccquant}) as
\be
\Lambda_{\tt total} = \bigl( \kappa_0^2 + 2 {\cal N} {\cal Q}_B \bigr)\Bigl( \frac{\Lambda_0}{{\cal M}^2} + \frac{{\cal Q}_A}{2} \bigl( N + 
\hat N \omega \bigr) \Bigr) \, . 
\ee
Because $\omega$ is not rational, it is straightforward to show that for any real number $\rho$, there exist integers $N, \hat N$
such that $N + \hat N \omega$ is arbitrarily close to $\rho$ \cite{Banks:1991mb,niven}. Therefore, there do exist 
integers $N, \hat N$ such that $N + \hat N \omega$ is arbitrarily close to $- \frac{2\Lambda_0}{{\cal Q}_A {\cal M}^2}$, for which
$\Lambda_{total}$ is arbitrarily close to zero! 

Crucially, this means, that there is a `discharge path' from any large value of $\Lambda$
to an arbitrarily small terminal value. Starting with any `initial value' of $\Lambda$, there exists a 
sequence of membrane discharges (with successive emission of positive or negative charges
${\cal Q}_A$ or ${\cal Q}_{\hat A}$, whichever it takes), whose end result yields a $\Lambda$ arbitrarily close to zero. 
This process will continue for as long as $\Lambda > 0$ at any intermediate charge. 

If a discharge in a sequence 
overshoots to $\Lambda < 0$, the sequence will stop. But if it comes close to zero, but $\Lambda$ is still positive, 
the evolution can always continue by an up-jump, with subsequent 
discharges bringing the later value of $\Lambda$ even closer to zero. Further, {\it \`a priori}, for any pair of $N, {\cal N}$ that lead to a tiny 
$\Lambda$, there is actually a very large number of degenerate `discharge paths': any 
order of discharges of ${\cal Q}_A$ and ${\cal Q}_{\hat A}$ which adjust $N$ and $\hat N$ to the correct 
terminal values, which bring $\Lambda_{\tt total}$ to the required terminal $\Lambda$, will produce 
the same answer irrespective of how the individual steps occur. The relaxation is Brownian drift, rather than
classical smooth evolution.   

At each step $\Lambda$ changes by 
$\Delta \Lambda \sim \kappa^2_{\tt eff} {\cal Q}_A$ or $\sim \kappa^2_{\tt eff} {\cal Q}_{\hat A}$, i.e. 
by a large value. Its small terminal value is achieved as a sum total of many such 
processes, due to the fact that $\omega$ of Eq. (\ref{irrcharge}) is
irrational. Finally note that in this case we are not using 
a scalar field which is `gauging' such irrationally discrete shifts
and so there is no danger of emerging global shift symmetries lurking around, 
a concern which was expressed in the
context of irrational axion \cite{Banks:1991mb,banksseiberg}. 

Instead, what has happened here is that the new charge sector $\hat A$, due to the irrational ratio
of charges (\ref{irrcharge}), has in fact mixed up {\it all} the previously separated 
superselection sectors depicted in Fig. (\ref{spectrumdisc}). They all mix now, 
transitioning between each other by utilizing both $A, \hat A$ charges. Since 
the nucleation processes are slow when 
$\Lambda$ slips well below the cutoff, the up-jumps which raise $\Lambda$ can also happen, and the superselection sectors will generically 
get shaken and stirred together into a very fine discretuum mesh, filling out the 
spectral bands in Fig. (\ref{spectrumpic}) densely. In particular there will be many states
with $\Lambda \simeq 0$. And also, with $\Lambda \simeq 10^{-120} \mpl^4$. They will be very long 
lived - the smaller the $\Lambda$, the more persistent the geometry. Ultimately, the trend for all the states with $\Lambda > 0$ to  
decay will remain (albeit {\it slowly}, when $q<1$ and $\hat q<1$, and using up-jumps occasionally). 

These stability arguments favor the value of $\Lambda = 0$. This results as a dynamical trend, where evolution
of an initial de Sitter via the discharge mediated by $(-+)$ instantons targets the attractor $\Lambda \rightarrow 0^+$,
precisely as indicated by the Euclidean partition function arguments. Indeed, we can consider
(\ref{partf}), or better yet, it's Euclidean magnetic dual 
\be
Z = \int \ldots {\cal DA} {\cal D} \hat {\cal A}  {\cal DB} {\cal D} \lambda {\cal D} \hat \lambda {\cal D} \kappa^2 {\cal D} g \ldots \,  e^{-{\cal S}_E} \, ,
\ee
which in the semiclassical, saddle point approximation reduces to
\be
Z = \sum_{\lambda, \kappa^2}  \int \ldots  {\cal D} g \ldots \, e^{-{\cal S}_E} \, .
\ee
The saddle point approximation implies that we sum over all classical configurations extremizing the action, which in our case
begins with summing over all Euclidean instantons with any number of membranes included, as long as 
they are allowed by Euclidean field equations which extremize the action (\ref{actionnewmemeu}). The $O(4)$ invariant
solutions should minimize the action, and so it seems this is a reasonable leading order approximation. 
Thus $Z$ is dominated by our instantons, 
\be
Z = \sum_{instantons} \sum_{\lambda, \hat \lambda, \kappa^2}  e^{-{\cal S}_E(instanton)} \, .
\label{sumZ}
\ee

Even at the cartoonish level, handling this sum is challenging. Summing over instantons means picking all the possible configurations with 
an arbitrary number of membranes included, and taking into account that both $A, \hat A$ processes contribute, which allows 
for a very fine structure of $\lambda, \hat \lambda, \kappa^2$ ranges of summation. Further, one needs to account for possible degeneracies
of a particular instanton configuration which includes different `discharge paths' as we noted above, as well as the possibility that some of the 
apparently different configurations are gauge transformations of those already included. Performing this sum is beyond the scope of 
this work. 

We can however get a feel for the individual terms in the sum. These terms reflect the evolution via membrane discharges. 
The individual terms -- 
representing ancestry trees of the evolution -- can be estimated using the definition of the bounce
action in Eq. (\ref{bounceact}), converting it to
\be
{\cal S}({\tt instanton}) = {\cal S}({\tt bounce}) + {\cal S}({\tt parent}) \, .
\label{instact}
\ee
If there is no offspring, the instanton action is given by the parent action, which is just the negative of the horizon area 
divided by $4G_N$ of the parent
de Sitter, 
\be
{\cal S}({\tt parent}) = - 24 \pi^2 \frac{\kappa^4_{\tt eff}}{\Lambda_{out}} \, . 
\ee
If the offspring is $n^{th}$ generation, 
we'd end up summing over the family tree, which we can try to approximate by imagining a `dilute gas' of membranes, 
added one by one as the matching conditions permit it,
\be
{\cal S}({\tt instanton}) = \sum_n {\cal S}({\tt offspring}, n) + {\cal S}({\tt progenitor}) \, ,
\ee
using successive iterations. The ``offspring" here refers to the geometric segments inside 
nested bubbles separated by the membranes. 
The ``progenitor" geometry is the primordial parent 
initiating the corresponding family tree. Note that the progeny can in principle 
be produced at the same Lorentzian time, as multiple membranes, but more importantly 
as a time ordered sequence of consecutive nucleations. 

In any case, the trees initiated by progenitors with any initial $\Lambda$ will evolve, by decreasing $\Lambda$ on the average,
as long as nucleations are possible. 
As per, e.g. Eq. (\ref{familiars2}), for a tree with two generations only,
${\cal S}({\tt instanton}) \simeq - 64 \pi^2 {\kappa_{\tt eff}^6}/{\cal T}_A^2$. When the offspring
cosmological constant is still large, another transition can happen, and so on,
with ${\cal S}({\tt instanton})$ growing, approximately by an amount of 
$\simeq - 64 \pi^2 {\kappa_{\tt eff}^6}/{\cal T}_j^2$ per step. This indicates an estimate for a family tree action, 
\be
{\cal S}({\tt instanton}) \rightarrow - 64 \pi^2 {\kappa_{\tt eff}^6} \Bigl(\frac{n_A}{ {\cal T}_A^2} + \frac{n_{\hat A}}{{\cal T}_{\hat A}^2} \Bigr) \, , 
\ee
which is bounded by $- 24 \pi^2 \frac{\kappa^4_{\tt eff}}{\Lambda}_{\tt terminal}$, 
for a terminal $\Lambda_{\tt terminal} \ga 0$, because membrane nucleations slow down, but can go on until 
$\Lambda_{\tt terminal} \rightarrow 0^+$. This implies that the sum (\ref{sumZ}) 
\be
Z \sim \sum e^{24 \pi^2 \frac{\kappa^4_{\tt eff}}{\Lambda} + \ldots} \, \, , 
\label{finsum}
\ee
will be heavily skewed toward small values of $\Lambda$. 
The emerging exponential bias may only benefit further from the 
degeneracies of specific instanton configurations which we noted above. 

Thus, the essential singularity of the bounce action at $\frac{\Lambda}{\kappa^4_{\tt eff}} \rightarrow 0^+ $ and the 
partition function behavior indeed conform with the dynamical trend that $\Lambda \rightarrow 0^+$ is an attractor, at least in 
the saddle point approximation, in full agreement with the discharge dynamics processes catalyzed by $(-+)$ instantons. 
We infer that the dynamics to leading
order in the saddle point approximation heavily prefers
\be
\frac{\Lambda}{\kappa^4_{\tt eff}} \rightarrow 0  \, .
\label{esssing}
\ee

It is difficult to see this outcome as anything but enticing and intriguing, to say the least. 
In our generalization of General Relativity, de Sitter is unstable. Quantum mechanics and
relativity prefer a huge hierarchy between $\kappa^2_{\tt eff}$ and the expected value of $\Lambda$. 
The terminal value of $\Lambda$ will be arbitrarily close to zero. Finally, as $\Lambda \rightarrow 0$, the 
processes cease and the resulting (near) Minkowski space is at least extremely long lived. 
This looks like a good approximation of reality. 

As this argument goes, we still need to explain the observed strength of gravity, with 
$G_N = \frac{1}{8\pi \mps} \simeq 10^{-38} \, ({\rm GeV})^{-2}$. Maybe this is 
really simply a lucky break. Alternatively, maybe we should interpret it as 
a manifestation of the `Weak Anthropic Principle'. 
If we fix chemistry, it does seem that this ensures
that our Earth is in the habitable zone in the Solar system, neither charred nor frozen, 
allowing us to ponder the problem. 
 
\section{Implications}

So if the cosmological constant is, most likely, extremely tiny compared to $\mpl^4$, why is the universe accelerating now? 
If the spacetime has been bubbling forever, there exist regions where cosmological constant is $10^{-120} \mpl^4$ in the
framework with the irrational ratio of charges. However they may not be 
typical, if Euclidean partition function is any indication of the likelihood of a value of $\Lambda$, strongly favoring 
$\Lambda \rightarrow 0$. In this context it also seems unlikely that anthropic
argument can help since the sum (\ref{finsum}) has an essential singularity at (\ref{esssing}) \cite{Hawking:1981gd}. 
Even in the context of string landscape it has been argued to be nontrivial to devise a weighting of probabilities which
allows the anthropic reasoning to produce the desired result of anthropic selection of $\Lambda/\mpl^4 \sim 10^{-120}$ 
\cite{Schwartz-Perlov:2006swo}. 

Among the possible options which might explain 
the current acceleration might be
\begin{itemize}
\item a blip of transient quintessence\footnote{This feels like a copout, but at least now it's out there. Maybe it is true.};
\item a late stage phase transition; perhaps the ``real" cosmological constant was cancelled early on,
but then a late phase transition in some gauge theory -- e.g. QCD -- occurred, leading to a nontrivial
vacuum structure thanks to gauge theory topology \cite{Weiss:1987ns}; 
this could lead to a cosmological constant
induced by a phase transition at late times, with values scanned by the vacuum $\theta$ parameter, 
and the terminal value selection might even be anthropic (sic!) \cite{Kaloper:2017fsa};
\item the ratio of charges ${\cal Q}_{\hat A}/{\cal Q}_A$ is rational, 
but it is a fraction of two very large\footnote{This is needed in order for the terminal value of cosmological constant to be close to zero; if the two mutual primes were comparable,
the theory might not even have an attractor with positive cosmological constant, since the possible values of the positive cosmological constant would be too large,
and the corresponding space-times too short lived. The only long-lived values of the cosmological 
constant would be negative.} mutually prime numbers; if
so there would be a state, which could be metastable and have a very small cosmological constant;
\item our accelerating universe seems atypical by the $Z$ counting, but 
may be typical by some other measure \cite{Garriga:2000cv}, 
which might have to do with inflation \cite{Linde:1991sk,Garriga:2000cv} and/or processes which set up 
``the initial state" \cite{Linde:1991sk} $\ldots$
\item $\ldots$
\end{itemize}
As interesting and urgent as it may be, answering this question more precisely, 
we fear, is beyond the scope of the present work.

Another question concerns the problem of the so called ``empty universe" \cite{Abbott:1984qf},
which may be an issue if the discharge of cosmological constant is slow and occurs 
in many extremely small steps. Or, by a classical slow roll. The 
end point will be an empty cold universe which has been dominated by cosmological constant throughout its history. 
Such a universe would be a barren wasteland because anything in it would be inflated away before
it had any chance to make its mark. However this may not be a problem in our case since the 
relaxation of the cosmological constant occurs in steps where $\Lambda$ changes by large amounts in each successive
step. Yet the end point is favored to be a local `vacuum'
with the final net $\Lambda$ much smaller than any of the individual charges. The terminal $\Lambda$ 
cancellation 
arises as a sum total of the sequence of emissions of charged membranes, with irrational ratio, and with the 
final result which is effectively weighted by $Z$ as $\Lambda \rightarrow 0$, due to an essential singularity of $Z$ there, 
rather than by a smooth gradual evolution. 
Thus the cosmological constant relaxation does not require the eternal cosmological constant domination 
on its path to zero. This is similar to how the empty universe problem
is avoided in \cite{Bousso:2000xa}. Basically, small $\Lambda$ is attained by Brownian drift, with the terminal 
value being a `mean' of many large jumps, instead of 
smooth evolution. 

Furthermore, since the up-jumps are also possible, it can happen that an empty 
universe with a nearly vanishing $\Lambda$ can 
`restart' itself by a rare quantum jump which increases the cosmological constant, and 
then in subsequent evolution back to $\Lambda \rightarrow 0$ 
an inflationary stage is stumbled upon \cite{Garriga:2000cv}. In this approach, 
inflation might seem to be {\it \`a priori} rare, 
but since the system can continue exploring the phase space, even a `rare' event will be found 
eventually \cite{Carroll:2004pn}. It has been noted that our universe may have been 
preceded by one such up-jump, but then it evolved to $\Lambda \rightarrow 0$. 
This can avoid potential problems with more likely smaller scale 
fluctuations dubbed `Boltzmann Brains' \cite{Carroll:2004pn,Bousso:2008hz,DeSimone:2008if,Susskind:2014rva}. 
Thus it appears that a conventional cosmology can be embedded in our
framework. 

It is clearly interesting to consider specific predictions and implications for observations \cite{Freivogel:2005vv}, 
among which might be a past record of colliding with 
other bubbleworlds \cite{kleban,aguirre}, applications to particle physics hierarchies, and maybe even late time variations
of cosmological parameters (leading to a fractal cosmology \cite{fractal}?), such as $H_0$ and/or the masses of particles. 
We will return to these issues at another time.

\section{Summary} 

In closing, our analysis in this article shows that we can view the standard formulation of General Relativity based 
on Einstein-Hilbert action \cite{Hilbert:1915tx,Einstein:1915ca} as a restriction of a much bigger theory to a single (huge) domain
of spacetime. The generalization is obtained by promoting dimensional parameters in the gravitational sector to
magnetic duals of $4$-forms and the introduction of membranes charged under those forms. Quantum-mechanically this
allows for the variation of the gravitational parameters by membrane emission. Thus, ordinary General Relativity is  
a restriction of Pancosmic Relativity to the confines of a single bubble in the multiverse. This implies that the multiverse was lurking 
over the shoulder of General Relativity all along, hiding in plain view. Perhaps this has already been divined in the formulation of the 
theory of eternal inflation\cite{Linde:1986fc,Guth:2007ng}. Our description of this multiverse might be even more basic. 

Finally, we can't resist drawing an analogy between our generalization of General Relativity, which we established here, 
and fluid flow. Consider fluid flow. At small Reynolds numbers it will be laminar, with each fluid streamline smoothly passing by each neighbor
streamline, without intersecting each other. As the Reynolds number goes up, being dialed by an external influence, 
the flow will turn turbulent, with the stream lines intersecting, breaking up, twisting around and mixing together. 

In some sense, we might think of Pancosmic General Relativity in this way. If we fix the 
gravitational ``couplings" $\lambda$ and $\kappa^2$, the full evolution of the geometry with a fixed matter contents is analogous
to a single laminar flow streamline. If we then dial $\lambda$ and $\kappa^2$ by hand, we move from one streamline to another, while they
remain separated. However when we turn on the membrane dynamics, the ``streamlines of geometry" start mixing up and transitioning from
one to another, just like they do in turbulent flow. There is no sense of stability in this regime, and certainly there is no global de Sitter anymore. 
The `fluid' will froth and bubble as long as it is kept in a small space, with a large Reynolds number, or a large cosmological constant. Reducing it 
may eventually restore laminar flow again, by for example allowing the fluid to flow into a larger vessel, or discharging the cosmological
constant to zero, making the resulting universe huge. 

Making this analogy sounds quite fantastic even to us. But given the ideas in, e.g. 
\cite{Jacobson:1995ab,Verlinde:2010hp,Jacobson:2018ahi,Jacobson:2019gco}, maybe it is not.

\vskip.5cm

{\bf Acknowledgments}: 
We would like to thank G. D'Amico and A. Westphal for valuable comments and discussions. 
NK is supported in part by the DOE Grant DE-SC0009999.

\end{document}